\documentclass[usegraphicx,usenatbib]{mn2e}

\begin{document}

\voffset -0.5in


\title[Interstellar Ti\,{\normalsize\it II}]{Interstellar Ti\,{\Large\bf II} in the Milky Way and Magellanic Clouds\thanks{Based in part on observations collected at the European Southern Observatory, Chile, under programmes 65.I-0526, 67.C-0281, 67.D-0238, 70.D-0164, 72.C-0682, and 74.D-0109.  Based in part on observations made with the NASA/ESA {\it Hubble Space Telescope}, obtained from the data archive at the Space Telescope Science Institute.  STScI is operated by the Association of Universities for Research in Astronomy, Inc.. under NASA contract NAS5-26555.}}

\author[Welty \& Crowther]{Daniel E. Welty$^1$\thanks{Visiting observer, European Southern Observatory and Kitt Peak National Observatory} and Paul A. Crowther$^2$ \\
$^1$University of Illinois at Urbana/Champaign, Dept. of Astronomy, 1002 W. Green St., Urbana, IL 61801, USA; dwelty@astro.illinois.edu \\
$^2$University of Sheffield, Dept. of Physics and Astronomy, Hicks Building, Hounsfield Rd., Sheffield 53 7RH, UK; Paul.crowther@sheffield.ac.uk}

\maketitle
\begin{abstract}
We discuss several sets of \mbox{Ti\,{\sc ii}} absorption-line data, which probe a variety of interstellar environments in our Galaxy and in the Magellanic Clouds.
Comparisons of high-resolution (FWHM $\sim$ 1.3--1.5 km~s$^{-1}$) \mbox{Ti\,{\sc ii}} spectra of Galactic targets with corresponding high-resolution spectra of \mbox{Na\,{\sc i}}, \mbox{K\,{\sc i}}, and \mbox{Ca\,{\sc ii}} reveal both similarities and differences in the detailed structure of the absorption-line profiles -- reflecting component-to-component differences in the ionization and depletion behaviour of those species.
Moderate-resolution (FWHM $\sim$ 3.4--4.5 km~s$^{-1}$) spectra of more heavily reddened Galactic stars provide more extensive information on the titanium depletion in colder, denser clouds -- where more than 99.9 per cent of the Ti may be in the dust phase.
Moderate-resolution (FWHM $\sim$ 4.5--8.7 km~s$^{-1}$) spectra of stars in the Magellanic Clouds suggest that the titanium depletion is generally much less severe in the LMC and SMC than in our Galaxy [for a given $N$(H$_{\rm tot}$), $E(B-V)$, or molecular fraction $f$(H$_2$)] -- providing additional evidence for differences in depletion patterns in those two lower-metallicity galaxies.
We briefly discuss possible implications of these results for the interpretation of gas-phase abundances in QSO absorption-line systems and of variations in the D/H ratio in the local Galactic ISM.
\end{abstract}


\section{Introduction}
\label{sec-intro}

Singly ionized titanium is unique among the small number of atomic species whose interstellar absorption features can be observed from the ground, in that its ionization potential is nearly coincident with that of neutral hydrogen -- so that \mbox{Ti\,{\sc ii}} is a dominant ion in \mbox{H\,{\sc i}} regions, with negligible contributions from ionized (\mbox{H\,{\sc ii}}) gas.
The column density ratio $N$(\mbox{Ti\,{\sc ii}})/$N$(H$_{\rm tot}$) -- where H$_{\rm tot}$ includes both neutral atomic and molecular hydrogen -- thus gives a direct measure of the gas phase interstellar abundance of titanium.
In principle, it is also fairly straightforward to obtain accurate column densities for \mbox{Ti\,{\sc ii}}.
Several lines of \mbox{Ti\,{\sc ii}}, with a range in strength, are available between 3060 and 3390 \AA\ -- so that any effects of saturation in the strongest line at 3383 \AA\ may be estimated.
Blending with stellar \mbox{Ti\,{\sc ii}} absorption features is not an issue for spectral types earlier than about B7 (Hobbs 1984). 

The earliest surveys of \mbox{Ti\,{\sc ii}} absorption in the relatively nearby Galactic ISM indicated that even the $\lambda$3383 line is often fairly weak, and that the gas phase titanium abundance is significantly sub-solar in most sight lines (Wallerstein \& Goldsmith 1974; Stokes 1978). 
Most of the titanium in the Galactic ISM appears to be depleted into dust grains, with less than 1 percent of the total (assumed solar) titanium in the gas in many sight lines.
The $N$(\mbox{Ti\,{\sc ii}})/$N$(H$_{\rm tot}$) ratio thus appears to be an excellent indicator of the overall level of depletion in a given line of sight -- at least in the local Galactic ISM (e.g., Jenkins 2004, 2009).
Because the \mbox{Ti\,{\sc ii}} lines are both weak (typically) and located in the near-UV, most subsequent surveys of interstellar \mbox{Ti\,{\sc ii}} have been undertaken at moderate spectral resolution (3--6 km~s$^{-1}$) toward relatively bright and generally lightly reddened background target stars (Albert 1983; Albert et al. 1993; Welsh et al. 1997); some more reddened targets are included in the recent survey by Hunter et al. (2006).

Comparisons of the absorption-line profiles and column densities of \mbox{Ti\,{\sc ii}} with those of \mbox{Na\,{\sc i}} and \mbox{Ca\,{\sc ii}} and examination of correlations with distance (to the target stars) have suggested that \mbox{Ti\,{\sc ii}} is distributed more similarly to \mbox{Ca\,{\sc ii}}, with both species primarily tracing the warmer, more smoothly distributed, more diffuse neutral component of the ISM (Edgar \& Savage 1989; Crinklaw, Federman, \& Joseph 1994).
Overall, the ratio $N$(\mbox{Ti\,{\sc ii}})/$N$(\mbox{Ca\,{\sc ii}}) appears to be relatively constant, with a value of 0.3--0.4 (Albert 1983; Hunter et al. 2006), even though both Ti and Ca can exhibit wide variations in depletion and even though \mbox{Ca\,{\sc ii}} can be either a dominant or a trace species in predominantly neutral regions.
The distributions of both \mbox{Ti\,{\sc ii}} and \mbox{Ca\,{\sc ii}} are characterized by scale heights greater than 1000 pc perpendicular to the Galactic disc -- significantly larger than the roughly 200 pc scale height of \mbox{H\,{\sc i}} (Albert 1983; Edgar \& Savage 1989; Albert et al. 1993; Albert, Welsh, \& Danly 1994; Lipman \& Pettini 1995).
Titanium is thus less severely depleted, by a factor of order 4, in the low Galactic halo at 250 pc $\la$ $|$z$|$ $\la$ 1000 pc, and is further enhanced in the gas phase in components at higher LSR velocities (Albert 1983).

Given the potential utility of \mbox{Ti\,{\sc ii}} as a depletion indicator, observations of \mbox{Ti\,{\sc ii}} have been employed in recent studies ranging from attempts to understand variations in the D/H ratio in the local ISM to interpretations of the gas phase abundances in QSO absorption-line systems.
One possible explanation for the local variations in D/H is that deuterium could be preferentially depleted into dust grains (Jura 1982; Tielens 1983; Draine 2006), in which case one would expect correlations between D/H and the depletions of various refractory elements, such as Ti (Prochaska, Tripp, \& Howk 2005; Ellison, Prochaska, \& Lopez 2007; Lallement, H\'{e}brard, \& Welsh 2008) or Si and Fe (Linsky et al. 2006).
While there does appear to be a correlation between the titanium depletion $\delta$(Ti) and D/H, the correlation is not as strong and the slope of the relationship is not as steep as might have been expected from the analogous relations seen for $\delta$(Si) and $\delta$(Fe) (Ellison et al. 2007).
Component-to-component differences between the depletions of Ti and D (Prochaska et al. 2005) may be a factor.

The gas phase abundances found for QSO absorption-line systems must reflect both the nucleosynthetic history of the absorbers (e.g., the relative contributions of Type I and Type II supernovae in enriching the ISM) and any (selective) depletion of various elements into dust -- but disentangling the effects of those two factors is often difficult (e.g., Lauroesch et al. 1996; Lu et al. 1996; Prochaska \& Wolfe 2002; Vladilo 2002; Dessauges-Zavadsky et al. 2004).
In principle, comparisons between \mbox{Ti\,{\sc ii}} and \mbox{Fe\,{\sc ii}} may help to separate the two effects, as Ti (an $\alpha$-element) is overabundant, relative to Fe and solar abundance ratios, in Galactic metal-poor stars, but is typically more severely depleted than Fe in the Galactic ISM (e.g., Dessauges-Zavadsky et al. 2002; Ledoux, Bergeron, \& Petitjean 2002).
A super-solar Ti/Fe ratio in a QSO absorber would thus seem to imply that more massive stars (and associated Type II SNe) have dominated the heavy element enrichment of that system (and that the effects of depletion are minimal) -- {\it if} the depletion patterns in low-metallicity systems are similar to those found in our Galaxy.

While the existing surveys of \mbox{Ti\,{\sc ii}} absorption have provided a general picture of the behaviour of titanium in the Galactic ISM, there are several aspects that remain poorly understood.
The moderate spectral resolution of most of the existing \mbox{Ti\,{\sc ii}} data generally has not allowed detailed investigations of the properties of the individual interstellar components discerned in higher resolution spectra of \mbox{Na\,{\sc i}}, \mbox{K\,{\sc i}}, and \mbox{Ca\,{\sc ii}} (e.g., Welty, Hobbs, \& Kulkarni 1994; Welty, Morton, \& Hobbs 1996; Welty \& Hobbs 2001; Price et al. 2001), which reveal variations in component properties (e.g., line widths, relative abundances) on velocity scales of 1--2 km~s$^{-1}$.
Moreover, while titanium is known to be severely depleted in some moderately reddened sight lines containing diffuse molecular material, little is known about the depletion behaviour of titanium in more heavily reddened sight lines (which may include some denser gas).
Significantly enhanced depletions were not found for iron in several recent samples of translucent sight lines (Snow, Rachford, \& Figoski 2002; Jensen \& Snow 2007; Miller et al. 2007), but enhanced depletions in individual dense components could have been masked by contributions from more diffuse gas in the relatively low-resolution {\it FUSE} spectra (FWHM $\sim$ 20 km~s$^{-1}$) used in the first two studies.

Very little is known regarding the depletion behaviour of titanium (and other elements) in lower metallicity systems.
For example, \mbox{Ti\,{\sc ii}} column densities have been reported for only a few sight lines in the LMC and SMC (Caulet \& Newell 1996; Welty et al. 1999a, 2001; Cox et al. 2006, 2007), where the current metallicities are typically 0.5 and 0.2 times solar, respectively (e.g., Russell \& Dopita 1992; various references in the appendices in Welty et al. 1997, 1999a).
The relatively nearby Magellanic Clouds provide an important venue for studies of the interstellar abundance/depletion patterns in low-metallicity galaxies, because the total interstellar (gas plus dust) abundances of many elements may be inferred from analyses of the abundances in stellar atmospheres and gaseous nebulae.
Studies of the interstellar abundances in a small number of sight lines have revealed possible differences in depletion behaviour for Mg, Si, and Ti (versus Fe and Ni) in at least some components in the LMC and SMC (Welty et al. 1997, 1999a, 2001, 2004, and in prep.; see also Sofia et al. 2006).

In this paper, we present several sets of \mbox{Ti\,{\sc ii}} absorption-line data, in order to further explore the behaviour of \mbox{Ti\,{\sc ii}} in different interstellar environments:
(1) high-resolution (FWHM = 1.3--1.5 km~s$^{-1}$) spectra of 17 relatively bright, lightly reddened Galactic stars -- to enable detailed comparisons with corresponding high-resolution spectra of \mbox{Na\,{\sc i}}, \mbox{K\,{\sc i}}, \mbox{Ca\,{\sc i}}, and \mbox{Ca\,{\sc ii}} -- and thus estimation of the properties of individual components in those sight lines;
(2) moderate resolution (FWHM = 3.4--4.5 km~s$^{-1}$) spectra of a number of fainter, more heavily reddened Galactic stars -- to investigate the depletion behaviour of Ti in denser gas; and 
(3) moderate resolution (FWHM = 4.5--8.7 km~s$^{-1}$) spectra of 43 stars in the Magellanic Clouds -- to examine the depletion behaviour of titanium in lower metallicity gas -- to aid in understanding the \mbox{Ti\,{\sc ii}} abundances in even lower metallicity absorption-line systems seen toward distant QSOs and gamma-ray bursts.
These data thus include both the highest resolution \mbox{Ti\,{\sc ii}} spectra available for the Galactic ISM and the most extensive \mbox{Ti\,{\sc ii}} spectra available for the ISM in the Magellanic Clouds.

\begin{table*}
\begin{minipage}{120mm}
\caption{Observing Runs.}
\label{tab:obs}
\begin{tabular}{@{}llcrrrl}
\hline
\multicolumn{1}{l}{Date}&
\multicolumn{1}{l}{Facility}&
\multicolumn{1}{c}{Resolution}&
\multicolumn{1}{l}{Galactic}&
\multicolumn{1}{c}{SMC}&
\multicolumn{1}{c}{LMC}&
\multicolumn{1}{c}{Run}\\
\multicolumn{1}{c}{ }&
\multicolumn{1}{c}{ }&
\multicolumn{1}{c}{(km s$^{-1}$)}&
\multicolumn{1}{c}{ }&
\multicolumn{1}{c}{ }&
\multicolumn{1}{c}{ }&
\multicolumn{1}{c}{ }\\
\hline
1994 Sep     & KPNO/coud\'{e} feed (cam 6) & 1.3     & 11 & 0 &  0 & K94\\
1995 Apr     & KPNO/coud\'{e} feed (cam 6) & 1.5     & 13 & 0 &  0 & K95\\
2000 Jun-Jul & ESO/VLT (UT2) + UVES        & 3.8     & 29 & 0 &  0 & V00\\
2000 Nov     & KPNO/coud\'{e} feed (cam 5) & 3.4     &  9 & 0 &  0 & K00\\
2001 Sep     & ESO/VLT (UT2) + UVES        & 5.0     &  0 & 1 &  0 & V01\\
2002 Nov-Dec & ESO/VLT (UT2) + UVES        & 8.7     &  0 & 3 & 10 & V02\\
2003 Nov     & ESO/VLT (UT2) + UVES        & 4.5     &  7 & 5 &  7 & V03\\
2004 Nov-Dec & ESO/VLT (UT2) + UVES        & 8.7     &  0 & 6 & 12 & V04\\
\hline
\end{tabular}
\medskip 
 ~ ~ \\
V00 data are from programme 65.I-0526 (E. Roueff, PI); V01 data are from programme 67.C-0281 (P. Ehrenfreund, PI).
\end{minipage}
\end{table*}

The following sections describe the various observational data (stellar samples, spectra, data reduction and analysis procedures), the derived \mbox{Ti\,{\sc ii}} column densities and titanium abundances (compared with similar data for other species), and the implications of those data for understanding the properties of interstellar clouds in various environments.
Several appendices contain plots of all the observed \mbox{Ti\,{\sc ii}} line profiles, lists of the individual component column densities derived from the highest resolution spectra, and two compilations of \mbox{Ti\,{\sc ii}} equivalent widths and column densities (from both current and previous work) used in exploring possible correlations between \mbox{Ti\,{\sc ii}} and other species.


\section{Data}
\label{sec-data}

The spectroscopic data collected in this survey were obtained using either the Kitt Peak 0.9m coud\'{e} feed telescope and echelle spectrograph or the ESO/VLT UT2 telescope and UVES spectrograph (Dekker et al. 2000).
The various observing runs are summarized in Table~\ref{tab:obs}, where we list for each run the telescope, spectrograph, resolution, number of sight lines, and a 3-character code used to identify data from the run in other tables and figures.


\subsection{Stellar sample}
\label{sec-stel}

Tables~\ref{tab:mwlos}, \ref{tab:smclos}, and \ref{tab:lmclos} list the 60 Galactic (MW), 15 SMC, and 28 LMC sight lines for which spectra are included in this survey.
For each sightline, the tables give Galactic (MW) or equatorial coordinates (MC), spectral type, $V$, $E(B-V)$ (both total and MC), either stellar distance (MW) or radial velocity (MC), and the observing run(s) in which the observations were obtained.
The sightlines are ordered by right ascension in each table.

The Milky Way targets range from very nearby, unreddened stars to distant, heavily reddened stars; nearly all the sight lines primarily probe gas within the Galactic disc.
The sample is not unbiased -- in that it includes both stars observed in previous high-resolution surveys of \mbox{Na\,{\sc i}}, \mbox{K\,{\sc i}}, and/or \mbox{Ca\,{\sc ii}} (Welty et al. 1994, 1996; Welty \& Hobbs 2001), stars observed in an ongoing survey of the diffuse interstellar bands (Thorburn et al. 2003), and a number of more reddened stars -- but the set of sight lines does explore a variety of regions and environments within the local Galactic ISM. 
Many of the spectral types and photometric data were taken from Thorburn et al. (2003); the rest were derived from values listed in the SIMBAD database.
Most of the targets are O or early-B stars, but there are a few late-B and A stars included as well.
Fifteen of the 60 stars are lightly reddened, with $E(B-V)$ $<$ 0.10; 29 others have $E(B-V)$ $\ge$ 0.35.
Stellar distances were taken from Thorburn et al. (2003) or Snow, Destree, \& Welty (2008; for Sco-Oph) or were estimated from the spectral types and photometry, using the absolute magnitudes of Blauuw (1963) and the intrinsic colours of Johnson (1963) and assuming total visual extinctions of 3.1 $\times$ $E(B-V)$.
Sixteen of the stars are within 200 pc of the sun; eleven are at distances $\ge$ 1000 pc; only one ($\rho$ Leo) lies more than 200 pc from the Galactic plane.

\begin{table*}
\begin{minipage}{130mm}
\caption{Galactic Sight lines.}
\label{tab:mwlos}
\begin{tabular}{@{}rlrrrrlrl}
\hline 
\multicolumn{1}{c}{HD}&
\multicolumn{1}{l}{Name}&
\multicolumn{1}{c}{$l$}&
\multicolumn{1}{c}{$b$}&
\multicolumn{1}{c}{$V$}&
\multicolumn{1}{c}{$E(B-V)$}&
\multicolumn{1}{l}{Type}&
\multicolumn{1}{c}{$d$}&
\multicolumn{1}{l}{Runs}\\
\multicolumn{1}{c}{ }& 
\multicolumn{1}{c}{ }& 
\multicolumn{1}{c}{(\degr~\arcmin)}& 
\multicolumn{1}{c}{(\degr~\arcmin) }& 
\multicolumn{1}{c}{ }& 
\multicolumn{1}{c}{ }& 
\multicolumn{1}{c}{ }& 
\multicolumn{1}{c}{(pc)}\\
\hline
 23180 & o Per          & 160 22 &$-$17 44 & 3.83 & 0.31 & B1 III     & 280 & K00 \\
 24398 & $\zeta$ Per    & 162 17 &$-$16 42 & 2.85 & 0.31 & B1 Ib      & 300 & K00 \\
 24534 & X Per          & 163 05 &$-$17 08 & 6.10 & 0.59 & O9.5pe     & 590 & K00 \\
 24760 & $\epsilon$ Per & 157 21 &$-$10 06 & 2.89 & 0.10 & B0.5V+A2   & 165 & K94 \\
 27778 & 62 Tau         & 172 46 &$-$17 23 & 6.36 & 0.37 & B3 V       & 225 & K00 \\
 32630 & $\eta$ Aur     & 165 21 &   00 16 & 3.17 & 0.02 & B3 V       &  67 & K95 \\ 
 35149 & 23 Ori         & 199 19 &$-$17 52 & 5.00 & 0.11 & B1 V       & 295 & K00,V03 \\
 36486 & $\delta$ Ori   & 203 51 &$-$17 45 & 2.23 & 0.08 & B0III+O9V  & 280 & K94,K95 \\
 37022 & $\theta^1$ OriC& 209 01 &$-$19 23 & 5.13 & 0.34 & O6-7p      & 450 & K00 \\
 37043 & $\iota$ Ori    & 209 32 &$-$19 36 & 2.77 & 0.07 & O9 III     & 450 & K95 \\
 37128 & $\epsilon$ Ori & 205 13 &$-$17 15 & 1.70 & 0.05 & B0 Iae     & 350 & K94 \\
 37742 & $\zeta$ Ori A  & 206 27 &$-$16 36 & 2.05 & 0.06 & O9.5 Ibe   & 370 & K94,K95 \\
 47839 & 15 Mon A       & 202 56 &   02 11 & 4.66 & 0.07 & O7 Ve      & 950 & K94,K95 \\
       & Walker 67      & 202 59 &   02 04 &10.79 & 0.86 & B2 V       & 950 & V03 \\ 
 62542 &                & 255 55 &$-$09 14 & 8.04 & 0.35 & B5 V       & 246 & V03 \\
 72127A&                & 262 34 &$-$03 22 & 5.20 & 0.10 & B2 III     & 480 & V03 \\ 
 72127B&                & 262 34 &$-$03 22 & 7.09 & 0.10 & B2.5 V     & 480 & V03 \\ 
 73882 &                & 260 11 &   00 38 & 7.22 & 0.70 & O8 V       &1100 & V03 \\
 91316 & $\rho$ Leo     & 234 53 &   52 46 & 3.85 & 0.05 & B1 Ib      & 760 & K95 \\
106625 & $\gamma$ Crv   & 290 59 &   44 30 & 2.59 & 0.00 & B8 III     &  51 & V00 \\ 
110432 &                & 301 58 &$-$00 12 & 5.31 & 0.51 & B2pe       & 300 & V00 \\
112244 &                & 303 33 &   06 02 & 5.32 & 0.30 & O9 Ibe     &1250 & V00 \\ 
113904 & $\theta$ Mus   & 304 40 &$-$02 29 & 5.51 & 0.25 & WC6+O6-7V  &2200 & V00 \\
114213 &                & 305 11 &   01 19 & 8.97 & 1.12 & B1 Ib      &1700 & V00 \\
116658 & $\alpha$ Vir   & 316 07 &   50 50 & 0.98 & 0.03 & B1 III     &  80 & K95 \\ 
141637 & 1 Sco          & 346 06 &   21 43 & 4.64 & 0.15 & B3 V       & 152 & K95 \\
143018 & $\pi$ Sco      & 347 12 &   20 14 & 2.89 & 0.04 & B1V+B2V    & 180 & K94,K95 \\
144217 & $\beta^1$ Sco  & 353 11 &   23 36 & 2.62 & 0.19 & B1 V       & 124 & K94,K95 \\
147165 & $\sigma$ Sco   & 351 19 &   17 00 & 2.89 & 0.41 & B2III+O9V  & 214 & K95 \\
147701 &                & 352 15 &   16 51 & 8.35 & 0.74 & B5 III     & 153 & V00 \\
147888 & $\rho$ Oph D   & 353 39 &   17 43 & 6.74 & 0.47 & B5 V       & 125 & V00 \\
147889 &                & 352 52 &   17 02 & 7.90 & 1.07 & B2 V       & 118 & V00 \\
147933 & $\rho$ Oph A   & 353 41 &   17 41 & 5.02 & 0.48 & B2 IV      & 111 & V00 \\
148184 & $\chi$ Oph     & 357 56 &   20 41 & 4.42 & 0.52 & B2 IVpe    & 161 & V00 \\
148379 &                & 337 15 &   01 35 & 5.36 & 0.72 & B2 Iab     & 770 & V00 \\ 
149404 &                & 340 32 &   03 01 & 5.47 & 0.68 & O9 Iae     & 820 & V00 \\
149757 & $\zeta$ Oph    &   6 17 &   23 36 & 2.56 & 0.32 & O9.5 V     & 140 & K94,K95,V00 \\
150136 &                & 336 43 &$-$01 34 & 5.60 & 0.48 & O6 III     &1270 & V00 \\ 
152236 & $\zeta^1$ Sco  & 343 02 &   00 52 & 4.73 & 0.68 & B1 Ia$^+$pe&1800 & V00 \\
154368 &                & 349 58 &   03 13 & 6.13 & 0.78 & O9 Ia      & 960 & V00 \\
159561 & $\alpha$ Oph   &  35 54 &   22 35 & 2.08 & 0.00 & A5 III     &  14 & K94,K95 \\
161056 &                &  18 41 &   11 47 & 6.30 & 0.62 & B1.5 V     & 310 & V00 \\ 
166734 &                &  18 55 &   03 38 & 8.41 & 1.39 & O8e        & 720 & V00 \\
167971 &                &  18 15 &   01 41 & 7.45 & 1.08 & O8e        & 730 & V00 \\
       & BD$-$14~5037   &  16 56 &$-$00 57 & 8.24 & 1.55 & B1.5 Ia    &1100 & V00 \\
169454 &                &  17 32 &$-$00 40 & 6.61 & 1.12 & B1.5 Ia    & 930 & V00 \\
170740 &                &  21 04 &$-$00 32 & 5.72 & 0.48 & B2 V       & 213 & V00 \\
172028 &                &  31 03 &   02 50 & 7.83 & 0.79 & B2 V       & 380 & V00 \\
182985 &                &  02 48 &$-$22 38 & 7.47 & 0.12 & A0 V       & 186 & V00 \\ 
183143 &                &  53 14 &   00 38 & 6.86 & 1.27 & B7 Iae     &1000 & V00 \\
188220 &                &  25 16 &$-$20 59 & 8.04 & 0.27 & A0 V       & 200 & V00 \\ 
197345 & $\alpha$ Cyg   &  84 17 &   02 00 & 1.25 & 0.09 & A2 Ia      & 540 & K94 \\
199579 &                &  85 42 &$-$00 18 & 5.96 & 0.37 & O6 Ve      &1200 & K00 \\
205637 & $\epsilon$ Cap &  31 56 &$-$44 59 & 4.65 & 0.02 & B3 V:p     & 203 & V00 \\ 
206267 &                &  99 17 &   03 44 & 5.62 & 0.53 & O6f        & 810 & K00 \\
207198 &                & 103 08 &   06 59 & 5.95 & 0.62 & O9 IIe     &1000 & K00 \\
210121 &                &  56 53 &$-$44 28 & 7.67 & 0.40 & B3 V       & 210 & V00 \\
212571 & $\pi$ Aqr      &  66 00 &$-$44 44 & 4.79 & 0.23 & B1 Ve      & 340 & V00 \\ 
217675 & o And          & 102 12 &$-$16 06 & 3.62 & 0.05 & B6 IIIp    & 212 & K94 \\ 
219688 & $\psi^2$ Aqr   &  67 37 &$-$61 32 & 4.40 & 0.01 & B5 V       &  99 & V03 \\ 
\hline
\end{tabular}
\end{minipage}
\end{table*}

\begin{table*}
\begin{minipage}{150mm}
\caption{SMC sight lines.}
\label{tab:smclos}
\begin{tabular}{@{}lrrrrrrlrrl}
\hline
\multicolumn{1}{l}{Star}&
\multicolumn{1}{l}{AzV}&
\multicolumn{2}{c}{RA (J2000) DEC}&
\multicolumn{1}{c}{$V$}&
\multicolumn{1}{c}{$B-V$}&
\multicolumn{1}{c}{$E(B-V)$}&
\multicolumn{1}{l}{Type}&
\multicolumn{1}{c}{Ref}&
\multicolumn{1}{c}{$v_{\rm rad}$}&
\multicolumn{1}{l}{Runs}\\
\multicolumn{1}{c}{ }& 
\multicolumn{1}{c}{ }& 
\multicolumn{1}{c}{($^{h~m~s}$)}& 
\multicolumn{1}{c}{(\degr~\arcmin~\arcsec) }& 
\multicolumn{1}{c}{ }& 
\multicolumn{1}{c}{ }&
\multicolumn{1}{c}{tot/SMC}& 
\multicolumn{1}{c}{ }& 
\multicolumn{1}{c}{ }&
\multicolumn{1}{c}{(km s$^{-1}$)}\\ 
\hline
Sk 13 & 18 & 00 47 12.2 & $-$73 06 33 & 12.44 &   0.03 & 0.20/0.16 & B2 Ia      & 1 & 138/147 & V03\\
Sk 18 & 26 & 00 47 50.0 & $-$73 08 21 & 12.46 &$-$0.17 & 0.15/0.11 & O7 III     & 1 & 142/149 & V03\\
      & 47 & 00 48 51.5 & $-$73 25 59 & 13.44 &$-$0.18 & 0.13/0.09 & O8 III     & 2 & 128/135 & V02\\
Sk 40 & 78 & 00 50 38.4 & $-$73 28 18 & 11.05 &$-$0.05 & 0.14/0.10 & B1 Ia+     & 3 & 164/155:& V03\\
      & 80 & 00 50 43.8 & $-$72 47 42 & 13.32 &$-$0.13 & 0.19/0.15 & O4-6n(f)p  & 2 & 131/133 & V04\\
      & 95 & 00 51 21.6 & $-$72 44 15 & 13.78 &$-$0.18 & 0.14/0.10 & O7 III     & 2 & 137/136 & V04\\
      &120 & 00 52 20.4 & $-$72 09 10 & 14.43 &$-$0.22 & 0.09/0.05 & O9 V       & 4 & .../197 & V04\\
      &207 & 00 58 33.2 & $-$71 55 47 & 14.25 &$-$0.20 & 0.12/0.08 & O7 V       & 4 & .../141 & V02\\
      &321 & 01 02 57.1 & $-$72 08 09 & 13.76 &$-$0.19 & 0.12/0.08 & O9 IInp    & 5 & .../ 95 & V04\\
      &388 & 01 05 39.5 & $-$72 29 27 & 14.09 &$-$0.21 & 0.11/0.07 & O4 V       & 4 & .../168 & V02\\
      &440 & 01 08 56.0 & $-$71 52 47 & 14.48 &$-$0.18 & 0.13/0.09 & O8 V       & 6 & .../191 & V04\\
Sk 143&456 & 01 10 55.8 & $-$72 42 56 & 12.83 &   0.10 & 0.36/0.32 & O9.7 Ib    & 7 & 169/... & E01\\
      &476 & 01 13 42.5 & $-$73 17 30 & 13.52 &$-$0.09 & 0.23/0.19 & O6.5 V     & 4 & .../... & V03\\
Sk 155&479 & 01 14 50.3 & $-$73 20 18 & 12.46 &$-$0.15 & 0.13/0.09 & O9 Ib      & 8 & 173/158 & V03\\
Sk 190&    & 01 31 28.0 & $-$73 22 14 & 13.54 &$-$0.18 & 0.11/0.07 & O8 Iaf     & 4 & .../176 & V04\\
\hline
\end{tabular}
\medskip
~ ~ \\
Photometry is from Massey 2002 (preferred) or from Dachs 1970, Azzopardi et al. 1975, Ardeberg \& Maurice 1977, Bouchet et al. 1985, and/or Gordon et al. 2003.
$E(B-V)$ is calculated using the intrinsic colours of FitzGerald 1970, assuming a Galactic contribution of 0.04 mag.
Spectral types are from 1 = Garmany et al. 1987; 2 = Walborn et al. 2000; 3 = Trundle \& Lennon 2005; 4 = Massey 2002; 5 = Walborn et al. 2009; 6 = Massey 2005; 7 = Fitzpatrick 1985; 8 = Lennon 1997.
Radial velocities are from Maurice et al. 1989, Evans et al. 2004 and/or Welty \& Crowther (in prep.); values with a colon are less well determined.
\end{minipage}
\end{table*}

\begin{table*}
\begin{minipage}{170mm}
\caption{LMC sight lines.}
\label{tab:lmclos}
\begin{tabular}{@{}lrrrrrrlrrl}
\hline 
\multicolumn{1}{l}{Star}& 
\multicolumn{1}{l}{HD}& 
\multicolumn{2}{c}{RA (J2000) DEC}& 
\multicolumn{1}{c}{$V$}& 
\multicolumn{1}{c}{$B-V$}&
\multicolumn{1}{c}{$E(B-V)$}& 
\multicolumn{1}{l}{Type}&
\multicolumn{1}{c}{Ref}&
\multicolumn{1}{c}{$v_{\rm rad}$}&
\multicolumn{1}{l}{Runs}\\ 
\multicolumn{1}{c}{ }&  
\multicolumn{1}{c}{ }&  
\multicolumn{1}{c}{($^{h~m~s}$)}&  
\multicolumn{1}{c}{(\degr~\arcmin~\arcsec)}&  
\multicolumn{1}{c}{ }&  
\multicolumn{1}{c}{ }&
\multicolumn{1}{c}{tot/LMC}&  
\multicolumn{1}{c}{ }&  
\multicolumn{1}{c}{ }&
\multicolumn{1}{c}{(km s$^{-1}$)}\\  
\hline
Sk$-$67 2   &270754 & 04 47 04.5 & $-$67 06 53 & 11.26 &   0.07 & 0.26/0.24   & B1 Ia+      & 1 & 309/... & V03\\ 
Sk$-$67 5   &268605 & 04 50 18.9 & $-$67 39 38 & 11.34 &$-$0.12 & 0.14/0.11   & O9.7 Ib     & 2 & 309/... & V03\\
Sk$-$66 18  &       & 04 55 59.8 & $-$65 58 30 & 13.50 &$-$0.20 & 0.12/0.10   & O6 V((f))   & 3 & .../292 & V04\\
LH 10-3061  &       & 04 56 42.5 & $-$66 25 18 & 13.68 &$-$0.01 & 0.31/0.28   & O2 III      & 4 & .../311 & V04\\
Sk$-$69 50  &       & 04 57 15.1 & $-$69 20 20 & 13.31 &$-$0.16 & 0.12/0.03   & O7(n)(f)p   & 5 & .../214 & V04\\ 
Sk$-$67 22  &       & 04 57 27.4 & $-$67 39 03 & 13.44 &$-$0.18 & 0.12/0.07   & O2If/WN5    & 6 & .../388 & V04\\ 
Sk$-$67 38  &       & 05 03 29.7 & $-$67 52 25 & 13.66 &$-$0.22 & 0.09/0.04   & O8 III      & 7 & 286/299 & V04\\ 
Sk$-$70 69  &       & 05 05 18.7 & $-$70 25 50 & 13.95 &$-$0.23 & 0.09/0.01   & O5 V        & 8 & .../262 & V02\\
Sk$-$68 41  &       & 05 05 27.1 & $-$68 10 03 & 12.01 &$-$0.14 & 0.08/0.02   & B0.5 Ia     & 9 & 239/237 & V02\\
Sk$-$68 52  &269050 & 05 07 20.4 & $-$68 32 09 & 11.63 &$-$0.07 & 0.18/0.12   & B0 Ia       &10 & 234/218:& V03\\
BI 128      &       & 05 18 19.8 & $-$65 49 15 & 13.82 &$-$0.25 & 0.06/0.01   & O9 V        & 3 & .../313 & V04\\ 
Sk$-$65 47  &       & 05 20 54.7 & $-$65 27 18 & 12.60 &$-$0.16 & 0.15/0.10   & O4 I(n)f+p  & 5 & .../227 & V04\\ 
Sk$-$68 73  &269445 & 05 22 59.8 & $-$68 01 47 & 11.45 &   0.27 & 0.40/0.34   & Of/WN       &11 & .../... & V03\\
Sk$-$67 101 &       & 05 25 56.3 & $-$67 30 29 & 12.67 &$-$0.23 & 0.08/0.03   & O8 II       & 2 & .../294 & V02\\
BI 170      &       & 05 26 47.7 & $-$69 06 12 & 13.06 &$-$0.20 & 0.07/0.01   & O9.5 Ib     & 2 & .../265 & V02\\
Sk$-$66 100 &       & 05 27 45.5 & $-$66 55 15 & 13.26 &$-$0.21 & 0.11/0.06   & O6 II       & 8 & 309/304 & V04\\
Sk$-$67 169 &       & 05 31 51.6 & $-$67 02 22 & 12.18 &$-$0.12 & 0.07/0.02   & B1 Ia       & 9 & .../289 & V02\\
Sk$-$67 168 &269702 & 05 31 52.1 & $-$67 34 20 & 12.09 &$-$0.17 & 0.12/0.07   & O8 Iaf      & 9 & .../242 & V04\\ 
Sk$-$67 211 &269810 & 05 35 13.9 & $-$67 33 28 & 12.29 &$-$0.21 & 0.10/0.04   & O2 III      & 4 & 264/270 & V02\\
BI 229      &       & 05 35 32.2 & $-$66 02 38 & 12.95 &$-$0.17 & 0.15/0.10   & O7 V-III    & 2 & .../305 & V02\\
BI 237      &       & 05 36 14.6 & $-$67 39 19 & 13.89 &$-$0.12 & 0.20/0.14   & O2 V        &12 & .../378 & V04\\
Sk$-$66 171 &       & 05 37 02.4 & $-$66 38 37 & 12.21 &$-$0.15 & 0.13/0.08   & O9 Ia       & 9 & .../400 & V04\\ 
BI 253      &       & 05 37 34.5 & $-$69 01 10 & 13.76 &$-$0.09 & 0.23/0.17   & O2 V        & 4 & .../271 & V04\\
Sk$-$68 135 &269896 & 05 37 49.1 & $-$68 55 02 & 11.36 &   0.00 & 0.26/0.20   & ON9.7 Ia+   &10 & 290/291 & V03\\
Melnick 42  &       & 05 38 42.2 & $-$69 05 54 & 12.71 &   0.00 & 0.30/0.24   & O2If/WN5    & 6 & .../254 & V02\\
Sk$-$69 246 & 38282 & 05 38 53.4 & $-$69 02 01 & 11.10 &$-$0.12 & 0.18/0.12   & WN6h        &13 & .../... & V03\\
BI 272      &       & 05 44 23.1 & $-$67 14 29 & 13.27 &$-$0.19 & 0.13/0.07   & O7 III-II   & 2 & .../365:& V02\\
Sk$-$70 115 &270145 & 05 48 49.7 & $-$70 03 58 & 12.24 &$-$0.10 & 0.20/0.14   & O6.5 Iaf    &14 & .../278:& V02,V03\\
\hline
\end{tabular}
\medskip
~ ~ \\
Photometry is from Massey 2002 or Parker 1993 (preferred) or from Ardeberg et al. 1972; Brunet et al. 1975; Isserstedt 1975, 1979, 1982; Feitzinger \& Isserstedt 1983; and/or Schmidt-Kaler et al. 1999.
$E(B-V)$ is calculated using the intrinsic colours of FitzGerald 1970; Galactic foreground extinction is estimated from fig.~13 of Staveley-Smith et al. 2003.
Spectral types are from 1 = Fitzpatrick 1991, 2 = Walborn et al. 2002b, 3 = Massey et al. 1995, 4 = Walborn et al. 2002c, 5 = Walborn et al. 2009, 6 = Crowther (priv. comm.), 7 = Massey 2002, 8 = Walborn et al. 1995, 9 = Fitzpatrick 1988, 10 = Walborn 1977, 11 = Crowther \& Smith 1997, 12 = Massey et al. 2005, 13 = Crowther \& Dessart 1998, 14 = Walborn et al. (in prep.).
Radial velocities are from Ardeberg et al. 1972, Fehrenbach \& Duflot 1982, Walborn et al. 2002 and/or Welty \& Crowther (in prep.).
\end{minipage}
\end{table*}

The sample of Magellanic Clouds sight lines also is not unbiased.
Most of the 43 stars were observed primarily for studies of stellar properties at lower metallicities (e.g., Crowther et al. 2002; Evans et al. 2004); 13 stars, with typically higher $E(B-V)$ and known strong absorption from interstellar \mbox{Na\,{\sc i}} and/or H$_2$, were observed in a search for interstellar molecular absorption lines (Welty et al. 2006).
All of the SMC and LMC targets are O, early B, or Wolf-Rayet stars, with 11.0 $\la$ $V$ $\la$ 14.6; some appear to be unresolved binary (or multiple) systems.
The stellar coordinates are from the catalogues painstakingly compiled by B. Skiff.\footnotemark
\footnotetext{ftp.lowell.edu/pub/bas/starcats}
References for the spectral types, $V$ magnitudes, and $B-V$ colours for the stars are given in footnotes to Tables~\ref{tab:smclos} and \ref{tab:lmclos}.
In general, we preferred the ccd photometry of Massey (2002) and Parker (1993), which should better account for blending, near neighbours, and local fluctuations in the background than the earlier studies based on aperture photometry.
In most cases, the total $E(B-V)$ colour excesses (first entries) were obtained using the intrinsic colours of FitzGerald (1970) or Walborn et al. (2002a); for the few Wolf-Rayet stars, the intrinsic colours were estimated from theoretical spectral energy distributions (e.g., Crowther 2007).
Only two SMC and six LMC sight lines have total $E(B-V)$ greater than 0.2 mag.
For the Galactic contributions to those total $E(B-V)$, we adopted a constant 0.04 mag for all the SMC sight lines (Schlegel, Finkbeiner, \& Davis 1998) and values ranging from 0.02--0.09 mag for the LMC sight lines [estimated from fig. 13 of Staveley-Smith et al. (2003)].
The second $E(B-V)$ entries give the corresponding contributions for the SMC or LMC ISM (total minus Galactic).
Considering that the uncertainties in the photometry, in the spectral types and intrinsic colours, and in the Galactic contribution to the colour excess could each be several hundredths of a magnitude, it is notable that $E(B-V)_{\rm MC}$ is not less than zero for any of the 43 sight lines.

Where possible, two stellar radial velocities are listed for the Magellanic Clouds targets.
The first value is from the literature, with references in footnotes to the tables. 
Many of those values were derived from objective prism spectra, and have an estimated accuracy of order 15 km~s$^{-1}$ (e.g., Fehrenbach \& Duflot 1982).
The second value is estimated from the spectra analyzed in this paper, using the stellar \mbox{H\,{\sc i}} lines; the \mbox{He\,{\sc i}} lines near 3926.5, 3964.7, and 5875.6~\AA\ (in air); and/or the \mbox{Ca\,{\sc ii}} lines near 3933.7 and 3968.5~\AA.
In most cases, the two values agree within $\pm$20 km~s$^{-1}$.


\subsection{Observations and data processing}
\label{sec-obs}

\subsubsection{KPNO coud\'{e} feed}
\label{sec-cf}

High-resolution spectra of 17 relatively bright ($V$ $\la$ 4.7) Galactic stars were obtained with the 0.9m KPNO coud\'{e} feed telescope and echelle spectrograph during two runs in 1994 (K94) and 1995 (K95).
Both runs employed setups similar to those used in previous high-resolution surveys of \mbox{K\,{\sc i}}, \mbox{Ca\,{\sc i}}, and \mbox{Ca\,{\sc ii}} (Welty et al. 1996; Welty, Hobbs, \& Morton 2003; Welty \& Hobbs 2001) -- with camera 6, grism 780-2 and the 3 degree wedge for cross dispersion, and BG3+CuSO$_4$ filters to isolate the spectral region from about 3270 to 3490 \AA.
The setups used a very narrow slit, corresponding to about 0.65 arcsec on the sky, to achieve resolutions of 1.3--1.5 km~s$^{-1}$ (similar to the values achieved in the previous high-resolution surveys), as determined from the widths of the thorium lines in the Th-Ar lamp spectra used for wavelength calibration.
Because no near-UV filter was available for the auto-guider and because the targets could not always be observed close to the meridian, slight offsets generally were required to position the slit on the centroid of the stellar flux near the \mbox{Ti\,{\sc ii}} $\lambda$3383 line.
Multiple exposures, of lengths typically 30--60 min, were obtained for most of the targets, with small night-to-night shifts of the spectral format on the detector to reduce the effects of any instrumental artefacts.
Standard routines within IRAF were used to remove bias and to divide sections of the 2-D spectral images containing the order(s) of interest by a normalized flat-field derived from quartz lamp exposures.
The 1-D spectra then were extracted from the flat-fielded image segments via the {\sc apextract} routines, using variance weighting.
Wavelength calibration was accomplished via Th-Ar lamp exposures, which were obtained several times during the course of each night, using the thorium rest wavelengths tabulated by Palmer \& Engelman (1983).
Multiple spectra for a given target were sinc interpolated to a common heliocentric wavelength grid, then summed.
The summed spectra were then normalized via Legendre polynomial fits to the continuum regions surrounding the interstellar (and stellar) absorption lines.
A signal-to-noise (S/N) ratio of about 220 (per half resolution element) was achieved near the $\lambda$3383 line in a total exposure time of 480 minutes for $\zeta$ Oph ($U$ = 1.7); the corresponding 3~$\sigma$ equivalent width limit (assuming equal contributions from photon noise and continuum placement uncertainties) for a single component with line width $b$ = 2 km~s$^{-1}$ would be about 0.5 m\AA.
Equivalent widths measured from these high-resolution spectra generally agree well with published values derived from lower resolution spectra (Table~\ref{tab:allwids}).
Detailed comparisons with the corresponding high-resolution \mbox{Ca\,{\sc ii}} profiles (e.g., Welty et al. 1996) indicate that the differences in velocity zero point are $\la$ 0.5 km~s$^{-1}$) in most cases.

Somewhat lower resolution spectra of nine (mostly) fainter, more reddened Galactic stars (5.0 $\la$ $V$ $\la$ 6.4) were obtained with the coud\'{e} feed telescope in 2000 (run K00), using camera 5, grism 780-2, the 3 degree wedge, and the BG3+CuSO$_4$ filters.
The wider slit used with the camera 5 setup, corresponding to 1.8 arcsec on the sky, yielded a resolution of about 3.4 km~s$^{-1}$ over the wavelength range from about 3240 to 3815 \AA.  
The S/N ratios in the summed spectra range from about 40 to 80, with corresponding 3~$\sigma$ equivalent width limits of 2--4 m\AA, for these fainter targets.
The $\lambda$3383 equivalent widths measured for o~Per and 23~Ori agree well with previous values; the equivalent width for $\zeta$~Per (based on a single exposure) lies at the low end of the range of published values (Table~\ref{tab:allwids}).

\subsubsection{ESO VLT + UVES}
\label{sec-vlt}

Spectra of seven Galactic, five SMC, and seven LMC stars were obtained in 2003 November using the ESO/VLT UT2 telescope and UVES spectrograph (run V03; Dekker et al. 2000), under programme 72.C-0682, which was aimed at detecting weak interstellar atomic and molecular absorption lines toward stars in the Magellanic Clouds with relatively high column densities of \mbox{Na\,{\sc i}} and/or H$_2$ (Welty et al. 2006).
The standard dichroic \#1 390/564 setting and a slit width corresponding to 0.7 arcsec were used to obtain nearly complete coverage of the wavelength range 3260--6680 \AA\ -- including lines from a number of interstellar atomic and molecular species.  
The spectral resolution, determined from the widths of the thorium lines in the Th-Ar lamp exposures used for wavelength calibration, was about 4.5 km~s$^{-1}$ near the \mbox{Ti\,{\sc ii}} $\lambda$3383 line.
Multiple exposures, generally of length 20--45 minutes, were obtained for most of the Magellanic Clouds targets over the three nights, with the camera tilted by $\pm$50 units on the second and third nights to ensure that the spectra fell on slightly different parts of the CCD.
The spectra were processed using standard routines within IRAF (as for the KPNO spectra), which produced extracted spectra essentially identical to those obtained with the UVES pipeline software for the observations obtained with the standard camera tilt.
The S/N in the final summed spectra were typically $\sim$ 150--215 near the \mbox{Ti\,{\sc ii}} $\lambda$3383 line; corresponding 3~$\sigma$ detection limits for weak, unresolved absorption lines are $\sim$ 1.0--1.5 m\AA.
Equivalent widths measured for the \mbox{Ti\,{\sc ii}} lines toward the Galactic stars $\psi^2$~Aqr and 23~Ori show very good agreement with previously reported values (Albert et al. 1993; Welty et al. 1999b; Table~\ref{tab:allwids}).

UVES spectra of the SMC star Sk~143, obtained using a very similar instrumental setup in 2001 under programme 67.C-0281 (run V01; P. Ehrenfreund, PI), were retrieved from the ESO archive and processed with our IRAF-based procedure.
This moderately reddened sight line has the highest known molecular fraction (Cartledge et al. 2005) and sole detection of CN absorption (Welty et al. 2006) in the SMC, and is the only SMC sight line known to exhibit a Milky Way-like extinction curve (with the 2175 \AA\ extinction bump; Gordon et al. 2003).
With one exception (see below), the equivalent widths and column densities (for various atomic and molecular species) derived from these spectra are consistent with those reported by Cox et al. (2007).

Spectra of nine SMC and 22 LMC stars were obtained in 2002 and 2004 (runs V02 and V04), under programmes 70.D-0164 and 74.D-0109, which were designed to investigate the atmospheric and wind properties of the target stars (e.g., Crowther et al. 2002; Evans et al. 2004).
Most of these sight lines thus have somewhat lower hydrogen column densities and $E(B-V)$ than those probed in runs V01 and V03.
The UVES pipeline reduction system (Ballester et al. 2000) was used to obtain the wavelength-calibrated extracted spectra used for analysis of the interstellar absorption lines.
Because these observations employed a wider slit and generally shorter total exposure times, both the spectral resolution and S/N ratios (25--75) are somewhat poorer than those achieved in runs V01 and V03.
Comparisons with higher resolution spectra, for several sight lines observed in runs V02 and V04, suggest that the actual achieved resolution is somewhat better than the nominal 8.7 km~s$^{-1}$ value cited in Table~\ref{tab:obs}, however.

Spectra of 29 Galactic stars were obtained in 2000 under programme 65.I-0526 (run V00; E. Roueff, PI), which was aimed at detecting the weak molecular absorption lines of OH and NH toward reddened stars in the disc. 
All but three of the sight lines have $E(B-V)$ $>$ 0.2 mag, and six have $E(B-V)$ $>$ 1.0 mag.
The raw spectral images were retrieved from the ESO archive and processed with the UVES pipeline reduction package (by A. Ritchey).
The use of a very narrow slit yielded a resolution of about 3.8 km~s$^{-1}$ between about 3060 and 3860 \AA; the S/N ratios range from about 225--455 (with 3~$\sigma$ equivalent width limits of 0.4--0.8 m\AA) near the $\lambda$3383 line.
Several weaker \mbox{Ti\,{\sc ii}} lines near 3066, 3072, 3229, and 3241 \AA\ are also included in these spectra.
The equivalent widths measured for the \mbox{Ti\,{\sc ii}} $\lambda$3383 line and the \mbox{Na\,{\sc i}} $\lambda$3302 lines are in good agreement with those reported by Hunter et al. (2006) for several sight lines in common.

The absorption features measured from the UVES spectra of the Magellanic Clouds targets from runs V01, V02, V03, and V04 for CH, CH$^+$, CN and several of the diffuse interstellar bands were discussed by Welty et al. (2006); the corresponding absorption lines from \mbox{Na\,{\sc i}}, \mbox{K\,{\sc i}}, \mbox{Ca\,{\sc i}}, \mbox{Ca\,{\sc ii}}, and \mbox{Fe\,{\sc i}} will be described by Welty \& Crowther (in prep.).
The molecular absorption features found in the spectra of the Galactic targets from runs V00 and V03 are included in a forthcoming survey of OH and NH absorption in translucent sight lines (Welty et al., in prep.).
 
\subsubsection{{\it HST} (GHRS and STIS)}
\label{sec-hst}

In order to perform detailed comparisons between the absorption-line profiles of \mbox{Ti\,{\sc ii}} and \mbox{Zn\,{\sc ii}} (typically at most very mildly depleted), high-resolution UV spectra including the \mbox{Zn\,{\sc ii}} lines at 2026 and/or 2062 \AA\ were obtained for four Galactic, one SMC, and two LMC sight lines.
The four Galactic sight lines were originally observed under programmes 1360, 3415 (D. Ebbets, PI), and 3472 (L. Hobbs, PI) with the {\it Hubble Space Telescope} Goddard High Resolution Spectrograph, at resolutions of 3.4--3.7 km~s$^{-1}$; the spectral data were reduced as described by Welty et al. (1999b).
The three Magellanic Clouds sight lines were observed under GO programmes 8145 and 9757 (D. Welty, PI) with the {\it HST} Space Telescope Imaging Spectrograph, at a resolution of about 2.7 km~s$^{-1}$ (Welty et al. 2001, 2004, and in prep.).
Velocity alignment of the \mbox{Zn\,{\sc ii}} profiles was accomplished via fits to the adjacent lines of \mbox{Mg\,{\sc i}} ($\lambda$2026) and \mbox{Cr\,{\sc ii}} ($\lambda$2062), using component structures derived from high-resolution optical spectra of similarly distributed \mbox{Na\,{\sc i}} and \mbox{Ca\,{\sc ii}}, respectively (e.g., Welty et al. 1999b).


\subsection{Spectra and equivalent widths}
\label{sec-ew}

The normalized Galactic \mbox{Ti\,{\sc ii}} $\lambda$3383 line profiles are displayed, together with the corresponding \mbox{Ca\,{\sc ii}} $\lambda$3933 and \mbox{K\,{\sc i}} $\lambda$7698 or \mbox{Na\,{\sc i}} ($\lambda$5895, $\lambda$5889, or $\lambda$3302) profiles, in Appendix Fig.~\ref{fig:specg1}. 
The \mbox{Ti\,{\sc ii}}, \mbox{Ca\,{\sc ii}}, and \mbox{Na\,{\sc i}} profiles for the SMC and LMC sight lines (including absorption from both Galactic and Magellanic Clouds components) are shown in Appendix Fig.~\ref{fig:specm1}.
The source of each spectrum is noted at the right, just above the continuum.
For the Galactic sight lines, the zero point for LSR velocities is indicated by a solid triangle near the bottom of the frame.
For the SMC and LMC targets, stellar radial velocities are indicated by asterisks (literature values) and/or the letter ``S'' (values determined from our data); stellar absorption lines estimated in fits to the profiles are shown by dotted lines for some of the \mbox{Ca\,{\sc ii}} profiles.
Individual interstellar components found in fits to the higher resolution Galactic line profiles are indicated by tick marks above the spectra; solid dots above the \mbox{Na\,{\sc i}} or \mbox{K\,{\sc i}} spectra show the velocities of components detected in CN absorption (an indicator of relatively dense gas).
Note the differences in velocity scale for the various Galactic sight lines; the vertical dotted lines are separated by 10 km~s$^{-1}$ in all cases.
Note also the expanded vertical scale for the weaker Galactic absorption lines; all lines toward the Magellanic Clouds targets are shown at the same (unexpanded) vertical scale, however.
For a number of the LMC sight lines, the LMC \mbox{Na\,{\sc i}} $\lambda$5889 (D2) absorption, at about $-$303 km~s$^{-1}$ relative to the $\lambda$5895 (D1) absorption, is blended with the Galactic D1 absorption near $v$ $\sim$ 0 km~s$^{-1}$.

As seen in previous surveys (e.g., Stokes 1978; Hunter et al. 2006), the \mbox{Ti\,{\sc ii}} profiles often are very similar to those of \mbox{Ca\,{\sc ii}}, but can be very different from those of \mbox{Na\,{\sc i}} and \mbox{K\,{\sc i}}.
In many sight lines (particularly in the LMC and SMC), the \mbox{Ti\,{\sc ii}} and \mbox{Ca\,{\sc ii}} absorption is spread over a wider range in velocity.
In many of the Magellanic Clouds sight lines, the absorption from \mbox{Ti\,{\sc ii}} (and \mbox{Ca\,{\sc ii}}) is significantly stronger, relative to the absorption from \mbox{Na\,{\sc i}}, than in most of the Galactic sight lines.
Even the highest resolution \mbox{Ti\,{\sc ii}} profiles appear smoother, with less distinct component structure and very little absorption in very narrow components, compared to the corresponding high-resolution \mbox{Na\,{\sc i}} and \mbox{K\,{\sc i}} profiles.
As discussed below, such differences among the line profiles of those atomic species presumably reflect differences in the ionization and depletion behaviour (and thus distribution) of those species -- and seem consistent with the \mbox{Ti\,{\sc ii}} primarily tracing warmer, more diffuse gas.
 
\begin{table*}
\begin{minipage}{110mm}
\caption{Galactic \mbox{Ti\,{\sc ii}} equivalent widths.}
\label{tab:mwew} 
\begin{tabular}{@{}lllrrrr}
\hline
\multicolumn{1}{c}{HD}& 
\multicolumn{1}{l}{Name}&
\multicolumn{1}{c}{Run}& 
\multicolumn{1}{c}{S/N}& 
\multicolumn{1}{c}{W$_{\lambda}$(3072)}&
\multicolumn{1}{c}{W$_{\lambda}$(3241)}&
\multicolumn{1}{c}{W$_{\lambda}$(3383)}\\  
\hline
 23180 & o Per            & K00 &  80 & --           & --           &  5.9$\pm$1.1 \\ 
 24398 & $\zeta$ Per      & K00 &  75 & --           & --           &  2.8$\pm$0.8 \\ 
 24534 & X Per            & K00 &  60 &  2.9$\pm$1.0 & --           &  7.8$\pm$1.6 \\ 
 24760 & $\epsilon$ Per   & K94 & 210 & --           & --           &  5.7$\pm$0.5 \\
 27778 & 62 Tau           & K00 &  40 & --           & --           &  5.0$\pm$2.0 \\ 
 32630 & $\eta$ Aur       & K95 &  95 & --           & --           &  1.9$\pm$0.7 \\
 35149 & 23 Ori           & K00 &  75 & --           & --           &  6.6$\pm$1.3 \\
       &                  & V03 & 125 & --           & --           &  5.8$\pm$1.0 \\
 36486 & $\delta$ Ori     & K95 & 215 & --           & --           &  3.9$\pm$0.8 \\
 37022 & $\theta^1$ Ori C & K00 &  55 & --           & --           & 10.2$\pm$2.4 \\
 37043 & $\iota$ Ori      & K95 & 215 & --           & --           &  8.0$\pm$0.6 \\
 37128 & $\epsilon$ Ori   & K94 & 280 & --           & --           &  8.6$\pm$0.8 \\
 37742 & $\zeta$ Ori      & K95 & 175 & --           & --           &  5.2$\pm$0.9 \\
 47839 & S Mon A          & K45 & 115 & --           & --           & 22.1$\pm$2.1 \\
       & Walker 67        & V03 &  55 & --           & --           & 10.9$\pm$2.5 \\
 62542 &                  & V03 &  85 & --           & --           & 12.6$\pm$2.4 \\
 72127A&                  & V03 & 130 & --           & --           & 17.0$\pm$2.6 \\
 72127B&                  & V03 & 105 & --           & --           & 17.2$\pm$1.5 \\
 73882 &                  & V03 &  90 & --           & --           & 51.6$\pm$3.2 \\
 91316 & $\rho$ Leo       & K95 & 335 & --           & --           & 29.7$\pm$0.7 \\
106625 & $\gamma$ Crv     & V00 & 435 & --           & --           &  1.4$\pm$0.1 \\
110432 &                  & V00 & 355 & --           &  4.9$\pm$0.4 & 10.4$\pm$0.4 \\
112244 &                  & V00 & 435 &  9.4$\pm$1.0 & 17.5$\pm$0.5 & 31.6$\pm$0.5 \\ 
113904 & $\theta$ Mus     & V00 & 265 & 19.1$\pm$2.2 & 36.0$\pm$1.2 & 66.7$\pm$1.4 \\ 
114213 &                  & V00 & 265 & 24.8$\pm$3.9 & 69.0$\pm$1.9 & 93.1$\pm$1.2 \\
116658 & $\alpha$ Vir     & K95 & 455 & --           & --           &  0.8$\pm$0.2 \\
141637 & 1 Sco            & K95 &  35 & --           & --           &  15.$\pm$4.  \\
143018 & $\pi$ Sco        & K95 & 245 & --           & --           &  4.0$\pm$0.7 \\
144217 & $\beta^1$ Sco    & K95 & 225 & --           & --           &  7.2$\pm$0.7 \\
147165 & $\sigma$ Sco     & K95 & 190 & --           & --           & 13.0$\pm$0.8 \\
147701 &                  & V00 & 250 & --           & 15.1$\pm$1.1 & 19.2$\pm$1.4 \\ 
147888 & $\rho$ Oph D     & V00 & 255 & --           &  9.1$\pm$0.8 & 13.0$\pm$0.6 \\
147889 &                  & V00 & 305 & --           &  5.0$\pm$0.6 &  9.5$\pm$0.7 \\
147933 & $\rho$ Oph A     & V00 & 325 &  5.7$\pm$1.3 &  9.1$\pm$0.7 & 14.0$\pm$0.6 \\
148184 & $\chi$ Oph       & V00 & 285 &  3.9$\pm$1.0 &  9.6$\pm$0.7 & 13.1$\pm$0.6 \\
148379 &                  & V00 & 315 & 21.6$\pm$1.8 & --           & 69.8$\pm$0.8 \\ 
149404 &                  & V00 & 225 & 18.7$\pm$1.8 & 29.4$\pm$0.8 & 52.5$\pm$1.0 \\
149757 & $\zeta$ Oph      & K45 & 220 & --           & --           &  9.1$\pm$0.8 \\ 
       &                  & V00 & 455 &  2.5$\pm$0.7 &  7.0$\pm$0.4 &  9.5$\pm$0.3 \\
150136 &                  & V00 & 305 & --           & 23.2$\pm$0.6 & 38.1$\pm$0.7 \\
152236 & $\zeta^1$ Sco    & V00 & 315 & 14.9$\pm$1.4 & --           & 56.9$\pm$1.0 \\ 
154368 &                  & V00 & 315 &  8.9$\pm$1.5 & 20.3$\pm$0.8 & 36.1$\pm$0.8 \\ 
159561 & $\alpha$ Oph     & K45 & 185 & --           & --           &  4.6$\pm$0.4 \\
161056 &                  & V00 & 270 &  4.9$\pm$1.3 & 11.9$\pm$0.7 & 20.9$\pm$0.7 \\ 
166734 &                  & V00 & 325 & --           & 53.2$\pm$0.7 & 87.0$\pm$0.9 \\ 
167971 &                  & V00 & 305 & 21.7$\pm$2.1 & 42.5$\pm$0.8 & 71.0$\pm$0.9 \\
       & BD$-$14~5037     & V00 & 370 & 33.4$\pm$2.1 & --           & 98.5$\pm$0.6 \\ 
169454 &                  & V00 & 255 & 34.2$\pm$1.9 & 64.6$\pm$1.2 &105.2$\pm$1.0 \\ 
170740 &                  & V00 & 295 & --           & --           & 16.0$\pm$0.5 \\
172028 &                  & V00 & 280 & --           & 15.6$\pm$1.0 & 23.8$\pm$0.7 \\
182985 &                  & V00 & 415 & --           &  1.7$\pm$0.3 &  4.9$\pm$0.4 \\ 
183143 &                  & V00 & 305 & --           & 44.5$\pm$0.8 & 67.4$\pm$0.9 \\
188220 &                  & V00 & 325 & --           &  4.3$\pm$0.5 &  5.9$\pm$0.4 \\
197345 & $\alpha$ Cyg     & K94 & 155 & --           & --           & 15.4$\pm$0.7 \\
199579 &                  & K00 &  65 & --           & --           & 17.9$\pm$2.1 \\
205637 & $\epsilon$ Cap   & V00 & 455 & --           &  6.1$\pm$0.4 & 13.4$\pm$0.4 \\
206267 &                  & K00 &  65 & --           & --           & 40.2$\pm$2.2 \\
207198 &                  & K00 &  60 & --           & --           & 47.1$\pm$2.9 \\
210121 &                  & V00 & 370 &  7.7$\pm$0.7 & 12.2$\pm$0.5 & 26.7$\pm$0.6 \\ 
212571 & $\pi$ Aqr        & V00 & 455 &  5.7$\pm$0.5 &  9.7$\pm$0.4 & 17.0$\pm$0.3 \\
217675 & o And            & K94 & 145 & --           & --           &  5.6$\pm$0.5 \\
219688 & $\psi^2$ Aqr     & V03 & 205 & --           & --           &  2.1$\pm$0.6 \\
\hline
\end{tabular}
\medskip
~ ~ \\
Entries are equivalent width $\pm$ 1$\sigma$ uncertainty (m\AA).  
S/N values are per half resolution element near 3383 \AA.
Run K45 refers to sum of spectra from runs K94 and K95.
W(3072) for X Per is from archival {\it HST} spectra.
\end{minipage}
\end{table*}

The equivalent widths $W$(3383) for the \mbox{Ti\,{\sc ii}} $\lambda$3383 lines measured from the normalized spectra are listed in Table~\ref{tab:mwew} (Milky Way sight lines) and Table~\ref{tab:mcew} (MC sight lines); corresponding values for the weaker $\lambda$3072 and $\lambda$3241 lines are included for some of the Galactic sight lines observed in run V00.
For the SMC and LMC sight lines, the absorption is separated into contributions from gas in our Galaxy (MW) and from gas in the Magellanic Clouds.
Toward the SMC, the division is made at a (heliocentric) velocity of 60 km~s$^{-1}$; toward the LMC, the division is made at $v$ = 100 km~s$^{-1}$ (e.g., Lehner, Staveley-Smith, \& Howk 2009).
For the \mbox{Ti\,{\sc ii}} absorption toward the LMC, separate values are listed for ``intermediate-velocity'' absorption (100 km~s$^{-1}$ $\la$ $v$ $\la$ 200 km~s$^{-1}$) and for the ``main LMC'' absorption ($v$ $\ga$ 200 km~s$^{-1}$).
The 1~$\sigma$ uncertainties on the equivalent widths include contributions from both photon noise (Jenkins et al. 1973) and continuum placement (Sembach \& Savage 1992); the latter generally dominates the total uncertainty for very broad absorption features.

Comparisons of $W$(3072) versus $W$(3383) and $W$(3241) versus $W$(3383) reveal essentially linear relationships, with slopes 0.31$\pm$0.02 and 0.60$\pm$0.02, respectively -- very close to the values 0.28 and 0.59 expected from the ratios of $f\lambda^2$ for the three \mbox{Ti\,{\sc ii}} lines [using the rest wavelengths and oscillator strengths listed by Morton (2003)].
Those observed relationships thus suggest that the relative $f$-values are accurate and that saturation effects generally are not very significant, even for the strongest $\lambda$3383 line.

\begin{table} 
\caption{Magellanic Clouds \mbox{Ti\,{\sc ii}} $\lambda$3383 equivalent widths.}
\label{tab:mcew}
\begin{tabular}{@{}lrrrr} 
\hline
\multicolumn{1}{l}{Star}&  
\multicolumn{1}{c}{S/N}&  
\multicolumn{1}{c}{MW}&
\multicolumn{1}{c}{IV}&   
\multicolumn{1}{c}{MC}\\   
\hline
Sk 13       & 165 &  22.8$\pm$1.8 &     --       & 281.0$\pm$4.2 \\
Sk 18       & 190 &  20.3$\pm$1.4 &     --       & 231.0$\pm$2.2 \\
AV 47       &  40 &  22.8$\pm$4.4 &     --       & 185.9$\pm$11.5\\
Sk 40       & 110 &  26.2$\pm$2.3 &     --       & 233.0$\pm$4.1 \\
AV 80       &  47 &  17.6$\pm$3.8 &     --       & 200.3$\pm$12.5\\
AV 95       &  39 &  22.9$\pm$5.2 &     --       & 209.1$\pm$17.7\\
AV 120      &  32 &  26.3$\pm$5.2 &     --       &  76.5$\pm$10.1\\
AV 207      &  25 &  21.1$\pm$6.1 &     --       & 100.8$\pm$17.1\\
AV 321      &  42 &  16.5$\pm$4.0 &     --       &  44.1$\pm$3.1 \\
AV 388      &  30 &  27.5$\pm$6.9 &     --       &  46.8$\pm$8.3 \\
AV 440      &  26 &  24.4$\pm$7.7 &     --       & 115.7$\pm$22.2\\
Sk 143      &  60 &  28.1$\pm$5.3 &     --       &  36.2$\pm$8.1 \\
AV 476      &  85 &  24.1$\pm$2.7 &     --       & 202.0$\pm$4.0 \\
Sk 155      & 195 &  32.9$\pm$1.8 &     --       & 121.2$\pm$3.6 \\
Sk 190      &  40 &  39.5$\pm$10.0&     --       &  48.0$\pm$9.1 \\
 & \\
Sk$-$67 2   & 150 &  28.5$\pm$1.4 &     --       &  45.6$\pm$2.2 \\
Sk$-$67 5   & 190 &  31.9$\pm$1.5 &     --       &  61.2$\pm$1.0 \\
Sk$-$66 18  &  45 &  26.8$\pm$4.2 &     --       &  53.6$\pm$6.2 \\
LH 10-3061  &  40 &  36.7$\pm$5.4 &     --       & 109.9$\pm$7.8 \\
Sk$-$69 50  &  47 &  32.1$\pm$3.8 & 11.8$\pm$2.5 &  55.9$\pm$5.6 \\
Sk$-$67 22  &  53 &  40.8$\pm$4.1 &     --       &  61.7$\pm$9.3 \\
Sk$-$67 38  &  44 &  36.5$\pm$5.0 &     --       &  24.1$\pm$11.6\\
Sk$-$70 69  &  43 &  44.3$\pm$3.4 &     --       &  13.8$\pm$4.8 \\
Sk$-$68 41  &  32 &  35.5$\pm$6.1 &     --       &  10.2$\pm$4.3 \\
Sk$-$68 52  & 190 &  33.1$\pm$1.7 &     --       & 115.0$\pm$2.4 \\
BI 128      &  28 &  29.3$\pm$5.0 &     --       &  38.9$\pm$12.0\\
Sk$-$65 47  &  57 &  45.9$\pm$8.3 &     --       &  32.2$\pm$8.3 \\
Sk$-$68 73  & 190 &  33.3$\pm$1.2 &  4.0$\pm$0.9 & 102.9$\pm$1.5 \\
Sk$-$67 101 &  55 &  29.0$\pm$2.2 &  5.8$\pm$3.3 &  58.2$\pm$5.1 \\
BI 170      &  45 &  37.9$\pm$3.9 &  9.1$\pm$3.2 &  90.9$\pm$6.6 \\
Sk$-$66 100 &  48 &  37.0$\pm$3.2 &     --       &  45.4$\pm$5.3 \\
Sk$-$67 169 &  31 &  31.1$\pm$3.9 &     --       &  17.9$\pm$6.4 \\
Sk$-$67 168 &  53 &  35.6$\pm$7.4 &     --       &  30.0$\pm$6.6 \\
Sk$-$67 211 &  70 &  30.6$\pm$2.1 &     --       &  76.1$\pm$4.6 \\
BI 229      &  48 &  36.6$\pm$3.4 &     --       &  17.9$\pm$4.4 \\
BI 237      &  38 &  39.1$\pm$9.9 &     --       &  97.6$\pm$9.4 \\
Sk$-$66 171 &  31 &  49.2$\pm$14.3&     --       &  35.7$\pm$11.8\\
BI 253      &  42 &  57.6$\pm$11.3&     --       & 169.6$\pm$7.0 \\
Sk$-$68 135 & 215 &  34.9$\pm$1.6 &  4.8$\pm$0.9 & 159.1$\pm$2.4 \\
Mk 42       &  38 &  39.0$\pm$4.4 &     --       & 154.7$\pm$10.9\\
Sk$-$69 246 & 215 &  45.2$\pm$2.4 &  1.1$\pm$0.6 & 136.1$\pm$1.8 \\
BI 272      &  54 &  34.8$\pm$3.1 &     --       &  59.9$\pm$5.5 \\
Sk$-$70 115 & 190 &  42.0$\pm$2.3 &     --       & 182.6$\pm$2.9 \\
            &  75 &  36.6$\pm$4.4 &     --       & 179.7$\pm$7.6 \\
\hline
\end{tabular}
\medskip
~ ~ \\
Entries are equivalent width $\pm$ 1$\sigma$ uncertainty (m\AA). 
For SMC sight lines, MW is for $v$ $\la$ 60 km s$^{-1}$; MC is for $v$ $\ga$ 60 km s$^{-1}$. 
For LMC sight lines, MW is for $v$ $\la$ 100 km s$^{-1}$; IV is for 100 km s$^{-1}$ $\la$ $v$ $\la$ 200 km s$^{-1}$; MC is for $v$ $\ga$ 200 km s$^{-1}$.
\end{table}


\subsection{Column densities}
\label{sec-cd}

Total sight line column densities were obtained from the normalized \mbox{Ti\,{\sc ii}} spectra both by integrating the apparent (instrumentally smeared) optical depth (AOD) over the line profile (Hobbs 1974; Savage \& Sembach 1991) and by performing multi-component fits to the line profiles (e.g., Welty et al. 2003).
The AOD method can be used to obtain the apparent column density $N_{\rm a}$($v$) as a function of velocity, but does not otherwise distinguish individual components in the profiles. 
Effects of saturation in the profiles may be estimated if several lines of different strength can be measured (Jenkins 1996).
Fits to absorption-line profiles can yield column densities ($N$), line widths ($b$ $\sim$ FWHM/1.665), and velocities ($v$) for individual components discernible in the spectra (given the resolution and S/N ratios characterizing the spectra and assuming that each component may be represented by a Voigt profile).
While saturation effects are treated explicitly in the fits, having several lines of different strength can provide important constraints on $N$ and $b$.

Fits to high-resolution spectra of \mbox{Na\,{\sc i}}, \mbox{K\,{\sc i}}, and \mbox{Ca\,{\sc ii}}, for Galactic sight line samples having significant overlap with the present \mbox{Ti\,{\sc ii}} sample, have suggested that most of the sight lines contain multiple, closely blended components, with median separation of order 1.2--1.3 km~s$^{-1}$ (Welty et al. 1994, 1996; Welty \& Hobbs 2001).
The median component $b$-values for those samples are $\sim$0.7 km~s$^{-1}$ for \mbox{Na\,{\sc i}} and \mbox{K\,{\sc i}} and $\sim$1.3 km~s$^{-1}$ for \mbox{Ca\,{\sc ii}}; the larger value for \mbox{Ca\,{\sc ii}} reflects a population of broader components (likely representing warmer, more diffuse gas) not detected in the other two species.
Similar complex structure is seen in the high-resolution \mbox{Na\,{\sc i}} spectra available for a small number of Magellanic Clouds targets (Pettini \& Gillingham 1988; Welty et al. 1999a; Welty \& Crowther, in prep.).

Because of the relative weakness and comparative smoothness of the \mbox{Ti\,{\sc ii}} absorption, it is generally difficult to discern such complex structure for \mbox{Ti\,{\sc ii}}, even in the highest resolution \mbox{Ti\,{\sc ii}} spectra.
The observed similarities between the \mbox{Ti\,{\sc ii}} and \mbox{Ca\,{\sc ii}} profiles and the roughly constant $N$(\mbox{Ti\,{\sc ii}})/$N$(\mbox{Ca\,{\sc ii}}) ratio (for total sight line values), however, suggest that those two species might (on the whole) be similarly distributed.
We therefore fitted the \mbox{Ti\,{\sc ii}} profiles using the component structures derived from fits to the corresponding \mbox{Ca\,{\sc ii}} profiles, but only for sight lines where high-resolution (FWHM $\la$ 2 km~s$^{-1}$) \mbox{Ca\,{\sc ii}} spectra were available (e.g., Welty et al. 1996; Crawford 1995; Pan et al. 2004).
In those fits, the relative velocities and most of the $b$-values were fixed to the values found for \mbox{Ca\,{\sc ii}}; the individual component \mbox{Ti\,{\sc ii}} column densities and an overall velocity offset were allowed to vary.
 
The adopted Galactic interstellar component parameters are listed in Appendix Table~\ref{tab:comps}; the individual components are noted by tick marks above the higher resolution spectra in Fig.~\ref{fig:specg1}.
For each line of sight, the first line in Table~\ref{tab:comps} gives the source of the spectrum and (in square braces) the approximate 3~$\sigma$ column density limit for a single component with $b$ = 2 km~s$^{-1}$. 
Subsequent lines give the component number and the heliocentric velocity, $b$-value, column density (in units of 10$^{10}$~cm$^{-2}$), and ratio $N$(\mbox{Ti\,{\sc ii}})/$N$(\mbox{Ca\,{\sc ii}}) for that component.  
Component $b$-values in square braces were either varied or fixed (but relatively well determined) in the \mbox{Ca\,{\sc ii}} fits; values in parentheses also were fixed, but are less well determined.  

For our Galactic sight lines, the total sight line column densities obtained from both AOD integrations and profile fits are included in Appendix Table~\ref{tab:allwids} (a compilation of Galactic \mbox{Ti\,{\sc ii}} $\lambda$3383 equivalent widths and column densities from the literature).
In most cases, the total \mbox{Ti\,{\sc ii}} column densities obtained from detailed profile fits (ours or others') are not more than 20 percent (0.08 dex) higher than the values obtained from the AOD integrations (or directly from the equivalent widths) -- a further indication that the \mbox{Ti\,{\sc ii}} $\lambda$3383 absorption is seldom significantly saturated.
The few cases of larger differences (HD~42933, HD~74575, HD~179406) appear to be due to the adoption of very small $b$-values in the fits (Welsh et al. 1997) -- which then require larger column densities to produce the observed absorption.

\begin{table}
\caption{Galactic column densities and Ti depletions.}
\label{tab:mwcd}  
\begin{tabular}{@{}lrrrrrr} 
\hline
\multicolumn{1}{l}{Star}& 
\multicolumn{1}{c}{H I}&  
\multicolumn{1}{c}{H$_2$}& 
\multicolumn{1}{c}{\mbox{Ca\,{\sc ii}}}&  
\multicolumn{1}{c}{Ref}&   
\multicolumn{1}{c}{\mbox{Ti\,{\sc ii}}}&
\multicolumn{1}{c}{D(Ti)}\\  
\hline
HD 23180  & 20.82 &   20.60 &   12.09 & 1/1 &   11.22 & $-$2.86 \\
HD 24398  & 20.80 &   20.67 &   11.96 & 1/1 &   11.09 & $-$3.02 \\
HD 24534  & 20.73 &   20.92 &   12.41 & 1/1 &   11.34 & $-$2.92 \\
HD 24760  & 20.45 &   19.52 &   11.59 & 1/1 &   11.20 & $-$2.26 \\
HD 27778  & 20.95 &   20.79 &   11.97 & 2/1 &   11.15 & $-$3.10 \\
HD 32630  &  --   &    --   &   11.22 &--/1 &   10.79 &    --   \\
HD 35149  & 20.56 &   18.30 &   12.10 & 1/1 &   11.22 & $-$2.26 \\
HD 36486  & 20.17 &   14.74 &   11.72 & 1/1 &   11.12 & $-$1.97 \\
HD 37022  & 21.54 &$<$17.55 &   12.10 & 1/1 &   11.46 & $-$3.00 \\
HD 37043  & 20.20 &   14.69 &   11.93 & 1/1 &   11.39 & $-$1.73 \\
HD 37128  & 20.48 &   16.28 &   12.04 & 1/1 &   11.41 & $-$1.99 \\
HD 37742  & 20.39 &   15.86 &   11.97 & 1/1 &   11.18 & $-$2.13 \\
HD 47839  & 20.31 &   15.54 &   12.33 & 1/1 &   11.79 & $-$1.44 \\
Walker 67 &  --   &    --   &   12.28 &--/1 &   11.49 &    --   \\
HD 62542  & 20.90 &   20.81 &   11.96 & 2/1 &   11.54 & $-$2.70 \\
HD 72127A &  --   &    --   &   12.97 &--/1 &   11.68 &    --   \\
HD 72127B &  --   &    --   &   13.06 &--/1 &   11.78 &    --   \\
HD 73882  & 21.11 &   21.11 &   12.66 & 3/1 &   12.19 & $-$2.32 \\
HD 91316  & 20.44 &   15.61 &   12.03 & 1/1 &   11.93 & $-$1.43 \\
HD106625  &  --   &    --   &   10.84 &--/1 &   10.59 &    --   \\
HD110432  & 20.85 &   20.64 &   11.96 & 4/2 &   11.43 & $-$2.69 \\
HD112244  & 21.08 &   20.14 &    --   & 1/--&   11.94 & $-$2.15 \\
HD113904  & 21.15 &   19.83 &    --   & 1/--&   12.27 & $-$1.84 \\
HD114213  &  --   &    --   &    --   &--/--&   12.48 &    --   \\
HD116658  & 19.00 &   12.95 &   10.76 & 5/1 &   10.36 & $-$1.56 \\
HD141637  & 21.18 &   19.23 &   11.51 & 1/1 &   11.72 & $-$2.39 \\
HD143018  & 20.66 &   19.32 &   10.93 & 1/1 &   11.04 & $-$2.58 \\
HD144217  & 21.03 &   19.83 &   11.73 & 1/1 &   11.31 & $-$2.69 \\
HD147165  & 21.38 &   19.79 &   11.86 & 1/1 &   11.58 & $-$2.74 \\
HD147701  &  --   &    --   &    --   &--/--&   11.79 &    --   \\
HD147888  & 21.60 &   20.47 &   12.22 & 2/1 &   11.61 & $-$2.97 \\
HD147889  &  --   &    --   &    --   &--/--&   11.40 &    --   \\
HD147933  & 21.63 &   20.57 &   12.15 & 1/1 &   11.61 & $-$3.01 \\
HD148184  & 21.13 &   20.63 &   11.99 & 2/1 &   11.58 & $-$2.68 \\
HD148379  &  --   &    --   &    --   &--/--&   12.32 &    --   \\
HD149404  & 21.40 &   20.79 &   12.81 & 1/1 &   12.19 & $-$2.30 \\
HD149757  & 20.69 &   20.64 &   11.79 & 1/1 &   11.43 & $-$2.62 \\
HD150136  &  --   &    --   &    --   &--/--&   12.05 &    --   \\
HD152236  & 21.77 &   20.73 &   13.08 & 1/1 &   12.23 & $-$2.53 \\
HD154368  & 21.00 &   21.16 &   12.66 & 6/1 &   12.02 & $-$2.49 \\
HD159561  &  --   &    --   &   11.53 &--/1 &   11.12 &    --   \\
HD161056  & 21.20 &    --   &    --   & 1/--&   11.77 & $-$2.35 \\
HD166734  &  --   &    --   &    --   &--/--&   12.43 &    --   \\
HD167971  & 21.60 &   20.85 &    --   & 7/--&   12.33 & $-$2.32 \\
BD$-$14~5037& --  &    --   &    --   &--/--&   12.50 &    --   \\
HD169454  &  --   &    --   &    --   &--/--&   12.54 &    --   \\
HD170740  & 21.04 &   20.86 &   12.20 & 1/1 &   11.66 & $-$2.67 \\
HD172028  &  --   &    --   &    --   &--/--&   11.85 &    --   \\
HD182985  &  --   &    --   &    --   &--/--&   11.15 &    --   \\
HD183143  &  --   &    --   &    --   &--/--&   12.32 &    --   \\
HD188220  &  --   &    --   &   11.56 &--/3 &   11.26 &    --   \\
HD197345  &  --   &    --   &   12.29 &--/1 &   11.68 &    --   \\
HD199579  & 21.04 &   20.53 &   12.61 & 1/1 &   11.71 & $-$2.46 \\
HD205637  &  --   &    --   &    --   &--/--&   11.54 &    --   \\
HD206267  & 21.30 &   20.86 &$>$12.56 & 7/4 &   12.09 & $-$2.37 \\
HD207198  & 21.34 &   20.83 &$>$12.64 & 1/4 &   12.16 & $-$2.31 \\
HD210121  & 20.63 &   20.75 &   12.49 & 8/1 &   11.83 & $-$2.28 \\
HD212571  & 20.50 &    --   &   12.10 & 2/1 &   11.69 & $-$1.73 \\
HD217675  &  --   &   19.67 &   11.73 &--/1 &   11.23 &    --   \\
HD219688  &  --   &    --   &   10.99 &--/1 &   10.76 &    --   \\
\hline
\end{tabular}
\medskip
~ ~ \\
Entries are log$N$ (cm$^{-2}$); limits are 3 $\sigma$.
References are for \mbox{H\,{\sc i}} and/or \mbox{Ca\,{\sc ii}}.
References for \mbox{H\,{\sc i}}: 1 = Diplas \& Savage 1994; 2 = this paper; 3 = Fitzpatrick \& Massa 1990; 4 = Rachford et al. 2001; 5 = York \& Rogerson 1976; 6 = Snow et al. 1996; 7 = Rachford et al. 2002; 8 = Welty \& Fowler 1992.
References for H$_2$: Savage et al. 1978; Rachford et al. 2002, 2009.
References for \mbox{Ca\,{\sc ii}}: 1 = Welty et al. 1996 and in prep.; 2 = Crawford 1995; 3 = Penprase 1993; 4 = Pan et al. 2004.
Depletions are relative to a solar Ti/H = $-$7.08 dex.
\end{table}

\begin{table*}
\begin{minipage}{150mm}
\caption{SMC column densities and Ti depletions.}
\label{tab:smccd}
\begin{tabular}{@{}lcrcrcrcrrc}   
\hline
\multicolumn{1}{l}{Star}&  
\multicolumn{1}{c}{$E(B-V)_{\rm MC}$}&
\multicolumn{1}{c}{\mbox{H\,{\sc i}}}&
\multicolumn{1}{c}{Ref}&  
\multicolumn{1}{c}{H$_2$}&   
\multicolumn{1}{c}{Ref}&
\multicolumn{1}{c}{\mbox{Ca\,{\sc ii}}}&  
\multicolumn{1}{c}{Ref}&   
\multicolumn{1}{c}{\mbox{Ti\,{\sc ii}}}&
\multicolumn{1}{c}{D(Ti)}&   
\multicolumn{1}{c}{Ref}\\   
\hline
Sk 13       & 0.16 &   22.04 &1&   20.36 &1&         13.04 &1&   12.98 & $-$1.36 &1\\
            &  --  &    --   & &    --   & &         12.93 &2&   12.95 &    --   &2\\
Sk 18       & 0.11 &   21.70 &1&   20.63 &2&         12.83 &1&   12.90 & $-$1.15 &1\\
AV 47       & 0.09 &   21.30 &1&   17.77 &2&         12.70 &1&   12.82 & $-$0.76 &1\\
Sk 33       & 0.07 &    --   & &    --   & &         13.11 &2&   13.24 &    --   &2\\
Sk 40       & 0.10 &   21.54 &2&    --   & &         12.65 &1&   12.92 & $-$0.90 &1\\
AV 80       & 0.15 &   21.81 &1&   20.15 &3&         12.82 &1&   12.81 & $-$1.30 &1\\
AV 95       & 0.10 &   21.49 &1&   19.40 &2&         12.71 &1&   12.82 & $-$0.96 &1\\
AV 120      & 0.05 &    --   & &    --   & &         12.69 &1&   12.35 &    --   &1\\
AV 207      & 0.08 &   21.43 &1&   19.40 &2&         12.77 &1&   12.49 & $-$1.23 &1\\
AV 214      & 0.25 &   21.40 &3&    --   & &         12.86 &2&   12.86 & $-$0.82 &2\\
Sk 78       & 0.03 &   21.02 &1&   15.66 &2&         12.68 &3&   12.38 & $-$0.92 &3\\
Sk 85       & 0.03 &   21.30 &1&   17.21 &2&         12.74 &2&   12.64 & $-$0.94 &2\\
AV 321      & 0.08 &    --   & &$<$14.56 &2&         12.46 &1&   12.10 &    --   &1\\
AV 388      & 0.07 &   21.15 &1&   19.40 &2&         12.33 &1&   12.14 & $-$1.31 &1\\
AV 398      & 0.29 &   21.90 &3&    --   & &         13.16 &2&   13.16 & $-$1.02 &2\\
AV 440      & 0.09 &   21.36 &1&    --   & &         12.59 &1&   12.57 & $-$1.07 &1\\
Sk 143      & 0.32 &   20.95 &1&   20.93 &1&         12.36 &1&   12.01 & $-$1.68 &1\\
            &  --  &    --   & &    --   & &         12.43 &2&   12.39 &    --   &2\\
AV 476      & 0.19 &   21.85 &1&   20.95 &3&         12.85 &1&   12.82 & $-$1.41 &1\\
Sk 155      & 0.09 &   21.40 &1&   19.15 &3&         12.69 &1&   12.57 & $-$1.11 &1\\
Sk 190      & 0.07 &    --   & &    --   & &         12.27 &1&   12.14 &    --   &1\\
Sk 191      & 0.10 &   21.51 &1&   20.65 &3&         12.21 &2&   12.17 & $-$1.73 &2\\
\hline
\end{tabular}
\medskip
~ ~ \\
Entries are log$N$ (cm$^{-2}$); limits are 3 $\sigma$.
SMC absorption is for $v$ $>$ 60 km~s$^{-1}$. \\
Depletions are relative to stellar Ti/H = $-$7.72.\\
References for \mbox{H\,{\sc i}}: 1 = Welty (in prep.); 2 = Fitzpatrick 1985a; 3 = Gordon et al. 2003.\\
References for H$_2$: 1 = Cartledge et al. 2005; 2 = Tumlinson et al. 2002; 3 = Welty et al. (in prep.).\\
References for \mbox{Ca\,{\sc ii}}: 1 = Welty \& Crowther (in prep.); 2 = Cox et al. 2007; 3 = Sembach et al. 1993 (see Welty \& Crowther, in prep.).\\
References for \mbox{Ti\,{\sc ii}}: 1 = this paper; 2 = Cox et al. 2007; 3 = Roth et al. 1995 (see Welty et al. 2001).
\end{minipage}
\end{table*}

\begin{table*}
\begin{minipage}{160mm}
\caption{LMC column densities and Ti depletions.}
\label{tab:lmccd}
\begin{tabular}{@{}lcrcrcrcrrc}
\hline
\multicolumn{1}{l}{Star}&
\multicolumn{1}{c}{$E(B-V)_{\rm MC}$}&
\multicolumn{1}{c}{\mbox{H\,{\sc i}}}&
\multicolumn{1}{c}{Ref}&
\multicolumn{1}{c}{H$_2$}&
\multicolumn{1}{c}{Ref}&
\multicolumn{1}{c}{\mbox{Ca\,{\sc ii}}}&
\multicolumn{1}{c}{Ref}&
\multicolumn{1}{c}{\mbox{Ti\,{\sc ii}}}&
\multicolumn{1}{c}{D(Ti)}&
\multicolumn{1}{c}{Ref}\\
\multicolumn{6}{c}{ }&
\multicolumn{1}{c}{IV/LMC}&
\multicolumn{1}{c}{ }&
\multicolumn{1}{c}{IV/LMC}&
\multicolumn{2}{c}{ }\\
\hline
Sk$-$67 2   & 0.24 &   21.00 &1&   20.95 &1&   11.00/12.12 &1&         12.13 & $-$2.07 &1\\
            &  --  &    --   & &    --   & &         12.32 &2&         12.16 &    --   &2\\
Sk$-$67 5   & 0.11 &   21.00 &2&   19.48 &2&   11.27/12.19 &1&         12.28 & $-$1.50 &1\\
            &  --  &    --   & &    --   & &         12.26 &2&         12.30 &    --   &2\\
Sk$-$66 18  & 0.10 &    --   & &    --   & &   11.04/12.32 &1&         12.20 &    --   &1\\
LH 10-3061  & 0.28 &    --   & &    --   & &   11.55/12.44 &1&         12.55 &    --   &1\\
Sk$-$69 50  & 0.03 &    --   & &    --   & &   12.18/12.31 &1&   11.51/12.13 &    --   &1\\
Sk$-$67 22  & 0.07 &   21.11 &2&    --   & &   11.08/12.11 &1&         12.26 & $-$1.61 &1\\
Sk$-$67 38  & 0.04 &    --   & &    --   & &   11.07/11.86 &1&         11.84 &    --   &1\\
Sk$-$70 69  & 0.01 &$<$20.50 &2&$<$14.41 &2&   11.44/11.96 &1&         11.59 &$>-$1.67 &1\\
Sk$-$68 41  & 0.02 &$<$20.50 &2&$<$14.15 &2&   10.86/12.02 &1&         11.46 &$>-$1.80 &1\\
Sk$-$68 52  & 0.12 &   21.40 &2&   19.47 &2&   11.11/12.58 &1&         12.54 & $-$1.63 &1\\
BI 128      & 0.01 &    --   & &    --   & &   11.20/11.98 &1&         12.05 &    --   &1\\
Sk$-$65 47  & 0.10 &$<$20.50 &2&    --   & &   10.89/12.11 &1&         11.96 &$>-$1.30 &1\\
Sk$-$68 73  & 0.34 &   21.60 &2&   20.09 &3&   11.96/12.55 &1&   11.04/12.52 & $-$1.86 &1\\
Sk$-$67 101 & 0.03 &$<$20.50 &2&$<$14.11 &2&   11.97/12.05 &1&   11.20/12.20 &$>-$1.02 &1\\
BI 170      & 0.01 &$<$20.90 &2&$<$13.95 &2&   11.71/12.46 &1&   11.40/12.39 &$>-$1.23 &1\\
Sk$-$66 100 & 0.06 &$<$20.50 &2&$<$14.25 &2&   11.32/12.08 &1&         12.12 &$>-$1.14 &1\\
Sk$-$67 169 & 0.02 &    --   & &$<$13.95 &2&   11.06/12.10 &1&         11.71 &    --   &1\\
Sk$-$67 168 & 0.07 &$<$20.70 &1&    --   & &   11.90/12.05 &1&         11.93 &$>-$1.53 &1\\
Sk$-$67 211 & 0.04 &   20.70 &2&$<$13.95 &2&   11.62/12.20 &1&         12.35 & $-$1.11 &1\\
Sk$-$69 202 & 0.13 &  (21.4) &2&    --   & &   11.86/12.59 &3&         12.34 & $-$1.82 &3\\
BI 229      & 0.10 &   20.70 &2&$<$14.08 &2&   11.38/11.87 &1&         11.69 & $-$1.77 &1\\
BI 237      & 0.14 &   21.62 &2&   20.05 &3&   11.72/12.44 &1&         12.49 & $-$1.89 &1\\
Sk$-$69 223 &  --  &    --   & &   20.07 &3&         12.68 &2&         12.77 &    --   &2\\
Sk$-$66 171 & 0.08 &$<$20.70 &2&    --   & &   10.80/12.12 &1&         12.02 &$>-$1.44 &1\\
BI 253      & 0.17 &   21.60 &2&   19.70 &3&   11.36/12.73 &1&         12.75 & $-$1.61 &1\\
Sk$-$68 135 & 0.20 &   21.60 &2&   19.86 &2&   11.89/12.72 &1&   11.12/12.70 & $-$1.67 &1\\
            &  --  &    --   & &    --   & &         12.68 &2&         12.73 &    --   &2\\
Mk 42       & 0.24 &   21.79 &2&    --   & &   11.55/12.81 &1&         12.69 & $-$1.86 &1\\
Sk$-$69 243 &  --  &   21.80 &3&    --   & &         12.88 &2&         12.76 & $-$1.80 &2\\
Sk$-$69 246 & 0.12 &   21.40 &2&   19.71 &2&   11.59/12.63 &1&   10.48/12.64 & $-$1.54 &1\\
Sk$-$69 274 & 0.13 &    --   & &    --   & &   11.74/12.58 &1&         12.76 &    --   &4\\
Sk$-$69 282 & 0.12 &   21.51 &1&    --   & &   11.78/12.71 &4&         12.74 & $-$1.53 &4\\
Sk$-$69 290 & 0.12 &    --   & &    --   & &   11.85/13.09 &4&         12.82 &    --   &4\\
BI 272      & 0.07 &$<$21.00 &2&$<$14.21 &2&   11.72/12.13 &1&         12.25 &$>-$1.51 &1\\
Sk$-$70 115 & 0.14 &   21.30 &2&   19.94 &2&   10.45/12.40 &1&         12.77 & $-$1.33 &1\\
            &  --  &    --   & &    --   & &         12.48 &4&         12.81 &    --   &4\\
Sk$-$70 120 & 0.06 &    --   & &    --   & &         12.59 &2&         12.63 &    --   &2\\
\hline
\end{tabular}
\medskip
~ ~ \\
Entries are log$N$ (cm$^{-2}$); limits are 3 $\sigma$.
LMC absorption is for $v$ $>$ 200 km~s$^{-1}$; IV absorption toward LMC is for 100 km~s$^{-1}$ $<$ $v$ $<$ 200 km~s$^{-1}$.
Depletions are relative to stellar Ti/H = $-$7.24.\\
References for \mbox{H\,{\sc i}}: 1 = Fitzpatrick 1985b, 1986; 2 = Welty (in prep.); 3 = Bluhm \& de Boer 2001.\\
References for H$_2$: 1 = Cartledge et al. 2005; 2 = Tumlinson et al. 2002; 3 = Welty et al. (in prep.).\\
References for \mbox{Ca\,{\sc ii}}: 1 = Welty \& Crowther (in prep.); 2 = Cox et al. 2006; 3 = Welty et al. 1999a; 4 = Caulet \& Newell 1996.\\
References for \mbox{Ti\,{\sc ii}}: 1 = this paper; 2 = Cox et al. 2006; 3 = Welty et al. 1999a; 4 = Caulet \& Newell 1996.
\end{minipage}
\end{table*}

The adopted total sight line \mbox{Ti\,{\sc ii}} column densities for all our Galactic, SMC, and LMC sight lines are listed in Tables~\ref{tab:mwcd}, \ref{tab:smccd}, and \ref{tab:lmccd}, respectively, together with the corresponding column densities of \mbox{H\,{\sc i}}, H$_2$, and \mbox{Ca\,{\sc ii}} and the Ti depletions obtained from $N$(\mbox{Ti\,{\sc ii}}) and $N$(H$_{\rm tot}$) = $N$(\mbox{H\,{\sc i}}) + 2$N$(H$_2$).
For several Galactic sight lines where the $\lambda$3383 line is relatively weak and/or our spectrum is of fairly low S/N ratio (e.g., $\zeta$~Per, $\delta$~Ori), the \mbox{Ti\,{\sc ii}} column density in Table~\ref{tab:mwcd} is based on data from all available references.
Where we have data for other \mbox{Ti\,{\sc ii}} lines (Table~\ref{tab:mwew}), the adopted $N$(\mbox{Ti\,{\sc ii}}) is a weighted average of the values derived from all the lines.
Tables~\ref{tab:smccd} and \ref{tab:lmccd} also include \mbox{Ca\,{\sc ii}} and \mbox{Ti\,{\sc ii}} column densities for a few Magellanic Clouds sight lines from previous studies (Caulet \& Newell 1996; Cox et al. 2006, 2007)\footnotemark.
\footnotetext{The Cox et al. (2007) \mbox{Ca\,{\sc ii}} values have not been adjusted for possible saturation effects; continuum placement may account for the differences in apparent column density derived from the H and K lines for some sight lines.}
For the LMC sight lines, separate values are given for the intermediate-velocity (100 km~s$^{-1}$ $\la$ $v$ $\la$ 200 lm~s$^{-1}$) and main LMC ($v$ $\ga$ 200 km~s$^{-1}$) absorption for both \mbox{Ca\,{\sc ii}} and \mbox{Ti\,{\sc ii}}; the two are combined in the various correlation plots discussed below.

Nearly all of the Galactic, SMC, and LMC \mbox{H\,{\sc i}} column densities were derived from fits to the (damped) \mbox{H\,{\sc i}} Lyman-$\alpha$ absorption-line profiles, with corrections for stellar Lyman-$\alpha$ absorption (for spectral types later than about B1) based on comparisons between stellar models and the Stromgren [c$_1$] photometric index (Diplas \& Savage 1994; Jenkins 2009). 
For HD~27778 (B3 V) and HD~147888 (B5 V), the corrections for stellar \mbox{H\,{\sc i}} are the mean values for the respective spectral types (Diplas \& Savage 1994); the adopted $N$(\mbox{H\,{\sc i}}) are consistent with the relationships observed in the Galactic ISM between $N$(H$_{\rm tot}$) and $E(B-V)$ (Bohlin, Savage, \& Drake 1978) and between $N$(\mbox{H\,{\sc i}}) and the equivalent width of the diffuse interstellar band at 5780\AA\ (Herbig 1993; Welty et al. 2006).
For Be stars [HD~148184 (B2 IVpe), HD 212571 (B1 Ve)], where [c$_1$] appears to be systematically lower than the values for the corresponding ``normal'' spectral types, the mean corrections for the ``normal'' types were used.
For HD~62542 (B5 V), $N$(\mbox{H\,{\sc i}}) was estimated via $N$(H$_{\rm tot}$) = 5.8$\times$10$^{21}$~$E(B-V)$ (Bohlin et al. 1978) and the observed $N$(H$_2$) (Rachford et al. 2002).
For the high-latitude star HD~210121 (B3 V), $N$(\mbox{H\,{\sc i}}) was estimated from 21 cm emission (Welty \& Fowler 1992). 
Except for Sk$-$69~202 (SN 1987A; Welty et al. 1999a), the \mbox{H\,{\sc i}} column densities for the SMC and LMC are based directly on Lyman-$\alpha$ data from {\it IUE} (Fitzpatrick 1985a, 1985b, 1986; Gordon et al. 2003) or {\it HST} (Welty, in prep.); the H$_2$ column densities are primarily from the {\it FUSE} surveys of Tumlinson et al. (2002) and Cartledge et al. (2005).

The Ti depletions are determined relative to the stellar reference abundances listed in Table~\ref{tab:refab}.  
For the Galactic sight lines, Ti/H = $-$7.08 dex (the solar photospheric value); for the SMC, Ti/H = $-$7.72 dex (an average for A, F, G, and K supergiants); for the LMC, Ti/H = $-$7.24 dex (an average for F, G, and K supergiants).

\begin{table}
\caption{Ti reference abundances.}
\label{tab:refab}
\begin{tabular}{@{}rll}
\hline
\multicolumn{1}{c}{Ti/H}&
\multicolumn{1}{c}{Objects}&
\multicolumn{1}{l}{Reference}\\
\hline
\multicolumn{3}{c}{Milky Way} \\ 
 $-$7.08 & solar photosphere & Lodders 2003 \\
\multicolumn{3}{c}{SMC} \\
 $-$7.71 & 10 A supergiants  & Venn 1999 \\
 $-$7.64 & 3 F-G supergiants & Spite et al. 1989 \\
 $-$7.59 & 8 F-G supergiants & Russell \& Bessell 1989 \\
 $-$7.61 & 7 F-G supergiants & Luck \& Lambert 1992 \\
 $-$7.65 & 6 F-G supergiants & Luck et al. 1998 \\
 $-$7.68 & 6 K supergiants   & Hill et al. 1997 \\
 $-$7.91 & 6 G-K supergiants & Hill 1999 \\
 $-$7.87 & 7 G-K supergiants & Gonzalez \& Wallerstein 1999 \\
 $-$7.72 & adopted value     & \\
\multicolumn{3}{c}{LMC} \\
 $-$7.24 & 8 F-G supergiants & Russell \& Bessell 1989 \\
 $-$7.28 & 7 F-G supergiants & Luck \& Lambert 1992 \\
 $-$7.13 & 9 F-G supergiants & Hill et al. 1995 \\
 $-$7.22 & 10 F-G supergiants& Luck et al. 1998 \\
 $-$7.31 & 1 K star          & Spite et al. 1993 \\
 $-$7.24 & adopted value     & \\
\hline
\end{tabular}
\end{table}


\section{Results}
\label{sec-res}

\subsection{Comparisons with previous work}
\label{sec-compar}

Some of the sight lines included in this study have previous measures of the \mbox{Ti\,{\sc ii}} $\lambda$3383 absorption, generally derived from spectra obtained at lower spectral resolution and/or lower S/N ratios.
[Apart from the observations of $\zeta$~Per by Hobbs (1979) and of $\kappa$~Vel by Dunkin \& Crawford (1999), all previously reported \mbox{Ti\,{\sc ii}} spectra were obtained with instrumental FWHM $\ga$ 3 km~s$^{-1}$ -- i.e., at resolutions at least a factor of 2 poorer than our KPNO coud\'{e} feed / camera 6 spectra (see Appendix Table~\ref{tab:allwids}).]
Sixteen of our Galactic sight lines were also observed by Stokes (1978), five were observed by Magnani \& Salzer (1989, 1991), four were examined by Hunter et al. (2006), and a few others appear in various other studies (Table~\ref{tab:allwids}).
In most cases, our equivalent widths and column densities agree with the previous values within the mutual 1 $\sigma$ uncertainties; weighted and unweighted fits to a comparison of our $\lambda$3383 equivalent widths with those of Stokes (1978), for example, yield slopes consistent with unity and intercepts consistent with zero.

Examination of the larger set of \mbox{Ti\,{\sc ii}} data in Table~\ref{tab:allwids}, however, reveals significant differences in the measured equivalent widths and/or derived column densities for some sight lines (e.g., toward $\beta$~Ori, $\mu$~Col, $\zeta$~Pup, $\gamma^2$~Vel, HD~72127AB, $\theta$~Car, 9~Sgr, $\kappa$~Aql, HD~206267).
In principle, some of those differences could be due to temporal variations in the \mbox{Ti\,{\sc ii}} absorption, as have been seen for the trace species \mbox{Na\,{\sc i}} and/or \mbox{Ca\,{\sc ii}} in a number of sight lines (Crawford 2003; Lauroesch 2007).
While temporal variations in the absorption from dominant ions like \mbox{Ti\,{\sc ii}} have seldom been observed (e.g., Danks et al. 2001; Welty 2007), they would not be unexpected in sight lines probing disturbed regions -- e.g., for HD~72127AB, which lies behind the Vela supernova remnant (Hobbs et al. 1991; Cha \& Sembach 2000; Welty, Simon, \& Hobbs 2008).
There appear to be differences among the \mbox{Ti\,{\sc ii}} profiles toward $\mu$~Col reported by Hobbs (1984), Welsh et al. (1997), and Lallement et al. (2008) -- particularly in the strength of the broad absorption feature from about 10 to 35 km~s$^{-1}$, relative to that of the narrower component near +40 km~s$^{-1}$.
As the \mbox{Ti\,{\sc ii}} absorption is generally relatively weak and often spread over a fairly wide velocity range, however, many of the differences in equivalent width might just be due to differences in locating and fitting the local continuum; blending with weak, broad stellar lines may add further difficulty in defining the appropriate continuum toward later-type targets (e.g., for $\beta$~Ori; Hobbs 1984; Welsh et al. 1997).

Five of the Magellanic Clouds sight lines listed in Tables~\ref{tab:smccd} and \ref{tab:lmccd} were also observed and analyzed by Cox et al. (2006, 2007).
For four of those five sight lines, the derived SMC or LMC \mbox{Ti\,{\sc ii}} column densities agree within about 10 per cent (0.04 dex).
The lone discrepancy -- the factor-of-2.4 difference in the SMC $N$(\mbox{Ti\,{\sc ii}}) toward Sk~143 -- is somewhat puzzling, as the raw spectral data used in the two studies are the same, and as there is much better agreement between the derived Galactic \mbox{Ti\,{\sc ii}} column densities. 
Comparison of our normalized \mbox{Ti\,{\sc ii}} profile (Fig.~\ref{fig:specm1}) with the AOD profile for the $\lambda$3383 line in Cox et al. (2007) suggests that differences in the adopted continuum at SMC velocities are likely responsible for the difference in derived $N$(\mbox{Ti\,{\sc ii}}) in this case.

Observations of the \mbox{Ti\,{\sc ii}} $\lambda$3072 line toward a few stars have been reported by Federman, Weber, \& Lambert (1996) and Felenbok \& Roueff (1996), in papers focused on the nearby OH absorption features. 
For several of those sight lines, the equivalent widths of the OH lines at 3078 and 3081 \AA\ are in good agreement with previously reported values.
The additional narrow blue-ward components seen in the \mbox{Ti\,{\sc ii}} line by Felenbok \& Roueff toward both $\zeta$~Per and HD~27778 are not present in the $\lambda$3383 profiles (Fig.~\ref{fig:specg1}), however; they may be instrumental artefacts.
The \mbox{Ti\,{\sc ii}} column densities obtained from the $\lambda$3072 equivalent widths are somewhat larger than those obtained from independent observations of the stronger $\lambda$3383 line (Table~\ref{tab:allwids}).

\begin{figure}   
\includegraphics[width=84mm]{fig1.eps}
\caption{$N$(\mbox{Ti\,{\sc ii}}) vs. $N$(\mbox{Ca\,{\sc ii}}) for both Galactic and Magellanic Clouds sight lines.
Green circles denote LMC sight lines and red triangles denote SMC sight lines (filled symbols for detections, open symbols for limits); 
blue open squares denote Sco-Oph sight lines; blue asterisks denote Orion Trapezium region sight lines; plus signs denote other Galactic sight lines (with size indicating $\pm$1$\sigma$ uncertainties).
The two (nearly indistinguishable) solid lines, with slopes $\sim$ 0.7, represent weighted and unweighted fits to the Galactic data (not including the Sco-Oph or Trapezium sight lines).}
\label{fig:tivsca}
\end{figure}

\subsection{\mbox{Ti\,{\sc ii}} vs. \mbox{Ca\,{\sc ii}}}
\label{sec-tica}

Titanium and calcium are two of the most severely depleted elements in diffuse Galactic disc clouds (e.g., Jenkins 1987).
Several previous studies have noted that the column density ratio $N$(\mbox{Ti\,{\sc ii}})/$N$(\mbox{Ca\,{\sc ii}}) appears to be roughly constant in the Galactic ISM, with a value of 0.3--0.4 for clouds in both the disc and low halo (Stokes 1978; Albert 1983; Welsh et al. 1997; Hunter et al. 2006).
Figure~\ref{fig:tivsca} shows the relationship between the column densities of \mbox{Ti\,{\sc ii}} and \mbox{Ca\,{\sc ii}} for sight lines in our Galaxy and in the Magellanic Clouds (including well-determined values from other references)\footnotemark.
\footnotetext{Compilations of column densities for Galactic and Magellanic Clouds sight lines (with uncertainties and references) are maintained at http://www.astro.illinois.edu/$\sim$dwelty/coldens.html and coldens\_mc.html, respectively.}
In this figure (and in subsequent such figures), sight lines in the SMC and LMC are denoted by red triangles and green circles, respectively, with filled symbols representing detections (of both species) and open symbols representing limits (for one or both species).
Galactic sight lines in the Orion Trapezium and Sco-Oph regions, which exhibit somewhat different abundances for some species (e.g., Welty \& Hobbs 2001), are denoted by blue asterisks and blue open squares, respectively.
The solid lines show the weighted and unweighted fits to the Galactic data (generally not including the Trapezium or Sco-Oph sight lines)\footnotemark.
\footnotetext{The Sco-Oph sight lines include 1~Sco, $\pi$~Sco, $\delta$~Sco, $\beta^1$~Sco, $\omega^1$~Sco, $\sigma$~Sco, $\nu$~Sco, $\tau$~Sco, o~Sco, 22~Sco, HD~147701, $\rho$~Oph (A-D), and HD~147889 -- but not $\zeta$~Oph, $\chi$~Oph, 67~Oph, $\zeta^1$~Sco, HD~154090, or HD~154368.
The Trapezium sight lines include $\theta^1$~Ori~B, $\theta^1$~Ori~C, $\theta^2$~Ori~A, and HD~37061.
The fits were performed with a slightly modified version of the subroutine {\sc regrwt.f}, obtained from the statistical software archive maintained at Penn State (http://www.astro.psu.edu/statcodes); see Welty \& Hobbs (2001).
The fits accounted for the uncertainties in both dependent and independent variables, but did not consider limits.} 
The points for the Galactic ISM define a roughly linear trend -- with slope $\sim$ 0.7, relatively small scatter (rms deviation $\sim$ 0.2 dex; correlation coefficient $r$ = 0.865), and a few outliers (e.g., HD~72127AB, with \mbox{Ti\,{\sc ii}}/\mbox{Ca\,{\sc ii}} $\la$ 0.05, and $\kappa$~Vel). 
The mean \mbox{Ti\,{\sc ii}}/\mbox{Ca\,{\sc ii}} ratio for this particular set of Galactic sight lines is $\sim$ 0.3, consistent with the values found in previous studies -- though the ratio is somewhat higher (on average) toward the Sco-Oph stars and does decrease with increasing column density and/or increasing $E(B-V)$.
The Magellanic Clouds sight lines appear to exhibit a somewhat steeper relationship, with slope $\ga$ 1 and a higher mean \mbox{Ti\,{\sc ii}}/\mbox{Ca\,{\sc ii}} ratio $\sim$ 0.9 for both the SMC and LMC (see also Cox et al. 2006, 2007). 
Most of the Galactic sight lines, however, primarily probe gas in the disc; the Galactic absorption seen toward the Magellanic Clouds targets (due to both disc and halo gas in those directions) falls somewhat above the mean Galactic trend, at (12.19$\pm$0.16, 11.89$\pm$0.13) toward the SMC and (12.12$\pm$0.16,12.05$\pm$0.10) toward the LMC.

\begin{figure} 
\includegraphics[width=84mm]{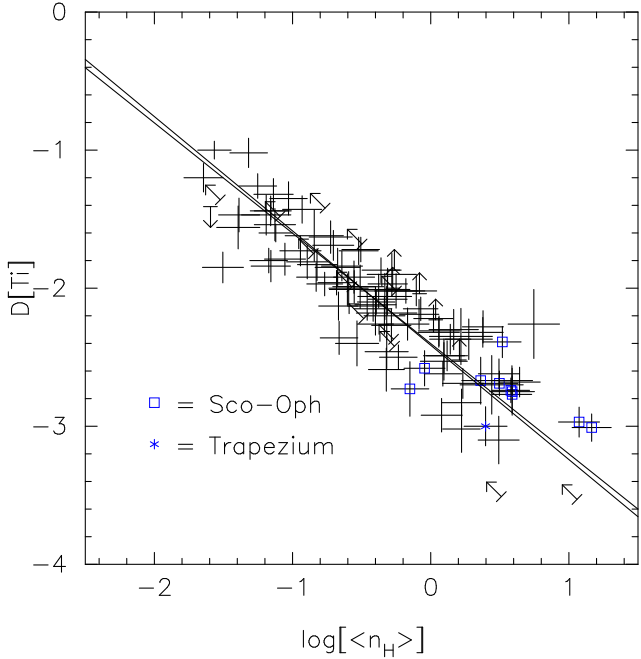}
\caption{D(Ti) vs. mean sight line density $<n_{\rm H}>$ for Galactic sight lines.
Blue open squares denote Sco-Oph sight lines; blue asterisks denote Orion Trapezium region sight lines; plus signs denote other Galactic sight lines (with size indicating $\pm$1$\sigma$ uncertainties).
The depletions are relative to the solar Ti abundance.
The two solid lines, with slopes $\sim$ $-$0.8, represent weighted and unweighted fits to the data (not including the Sco-Oph or Trapezium sight lines).}
\label{fig:tivsnh}
\end{figure}
  
\begin{figure}   
\includegraphics[width=84mm]{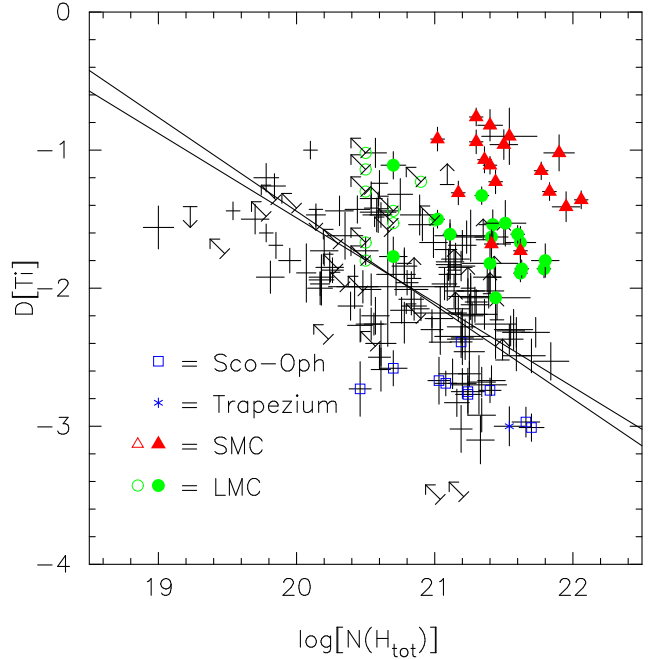}
\caption{D(Ti) vs. $N$(H) for both Galactic and Magellanic Clouds sight lines.
Green circles denote LMC sight lines and red triangles denote SMC sight lines (filled symbols for detections, open symbols for limits);
blue open squares denote Sco-Oph sight lines; blue asterisks denote Orion Trapezium region sight lines; plus signs denote other Galactic sight lines (with size indicating $\pm$1$\sigma$ uncertainties).
The depletions are relative to the solar Ti abundance for the Galactic sight lines and to the stellar Ti abundances for the SMC and LMC sight lines.
The two solid lines, with slopes $\sim$ $-$0.65, represent weighted and unweighted fits to the Galactic data (not including the Sco-Oph or Trapezium sight lines).}
\label{fig:tivsh}
\end{figure}
  
As also noted previously (Stokes 1978; Albert et al. 1993), there are indications of a somewhat broader distribution of the \mbox{Ti\,{\sc ii}}/\mbox{Ca\,{\sc ii}} ratio for individual velocity components, compared to the range seen for integrated sight line values.
The fits to the highest resolution Galactic \mbox{Ti\,{\sc ii}} profiles (using the component structures determined from high-resolution spectra of \mbox{Ca\,{\sc ii}}) yield \mbox{Ti\,{\sc ii}}/\mbox{Ca\,{\sc ii}} ratios ranging from $\sim$ 0.05 to $\sim$ 5 (last column in Appendix Table~\ref{tab:comps}).
The lowest values are typically found for components with small $b$-values (often with corresponding strong, narrow \mbox{Na\,{\sc i}} and/or \mbox{K\,{\sc i}} lines) and/or for components also detected in CH and CN -- which likely arise in relatively cold, dense gas.
Even lower values, however, are found for some intermediate-velocity components toward HD~72127AB -- where high \mbox{Fe\,{\sc ii}}/\mbox{Zn\,{\sc ii}} ratios suggest significant disruption of the dust grain populations (Welty et al. 2008).
The highest Galactic \mbox{Ti\,{\sc ii}}/\mbox{Ca\,{\sc ii}} values in our sample are found for some of the higher velocity components ($-$35 km~s$^{-1}$ $\la$ $v$ $\la$ $-$20 km~s$^{-1}$) toward stars in the Sco-Oph region and for some components toward the low halo star $\rho$ Leo.

Ratios of the apparent optical depths for \mbox{Ti\,{\sc ii}} and \mbox{Ca\,{\sc ii}} in the lower resolution UVES spectra of the Magellanic Clouds targets yield analogous information on the variation of \mbox{Ti\,{\sc ii}}/\mbox{Ca\,{\sc ii}} with velocity in each sight line.
In both the SMC and LMC, the overall distributions of the \mbox{Ti\,{\sc ii}}/\mbox{Ca\,{\sc ii}} ratios peak at about 1.25; the individual ratios (for velocity intervals of 2--4 km~s$^{-1}$) range from about 0.08 to 8.0.

\begin{figure*}   
\begin{minipage}{180mm}
\includegraphics[width=160mm]{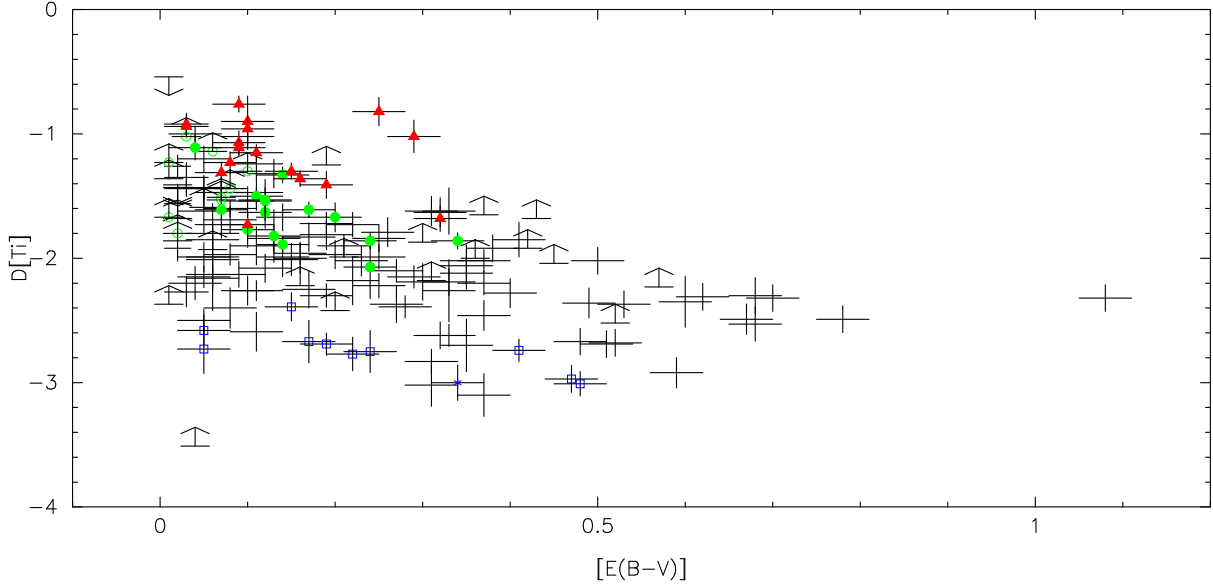}
\caption{D(Ti) vs. $E(B-V)$ for both Galactic and Magellanic Clouds sight lines.
Green circles denote LMC sight lines and red triangles denote SMC sight lines (filled symbols for detections, open symbols for limits);
blue open squares denote Sco-Oph sight lines; blue asterisks denote Orion Trapezium region sight lines; plus signs denote other Galactic sight lines (with size indicating $\pm$1$\sigma$ uncertainties).
The depletions are relative to the solar Ti abundance in our Galaxy and to the stellar Ti abundances in the SMC and LMC.}
\label{fig:tivsebv}
\end{minipage}
\end{figure*}
  
\begin{figure*}   
\begin{minipage}{180mm}
\includegraphics[width=160mm]{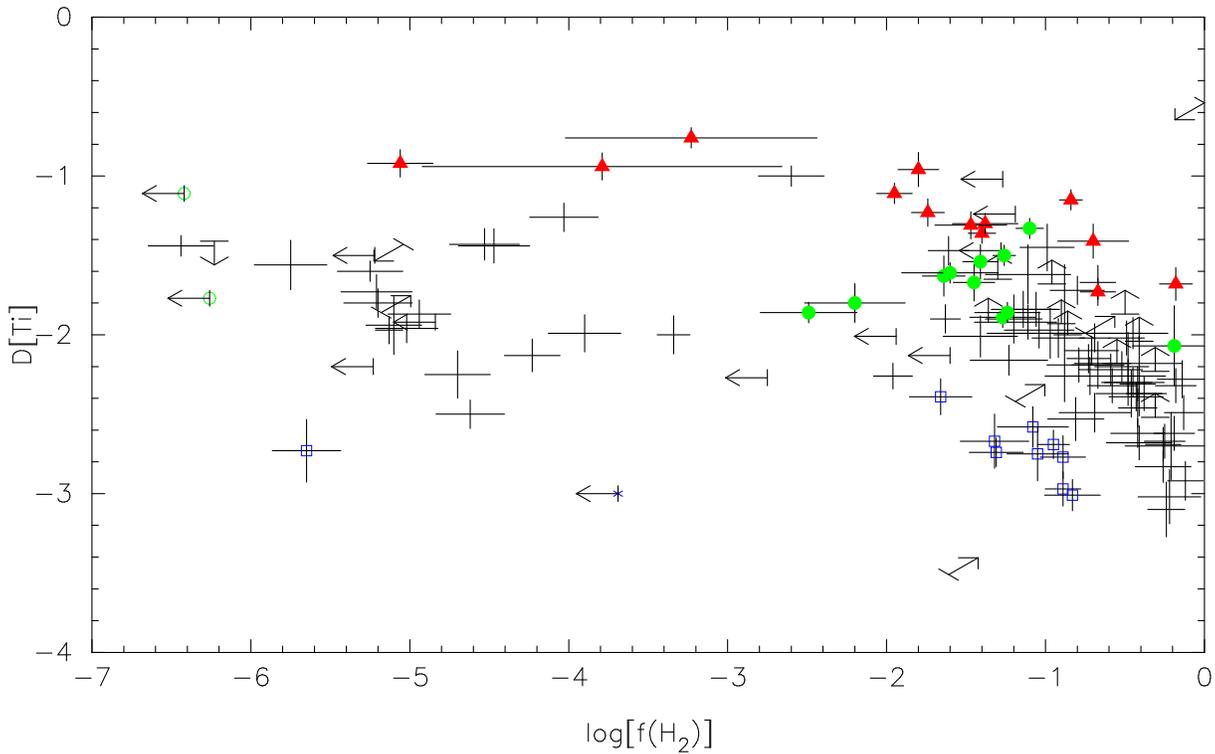}
\caption{D(Ti) vs. $f$(H$_2$) for both Galactic and Magellanic Clouds sight lines.
Green circles denote LMC sight lines and red triangles denote SMC sight lines (filled symbols for detections, open symbols for limits);
blue open squares denote Sco-Oph sight lines; blue asterisks denote Orion Trapezium region sight lines; plus signs denote other Galactic sight lines (with size indicating $\pm$1$\sigma$ uncertainties).
The depletions are relative to the solar Ti abundance in our Galaxy and to the stellar Ti abundances in the SMC and LMC.}
\label{fig:tivsfh2}
\end{minipage}
\end{figure*}
  
\subsection{Ti depletion}
\label{sec-depl}

Previous surveys of \mbox{Ti\,{\sc ii}} absorption have explored the behaviour of the titanium depletions D(Ti) = log[$\delta$(Ti)] in the disc and low halo of our Galaxy (e.g., Stokes 1978; Albert et al. 1993; Welsh et al. 1997; Hunter et al. 2006).
Figure~\ref{fig:tivsnh} gives an updated version of the relationship between D(Ti) and the average sight line density $<n_{\rm H}>$ = $N$(H$_{\rm tot}$)/$d$ (where $d$ is the distance to the target star) for the Galactic sample (e.g., Albert et al. 1993).
Figures~\ref{fig:tivsh}, \ref{fig:tivsebv}, and \ref{fig:tivsfh2} show the variations of D(Ti) with respect to the total hydrogen column density $N$(H$_{\rm tot}$), the colour excess $E(B-V)$, and the molecular fraction $f$(H$_2$), respectively, for the Milky Way, LMC, and SMC.
These figures are based on \mbox{Ti\,{\sc ii}} column density data (from this study and from other published studies) for 204 Galactic sight lines, 35 LMC sight lines, and 21 SMC sight lines (though not all those sight lines have corresponding data for hydrogen).
The titanium depletions in the Galactic sight lines range from about $-$1.0 dex (toward the distant, low halo star HD~93521) to about $-$3.1 dex (toward the relatively nearby, moderately reddened HD~27778) -- very similar to the range seen in the less extensive compilation of Jenkins (2009).
The depletions for the current sample of sight lines in the Magellanic Clouds (Tables~\ref{tab:smccd} and \ref{tab:lmccd}) are generally less severe, ranging from $-$1.1 to $-$2.1 dex in the LMC and from $-$0.8 to $-$1.7 dex in the SMC -- consistent with the Ti/H ratios found for smaller samples by Cox et al. (2006, 2007)\footnotemark.
\footnotetext{The Ti depletions listed by Cox et al. (2007) for the SMC sight lines, however, are with respect to a Galactic reference abundance ($-$6.9 dex).}
The most severe depletions in the Magellanic Clouds are found toward the moderately reddened stars Sk~143 (SMC) and Sk$-$67~2 (LMC) -- the two sight lines with the highest molecular fractions and the only detections of CN absorption in the currently available sample (Cartledge et al. 2005; Welty et al. 2006).\footnotemark
\footnotetext{While D(Ti) is listed as $-$1.73 dex for Sk~191 in Table~\ref{tab:smccd}, that sight line may be characterized by a somewhat lower metallicity than is typical for the SMC -- in which case the depletion would be less severe.}

While the mean sight line density $<n_{\rm H}>$ is an imperfect indicator of the local hydrogen density, the relationship between D(Ti) and $<n_{\rm H}>$ is remarkably tight in the Galactic ISM, with slope $\sim$ $-$0.8 and rms scatter $\sim$ 0.17 dex -- suggestive of more severe Ti depletion at higher $n_{\rm H}$ (though see Sec.~\ref{sec-dens} below).
All three galaxies exhibit generally increasingly severe titanium depletions with increasing $N$(H$_{\rm tot}$) and $E(B-V)$, with similar slopes ($\sim$ $-$0.65) versus $N$(H$_{\rm tot}$) and significant scatter (rms $\sim$ 0.3 dex for the Galactic sight lines) at any given $N$(H$_{\rm tot}$).
On average, the titanium depletions toward the Trapezium and Sco-Oph targets are more severe than the average Galactic values at the corresponding $N$(H$_{\rm tot}$), by a factor $\sim$ 3 (though they appear more ``normal'' versus $<n_{\rm H}>$); inclusion of those sight lines yields slopes $\sim$ $-$0.7 versus $N$(H$_{\rm tot}$), consistent with the value found by Wakker \& Mathis (2000). 
A number of the Galactic sight lines with the least severe depletions [at a given $N$(H$_{\rm tot}$)] probe either relatively long paths through the lower Galactic halo (HD~93521, HD~149881, HD~215733) or somewhat disturbed regions (e.g., HD~93205 and HD~303308 in Carina).
The depletions are typically even less severe in the Magellanic Clouds, by factors of $\sim$ 5 in the LMC and $\sim$ 20 in the SMC [at any given $N$(H$_{\rm tot}$)].
Although the differences among the three galaxies are smaller when viewed versus $E(B-V)$ (a consequence of the smaller dust-to-gas ratios in the Magellanic Clouds), the titanium depletions (on average) still are less severe in the LMC and SMC than in our sample of Galactic sight lines.
While H$_2$ column densities are not known for some of the Magellanic Clouds sight lines in Fig.~\ref{fig:tivsh}, examination of the available {\it FUSE} spectra indicates that $f$(H$_2$) is generally less than about 0.1 in those cases (Welty et al., in prep.).

For molecular fractions less than 0.1, the Galactic titanium depletions appear to be roughly constant, with mean D(Ti) $\sim$ $-$1.8 dex [and with considerable scatter at any given $f$(H$_2$)].
For $f$(H$_2$) $\ga$ 0.1, the mean D(Ti) is about $-$2.3 dex, and the depletions become increasingly severe with increasing $f$(H$_2$); several sight lines with more than half the hydrogen in molecular form have D(Ti) $\la$ $-$3.0 dex.
The Sco-Oph and Trapezium sight lines stand well apart from the general Galactic trend -- reflecting a combination of fairly severe depletions [mean D(Ti) $\sim$ $-$2.8 dex] and somewhat lower molecular fractions [versus other sight lines with comparable $N$(H$_{\rm tot}$) or depletions].
While the depletions are again less severe for the LMC and SMC [at any given $f$(H$_2$)], the small sample size renders more quantitative estimates of the differences rather uncertain.

\begin{figure*}   
\begin{minipage}{180mm}
\includegraphics[width=160mm]{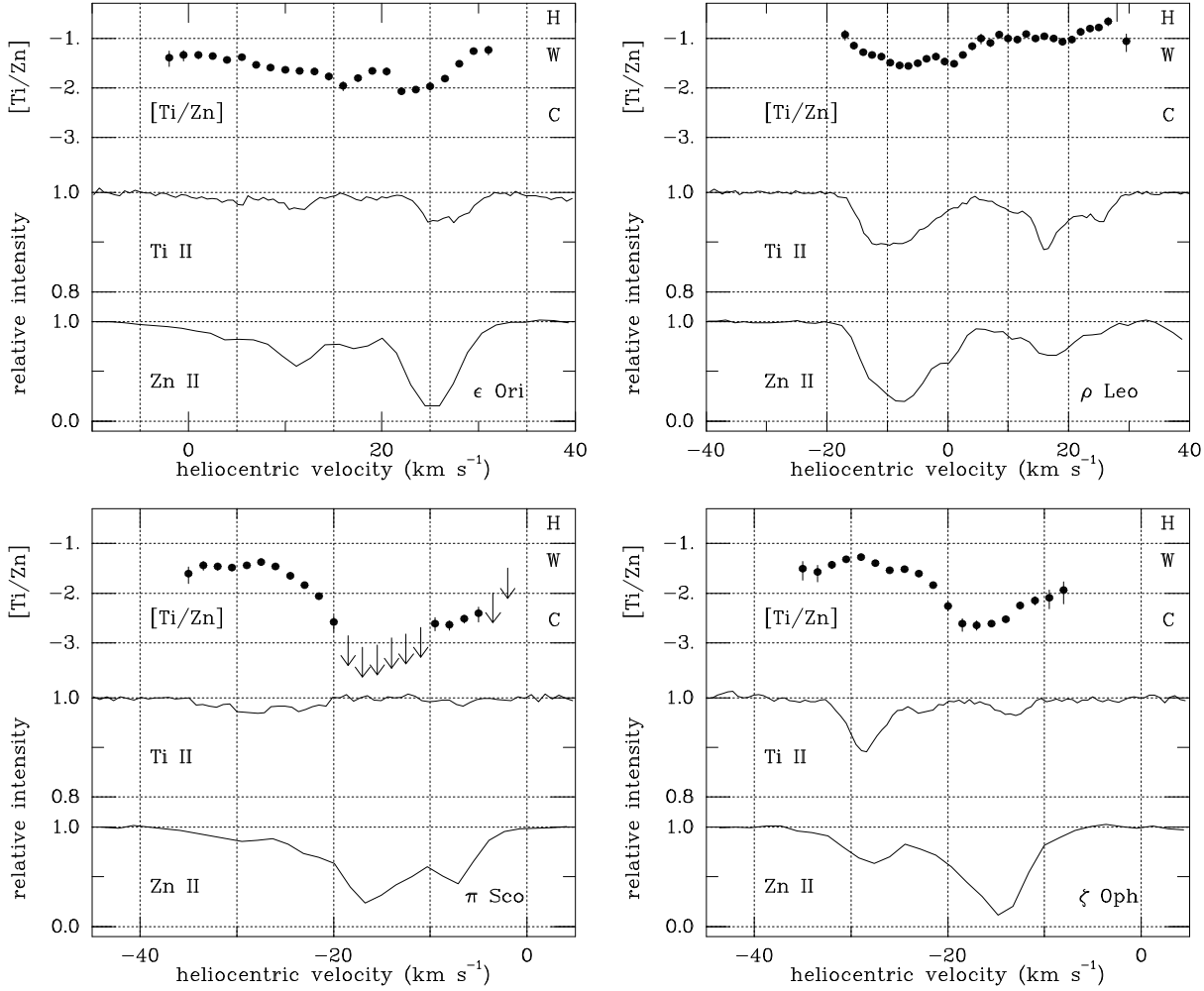}
\caption{Spectra of \mbox{Ti\,{\sc ii}} ($\lambda$3383) and \mbox{Zn\,{\sc ii}} ($\lambda$2026 or $\lambda$2062) for four Galactic sight lines.
The \mbox{Ti\,{\sc ii}} spectra are from the KPNO coud\'{e} feed (FWHM $\sim$ 1.3--1.5 km~s$^{-1}$); the \mbox{Zn\,{\sc ii}} spectra are from the {\it HST} GHRS (FWHM $\sim$ 3.4--3.7 km~s$^{-1}$).
The points at the top of each panel show the relative abundance [Ti/Zn] (indicative of the titanium depletion), obtained from ratios of the ``apparent'' optical depths, sampled at intervals of 1.5 km~s$^{-1}$.
Uncertainties in [Ti/Zn] are 1~$\sigma$; limits are 3~$\sigma$.
The letters H, W, and C (at the upper right in each panel) give representative values for [Ti/Zn] for clouds in the low Galactic halo; warm, diffuse clouds in the disc; and colder, denser disc clouds.}
\label{fig:mwzn}
\end{minipage}
\end{figure*}
  
Because hydrogen column densities, colour excesses, and molecular fractions cannot be determined directly for the individual components comprising the complex interstellar line profiles seen in most sight lines, it is more difficult to determine the depletion behaviour for individual interstellar clouds.
For a small number of sight lines, however, high-resolution (FWHM $\sim$ 2--4 km~s$^{-1}$) UV spectra of the absorption lines from relatively undepleted dominant ions (e.g, \mbox{O\,{\sc i}}, \mbox{Zn\,{\sc ii}}, \mbox{Kr\,{\sc i}}) may be used to estimate the total hydrogen column densities (and then the titanium depletions) in those individual components/clouds, via fits to the various absorption-line profiles.
Alternatively, if the various absorption line profiles are placed on a common velocity grid, then ratios of the apparent optical depths [e.g., $\tau_{\rm a}$(\mbox{Ti\,{\sc ii}})/$\tau_{\rm a}$(\mbox{Zn\,{\sc ii}})] may be used to gauge the Ti depletion as a function of velocity.

\begin{figure}   
\includegraphics[width=75mm]{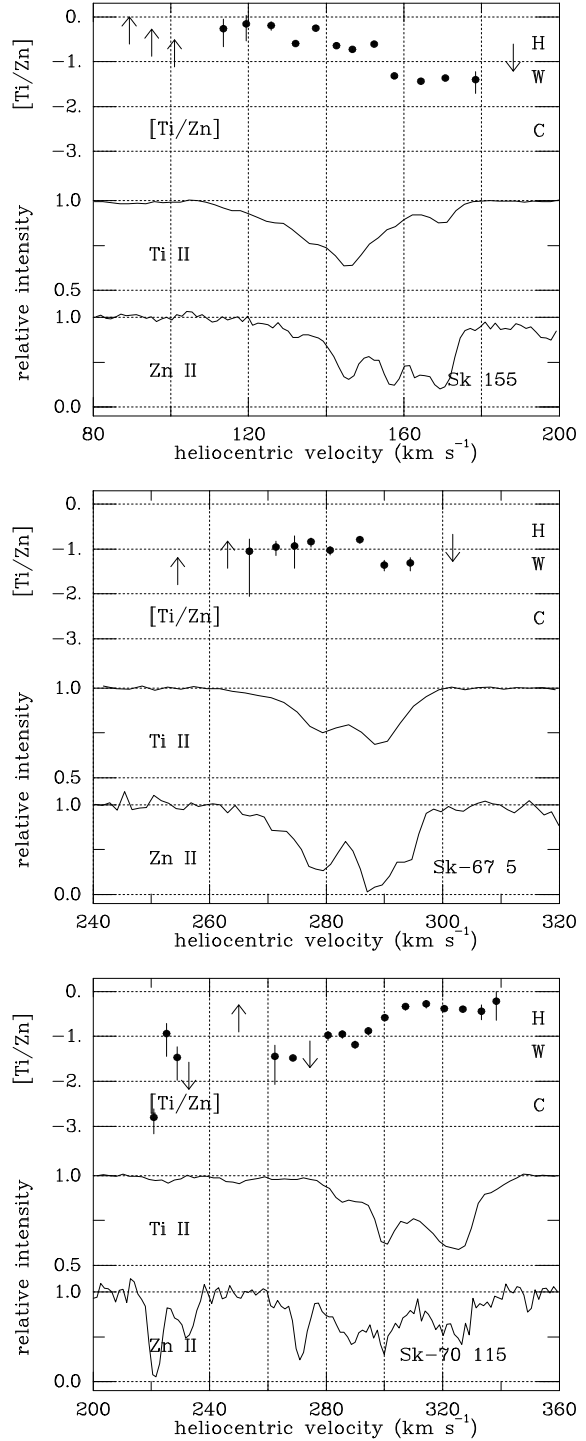}
\caption{Spectra of \mbox{Ti\,{\sc ii}} ($\lambda$3383) and \mbox{Zn\,{\sc ii}} ($\lambda$2026) for one SMC sight line (Sk~155) and two LMC sight lines (Sk$-$67~5, Sk$-$70~115); only the Magellanic Clouds absorption is shown.
The \mbox{Ti\,{\sc ii}} spectra are from UVES (FWHM $\sim$ 4.5 km~s$^{-1}$); the \mbox{Zn\,{\sc ii}} spectra are from the {\it HST} STIS (FWHM $\sim$ 2.8 km~s$^{-1}$).
The points at the top of each panel show the relative abundance [Ti/Zn] (indicative of the titanium depletion), for components identified in simultaneous fits to the profiles of lines from many dominant species.
Uncertainties in [Ti/Zn] are 1~$\sigma$; limits are 3~$\sigma$.
The letters H, W, and C (at the upper right in each panel) give representative values for [Ti/Zn] for clouds in the low Galactic halo; warm, diffuse clouds in the disc; and colder, denser disc clouds.
The narrow absorption features in the \mbox{Zn\,{\sc ii}} spectrum of Sk$-$70~115 near 270 and 350 km~s$^{-1}$ are due to \mbox{Mg\,{\sc i}}.}
\label{fig:mczn}
\end{figure}

Figures~\ref{fig:mwzn} and \ref{fig:mczn} compare high-resolution \mbox{Ti\,{\sc ii}} ($\lambda$3383) and \mbox{Zn\,{\sc ii}} ($\lambda$2026 or $\lambda$2062) profiles for four Galactic and three Magellanic Clouds sight lines, respectively, sampling different regions/environments in each of the three galaxies:
$\epsilon$~Ori is one of the belt stars in Orion,
$\rho$~Leo lies about 600 pc above the Galactic plane in the low halo,
$\pi$~Sco is one of the ``Sco-Oph'' stars (exhibiting weak absorption from trace neutral and molecular species),
$\zeta$~Oph shows relatively strong absorption from trace neutral and molecular species,
Sk~155 is in the near ``wing'' region of the SMC,
Sk$-$67~5 is at the northwestern end of the main bar of the LMC, and
Sk$-$70~115 lies within the LMC2 supershell (southeast of the 30 Doradus star-forming region).
In each panel, the points in the top section show the ratio [Ti/Zn], which should track any variations in the Ti depletion.\footnotemark
\footnotetext{[Ti/Zn] = log[$N$(\mbox{Ti\,{\sc ii}})/$N$(\mbox{Zn\,{\sc ii}})] $-$ log(Ti/Zn)$_0$, where (Ti/Zn)$_0$ is the solar (0.29 dex) or stellar reference ratio.  Since the corresponding [Zn/H] ratios are typically $\ga-$0.4 dex, [Ti/Zn] is generally a reasonably good indicator of the Ti depletion in predominantly neutral gas.}
For the Galactic sight lines, the [Ti/Zn] ratios were determined from corresponding ratios of the apparent optical depths in the line profiles (after smoothing the \mbox{Ti\,{\sc ii}} spectra to the lower resolution of the \mbox{Zn\,{\sc ii}} spectra).
For the Magellanic Clouds sight lines, the [Ti/Zn] ratios were determined from detailed, simultaneous, multi-component fits to the profiles of lines from a number of dominant ions (Welty et al. 2001, 2004, and in prep.).
The letters H, W, and C (at the upper right in each panel) give representative values for [Ti/Zn] for clouds in the low Galactic halo ($-$0.6 dex); warm, diffuse clouds in the disc ($-$1.35 dex); and colder, denser disc clouds ($-$2.55 dex) (Savage \& Sembach 1996; Welty et al. 1999b; updated for more recent solar and interstellar abundance data; see also Jenkins 2009).

In general, the \mbox{Ti\,{\sc ii}} and \mbox{Zn\,{\sc ii}} absorption-line profiles span very similar velocity ranges, but the profiles may be quite different in detail.
While the \mbox{Ti\,{\sc ii}} and \mbox{Zn\,{\sc ii}} profiles are broadly similar toward both $\epsilon$~Ori and $\rho$~Leo, for example, there are still differences of nearly an order of magnitude in [Ti/Zn] for different velocity intervals.
The profiles of the two species are very different toward $\pi$~Sco and $\zeta$~Oph, however, with correspondingly larger ranges in [Ti/Zn] or D(Ti).
Toward $\pi$~Sco, [Ti/Zn] ranges from about $-$1.4 to $<-$3.2 dex (versus $-$2.4 dex for the sight line as a whole); toward $\zeta$~Oph, [Ti/Zn] ranges from about $-$1.3 to $\la-$2.7 dex (the strongest \mbox{Zn\,{\sc ii}} components may be somewhat saturated).
Estimates for $N$(H$_{\rm tot}$) based on detailed fits to the \mbox{Zn\,{\sc ii}} profiles toward $\zeta$~Oph suggest that D(Ti) ranges from $-$1.3 dex for the ``group A'' components ($-$36 km~s$^{-1}$ $\la$ $v$ $\la$ $-$25 km~s$^{-1}$; log[$N$(H$_{\rm tot}$)] $\sim$ 19.6) to $-$3.1 dex for the ``group B'' components ($-$24 km~s$^{-1}$ $\la$ $v$ $\la$ $-$5 km~s$^{-1}$; log[$N$(H$_{\rm tot}$)] = 21.14), where the molecular material is located (e.g., Crawford et al. 1994).
In many cases, the variations in [Ti/Zn] (and thus the titanium depletion) with velocity appear to be relatively smooth.

The [Ti/Zn] ratios in the three Magellanic Clouds sight lines in Fig.~\ref{fig:mczn} also exhibit a variety of behaviours.
Toward the LMC star Sk$-$67~5, most of the components identified in fits to the line profiles have rather similar [Ti/Zn] ratios, between about $-$0.8 and $-$1.3 dex.
A much larger range in [Ti/Zn] -- from about $-$0.3 to about $-$2.8 dex -- is seen in the LMC2 region toward Sk$-$70~115, however.
The most severe titanium depletion there is in the component at 220 km~s$^{-1}$, where the bulk of the neutral atomic and molecular material is found (Tumlinson et al. 2002; Welty et al. 2006).
Toward Sk~155 (in the SMC ``wing'' region), [Ti/Zn] ranges over a factor of 20, from about $-$0.1 to $-$1.4 dex.
The Ti depletions in the components between about 162 and 174 km~s$^{-1}$ are much less severe than would have been expected from the depletions found for Fe and Ni there -- similar to (but not quite as extreme as) the remarkably mild depletions found for Si and Mg in those components (Welty et al. 2001 and in prep.; Sofia et al. 2006).

\section{Discussion}
\label{sec-disc}

\subsection{Differences in spatial distribution}
\label{sec-dist}

In principle, differences in the ionization and/or depletion behaviour of the various species observed in the predominantly neutral (\mbox{H\,{\sc i}}) gas may lead to corresponding differences in the distribution of those species, depending on specific environmental conditions. 
While some of those differences may be inferred from total sight line column densities, comparisons of the respective high-resolution absorption-line profiles can provide more direct indications of the degree to which those species coexist.
With the notable exception of some of the sight lines in the Sco-Oph region (e.g., $\pi$~Sco; Joseph \& Jenkins 1991), the \mbox{Ti\,{\sc ii}} profiles shown in Figs.~\ref{fig:specg1} and \ref{fig:specm1} most closely resemble those of \mbox{Ca\,{\sc ii}} and of various dominant ions of other significantly depleted elements (e.g., \mbox{Fe\,{\sc ii}}, \mbox{Ni\,{\sc ii}}), with the absorption often spread fairly smoothly over a relatively wide range in velocity.
There is generally little sign of the distinct narrow components (with $b$ $\la$ 0.8 km~s$^{-1}$) seen for both \mbox{Ca\,{\sc ii}} and various trace neutral species (e.g., \mbox{Na\,{\sc i}}, \mbox{K\,{\sc i}}, \mbox{Ca\,{\sc i}}) in spectra of comparable resolution -- whether that narrow absorption is relatively weak (e.g., for some components seen toward $\rho$~Leo and several of the stars in Orion) or very strong (e.g., for the main components seen toward $\zeta$~Oph, $\chi$~Oph, and other more heavily reddened stars). 
Fits to the \mbox{Ti\,{\sc ii}} profiles using component structures derived from \mbox{Ca\,{\sc ii}} often yield low \mbox{Ti\,{\sc ii}}/\mbox{Ca\,{\sc ii}} ratios for components with such small $b$-values (when $b$ is well-determined), and the \mbox{Ti\,{\sc ii}} absorption is often relatively weak for components with strong absorption from the trace neutral species -- especially for components also detected in CH and CN.
Offsets between the strongest \mbox{Ti\,{\sc ii}} absorption and the strongest molecular absorption are not uncommon -- even when only lower resolution spectra are available.
All these tendencies are consistent with the view that the observed \mbox{Ti\,{\sc ii}} absorption traces primarily the warmer, more diffuse gas -- where the titanium depletions are much less severe than in the colder, denser clouds where the various trace neutral species and molecules are concentrated (e.g., Crinklaw et al. 1994).
The absorption from \mbox{Ca\,{\sc ii}} traces both the colder, denser gas (where \mbox{Ca\,{\sc ii}} can be a dominant ion but calcium is severely depleted) and the warmer, more diffuse gas (where \mbox{Ca\,{\sc ii}} is a trace species but calcium is much less severely depleted) (Welty et al. 2003).
The general decrease in the \mbox{Ti\,{\sc ii}}/\mbox{Ca\,{\sc ii}} ratio at higher column densities and/or higher $E(B-V)$ thus may primarily reflect changes in the calcium ionization balance (i.e., an increase in \mbox{Ca\,{\sc ii}}/Ca$_{\rm tot}$) at higher average densities.

\subsection{Depletions in individual components}
\label{sec-icdepl}

Spitzer (1985) suggested that the correlations observed between the total sight line depletions of various elements and the mean sight line density $<n_{\rm H}>$ could arise if individual sight lines contained different proportions of moderate density ``standard'' cold clouds, higher density ``large'' cold clouds, and more smoothly distributed warmer, lower density neutral gas.
Jenkins, Savage, \& Spitzer (1986) used that simple model, with the further assumption of constant depletions for each element in the cold and warm gas (more severe in the cold gas, less severe in the warm gas), to interpret a large body of abundance data derived from UV spectra obtained with {\it Copernicus} (with FWHM $\sim$ 15 km~s$^{-1}$).
More recent detailed studies based on higher resolution optical and UV spectra of individual sight lines (e.g., Fitzpatrick \& Spitzer 1997; Welty et al. 1999b; Howk, Savage, \& Fabian 1999), however, suggest that there is a continuum of depletion behaviour in individual interstellar clouds -- likely depending on both current physical conditions and past history -- and that the commonly cited ``cold cloud'', ``warm cloud'', and ``halo cloud'' depletion patterns (e.g., Savage \& Sembach 1996) may be viewed as useful representative values within that continuum.

The [Ti/Zn] ratios shown in Figs.~\ref{fig:mwzn} and \ref{fig:mczn}, together with estimates of [Ti/Zn], [Ti/S], or [Ti/H] for individual components or component groups in additional sight lines, indicate that the titanium depletion may range (essentially continuously) from less than a factor of 2 to more than a factor of 3000 in individual interstellar clouds.
The least severe Ti depletions [D(Ti) $\ga$ $-$0.5 dex; assuming D(Zn) $\sim$ $-$0.1 dex] are found for several components toward the Galactic halo stars HD~93521 and HD~215733 (Albert 1983; Spitzer \& Fitzpatrick 1993; Fitzpatrick \& Spitzer 1997; Jenkins 2009) and for some of the components toward Sk~155 (SMC) and Sk$-$70~115 (LMC).
The most severe Ti depletions [D(Ti) $\la$ $-$3.0 dex; assuming D(Zn) $\sim$ $-$0.4 dex] are found for the main components ($-$20 km~s$^{-1}$ $\la$ $v$ $\la$ $-$5 km~s$^{-1}$) toward $\zeta$~Oph, $\chi$~Oph, and several of the Sco-Oph stars; for the main components toward HD~62542 and several stars in the Taurus-Perseus region ($\zeta$~Per, HD~27778, X~Per); and for the component near 220 km~s$^{-1}$ toward Sk$-$70~115 (LMC).
These ``main'' components typically contain most of the trace neutral species and most of the molecular material in the respective lines of sight, and (presumably) arise in relatively cold, dense gas.
Plots of the [Ti/Zn] ratios versus $N$(\mbox{Zn\,{\sc ii}}) reveal trends very similar to those seen for D(Ti) versus $N$(H$_{\rm tot}$) (Fig.~\ref{fig:tivsh}), but with slightly steeper slope.

Much of the scatter in Figs.~\ref{fig:tivsh}, ~\ref{fig:tivsebv}, and ~\ref{fig:tivsfh2} is likely due to differences in the relative amounts of cold, moderately dense gas (with more severe depletions) and warmer, more diffuse gas (with milder depletions) for sight lines with similar overall $N$(H$_{\rm tot}$), $E(B-V)$, or $f$(H$_2$).
A number of the sight lines with the least severe overall depletions, for a given $N$(H$_{\rm tot}$) or $E(B-V)$, are quite complex -- sampling either long paths dominated by relatively diffuse gas through the Galactic disc (or lower halo) or disturbed regions (e.g., near the Carina nebula; Walborn et al. 2002a) where significant dust processing has occurred.
Conversely, a number of the sight lines with the most severe depletions are relatively nearby, with fewer components and a larger fraction of colder, denser gas.
For example, the complex sight line toward HD~167971 [d = 730 pc, $E(B-V)$ = 1.08, $N$(H$_2$) = 7 $\times$ 10$^{20}$ cm$^{-2}$] has D(Ti) = $-$2.32 dex, while the much simpler sight line toward HD~27778 [d = 225 pc, $E(B-V)$ = 0.37, $N$(H$_2$) = 6.2 $\times$ 10$^{20}$ cm$^{-2}$] -- with lower reddening but a higher molecular fraction -- has D(Ti) = $-$3.10 dex.
Severe depletions in a single component can easily be ``hidden'' in overall average sight line values by much less severe depletions in other more diffuse components along the line of sight -- even if those other components have lower aggregate $N$(H$_{\rm tot}$).

\begin{table}
\caption{D(Ti) vs. $n_{\rm H}$}
\label{tab:dens}
\begin{tabular}{@{}lrrccl}
\hline
\multicolumn{1}{c}{Star}&
\multicolumn{1}{c}{D(Ti)}&
\multicolumn{1}{c}{D(Ti)}&
\multicolumn{1}{c}{$n_{\rm H}$(\mbox{C\,{\sc i}})}&
\multicolumn{1}{c}{$n_{\rm H}$(C$_2$)}&
\multicolumn{1}{l}{Ref}\\
\multicolumn{1}{c}{ }&
\multicolumn{1}{c}{all}&
\multicolumn{1}{c}{main}&
\multicolumn{1}{c}{main}&
\multicolumn{1}{c}{main}\\
\hline
HD 23180        & $-$2.86&   --    & 45--70     & 265$^{+335}_{-65}$  & 1,2 \\ 
HD 24398        & $-$3.02& $-$3.3  & 10--16     & 215$^{+285}_{-70}$  & 1,2 \\ 
HD 24534        & $-$2.92&   --    & 45--110    & 325$^{+80}_{-80}$   & 3,2 \\ 
HD 24760        & $-$2.26& $-$2.3  & 12--33     &    --               & 4 \\
HD 27778        & $-$3.10&   --    &     --     & 280$^{+35}_{-35}$   & 2 \\ 
HD 35149        & $-$2.26& $-$2.3  & 10--16     &    --               & 1 \\ 
HD 37742        & $-$2.13& $-$2.2  & 16--40     &    --               & 1 \\ 
HD 62542        & $-$2.70& $-$3.0  &     --     & 500$^{+570}_{-145}$ & 2 \\
HD 73882        & $-$2.32& $-$2.7  &     --     & 520$^{+745}_{-225}$ & 2 \\
HD 110432       & $-$2.69& $-$3.1  &     --     & 140$^{+35}_{-35}$   & 2 \\
HD 112244 A     & $-$2.15&   --    & 14--28     &    --               & 4 \\
HD 112244 B     & $-$2.15&   --    & 17--48     &    --               & 4 \\
HD 141637       & $-$2.39& $-$2.8  & 200--800   &    --               & 4 \\ 
HD 143018       & $-$2.58& $<-$3.0 & 13--33     &    --               & 4 \\ 
HD 144217       & $-$2.69& $-$3.2  & 35--205    &    --               & 4 \\ 
HD 147165       & $-$2.74& $-$2.8  & 137--345   &    --               & 4 \\ 
HD 147888       & $-$2.97& $-$3.2  &     --     & 215$^{+55}_{-30}$   & 2 \\ 
HD 148184       & $-$2.68& $-$3.6  &     --     & 185$^{+125}_{-30}$  & 2 \\
HD 149757       & $-$2.62& $-$3.1  & 60--150    & 215$^{+35}_{-35}$   & 1,2 \\ 
HD 154368       & $-$2.49& $-$2.8  &     --     & 240$^{+80}_{-40}$   & 2 \\ 
HD 206267       & $-$2.37&   --    & 40--95     & 315$^{+125}_{-30}$  & 3,2 \\ 
HD 207198       & $-$2.31&   --    &     --     & 245$^{+30}_{-60}$   & 2 \\ 
HD 210121       & $-$2.28& $-$2.8  &     --     & 315$^{+80}_{-80}$   & 2 \\
\hline
\end{tabular}
\medskip
~ ~ \\
The first entry for D(Ti) is for the whole sight line; the second is for the ``main'' component(s) -- where the \mbox{C\,{\sc i}} and/or C$_2$ are concentrated.\\
References for $n_{\rm H}$: 1 = Welty et al. 2003; 2 = Sonnentrucker et al. 2007; 3 = this paper; 4 = Zsarg\'{o} \& Federman 2003.
\end{table}

\subsection{Do depletions depend on local densities?}
\label{sec-dens}

Both the observed correlations between the depletions of various refractory elements and the mean sight line density $<n_{\rm H}>$ (Fig.~\ref{fig:tivsnh}) and models of grain growth and destruction have suggested that the depletions in individual clouds should depend on the local hydrogen density $n_{\rm H}$ -- but direct evidence for that dependence has been rather scanty.
Estimates of $n_{\rm H}$ have recently been obtained for the main neutral components in a number of the sight lines in this \mbox{Ti\,{\sc ii}} survey, however.
Where possible, Table~\ref{tab:dens} lists two values for the Ti depletions (for the sight line as a whole and for just the main neutral components) and two estimates for the local hydrogen densities.
The first set of $n_{\rm H}$ is based on values of the thermal pressures ($n_{\rm H}T$) derived from the fine-structure excitation of \mbox{C\,{\sc i}} (Jenkins \& Tripp 2001; Zsarg\'{o} \& Federman 2003; Welty et al. 2003) and the corresponding kinetic temperatures derived from the rotational excitation of H$_2$ (assuming the \mbox{C\,{\sc i}} and H$_2$ to be coextensive); the second set of $n_{\rm H}$ (for more heavily reddened sight lines) was obtained from analyses of the rotational excitation of C$_2$ (van Dishoeck \& Black 1982; Sonnentrucker et al. 2007).
The two estimates for $n_{\rm H}$ are not directly comparable, as C$_2$ appears to trace somewhat denser, colder regions than \mbox{C\,{\sc i}} and H$_2$ (e.g., Sonnentrucker et al. 2007).
The Ti depletions in the main neutral components are generally estimated from the \mbox{Ti\,{\sc ii}} absorption over the velocity interval emcompassing the strongest components seen in \mbox{Na\,{\sc i}} and/or \mbox{K\,{\sc i}}, which are assumed to contain most of the total hydrogen, \mbox{C\,{\sc i}}, and C$_2$ in the sight line.
These main component D(Ti) are typically factors of 2--3 more severe than the integrated sight line values.

For this rather small sample, there does not appear to be any clear relationship between D(Ti) (for the total sight line or for the main components) and either set of $n_{\rm H}$, however.
It is intriguing, for example, that the Ti depletions toward $\pi$~Sco and $\zeta$~Oph are very similar (both overall and as functions of velocity) -- even though the density and molecular fraction are much higher toward $\zeta$~Oph.
Unfortunately, the estimates for the ``main component'' depletion can be rather uncertain, as there could still be some warmer, more diffuse gas (with milder depletions) at velocities similar to those of the cooler gas traced by \mbox{C\,{\sc i}} and C$_2$.
Unresolved blends of gas with different properties may also affect the densities derived from \mbox{C\,{\sc i}} (e.g., Jenkins \& Tripp 2001), and the dynamic range sampled by the densities derived from C$_2$ excitation is not very large.
More detailed component analyses of higher resolution spectra -- to more reliably identify and characterize ``corresponding'' components in \mbox{Ti\,{\sc ii}}, \mbox{Zn\,{\sc ii}}, and \mbox{C\,{\sc i}} -- may yield a clearer picture of the relationship between D(Ti) and $n_{\rm H}$.

\subsection{Electron densities from \mbox{Ca\,{\sc ii}}/\mbox{Ti\,{\sc ii}}}
\label{sec-ne}

Because \mbox{Ca\,{\sc ii}} is typically a trace ion in warmer, more diffuse clouds, the \mbox{Ca\,{\sc ii}}/\mbox{Ti\,{\sc ii}} ratio might, in principle, provide a useful measure of the electron density there (Stokes 1978) -- if that ratio can be calibrated via independent determinations of $n_e$.
Stokes \& Hobbs (1976) derived the approximation
\begin{equation}
n_e \simeq \frac{\Gamma}{\alpha} [R \frac{N(\mbox{Ti\,{\sc ii}})}{N(\mbox{Ca\,{\sc ii}})} - 1 ]^{-1},
\end{equation}
where $R$, the ratio of the gas-phase abundances of Ca and Ti (in all ionization states), is assumed to not vary strongly and $\Gamma$/$\alpha$ is the ratio of the \mbox{Ca\,{\sc ii}} photoionization rate and the \mbox{Ca\,{\sc iii}} recombination rate coefficient. 
Stokes (1978) then estimated $R$ = 9.2$\pm$3.1, using \mbox{Ti\,{\sc ii}} and \mbox{Ca\,{\sc ii}} column densities for the main component groups in 10 sight lines, together with values for $n_e$ derived from the corresponding \mbox{Ca\,{\sc i}}/\mbox{Ca\,{\sc ii}} ratios.
Combining individual component \mbox{Ti\,{\sc ii}} column densities derived from the higher resolution spectra presented in this paper with corresponding values for \mbox{Ca\,{\sc ii}} and \mbox{Ca\,{\sc i}} from Welty et al. (1996, 2003), for 19 components in seven sight lines, yields the same mean $R$ but twice the standard deviation (the individual values for $R$ range from 2 to 20).
The differences in distribution of the three species complicate the calibration of $R$, as \mbox{Ca\,{\sc i}} is detected primarily in the colder, denser gas (where the \mbox{Ti\,{\sc ii}}/\mbox{Ca\,{\sc ii}} ratio appears to be lower, on average, than in warmer, more diffuse clouds and where \mbox{Ca\,{\sc ii}} can be a dominant ion) and as the recombination rate coefficient $\alpha$ depends on the temperature. 
Moreover, estimates of $n_e$ derived from \mbox{Ca\,{\sc i}}/\mbox{Ca\,{\sc ii}} ratios are often significantly higher than values derived from other trace-dominant ratios (Welty et al. 2003) -- so that $R$ could be somewhat larger.
Until these issues are resolved, values of $n_e$ estimated from eqn.~1 must be considered to be very uncertain.

\subsection{Scale height of Galactic \mbox{Ti\,{\sc ii}} absorption}
\label{sec-scale}

Different constituents of the ISM can exhibit quite different distributions perpendicular to the Galactic plane -- probing different levels of the Galactic halo and providing clues to the relationship between disc and halo gas.
For example, while observations of Lyman-$\alpha$ absorption suggest an exponential scale height of about 200~pc for neutral hydrogen (Diplas \& Savage 1994), highly ionized species such as \mbox{C\,{\sc iv}} and \mbox{Si\,{\sc iv}} can extend much farther from the plane, with scale heights of order 3.5~kpc (e.g., Savage \& Wakker 2009).
Because the titanium depletion is typically less severe in more diffuse gas, the gas-phase \mbox{Ti\,{\sc ii}}/\mbox{H\,{\sc i}} ratio would be expected to increase away from the Galactic plane (on average, as $<n_{\rm H}>$ decreases), so that observations of \mbox{Ti\,{\sc ii}} absorption might be used to trace the neutral material in the Galactic halo well beyond the 200~pc scale height of the \mbox{H\,{\sc i}}.
Based on \mbox{Ti\,{\sc ii}} detections toward 15 stars at distances $|z|$ $\ga$ 1~kpc from the plane (and many stars below 1~kpc), Albert et al. (1994) concluded that the \mbox{Ti\,{\sc ii}} scale height could not yet be determined, but that it might be greater than 2--3~kpc.
Lipman \& Pettini (1995), however, using eight additional \mbox{Ti\,{\sc ii}} detections toward stars at 1.0~kpc $\la$ $|z|$ $\la$ 4.3~kpc (plus four toward targets in the Magellanic Clouds), derived a scale height of 1.5$\pm$0.2~kpc.
Observations of the Galactic \mbox{Ti\,{\sc ii}} absorption toward 20 SMC and 34 LMC stars (Caulet \& Newell 1996; Cox et al. 2006, 2007; this paper) yield log[$N$(\mbox{Ti\,{\sc ii}})~sin$|b|$] = 11.74$\pm$0.13 dex and 11.80$\pm$0.10 dex, respectively -- consistent with the values found by Lipman \& Pettini (1995) -- providing further support for a \mbox{Ti\,{\sc ii}} scale height $\la$ 1.5~kpc.

Lipman \& Pettini (1995) also conjectured that titanium might be undepleted in the Galactic halo beyond $|z|$ $\sim$ 1~kpc.
If gas in the low halo corotates with the disc, then toward the LMC its velocity would increase monotonically with $|z|$, reaching $v_{\odot}$ $\sim$ 100 km~s$^{-1}$ at $|z|$ $\sim$ 7.5 kpc; toward the SMC, the velocity of the gas would decrease out to $|z|$ $\sim$ 4 kpc, then increase steadily beyond that point (Savage \& de Boer 1981).
Given the non-zero dispersion in velocity for gas within the Galactic disc, identification of halo gas by velocity should thus be easier toward the LMC.
The \mbox{Ti\,{\sc ii}} spectra of LMC targets in Fig.~\ref{fig:specm1} all show absorption at 0 km~s$^{-1}$ $\la$ $v$ $\la$ 30 km~s$^{-1}$, due primarily to gas in the disc (with some contribution from gas in the low halo).
For 40 km~s$^{-1}$ $\la$ $v$ $\la$ 100 km~s$^{-1}$ (corresponding to gas at 3.5 kpc $\la$ $|z|$ $\la$ 7.5 kpc), however, very little (if any) \mbox{Ti\,{\sc ii}} is detected, and it will be difficult to determine accurate titanium abundances.
We note, however, that the relative abundances [X/Zn] (for X = Si, Cr, Mn, Fe, and Ni) in the halo components at 56 km~s$^{-1}$ $\la$ $v$ $\la$ 91 km~s$^{-1}$ toward the LMC SN 1987A are consistent with no depletion of those refractory elements (Welty et al. 1999a).

\subsection{D/H variations and deuterium depletions}
\label{sec-dh}

In principle, the current abundance of deuterium (D/H) is dependent only on its initial (``primordial'') abundance produced in the Big Bang (with a fairly sensitive dependence on the baryon density $\Omega_{\rm b}$) and its gradual destruction (via conversion to $^3$He and $^4$He) in subsequent generations of stars (``astration'').
Measurements of the current local D/H ratio thus yield a firm lower limit to the primordial deuterium abundance (and a corresponding upper limit to $\Omega_{\rm b}$).
If the actual primordial D abundance can be estimated independently, then the current D/H ratio may be used to constrain models of galactic chemical evolution (see, e.g., Pettini 2006, Steigman 2009, and Linsky 2009 for recent summaries of some of these issues).
The Galactic interstellar gas phase \mbox{D\,{\sc i}}/\mbox{H\,{\sc i}} ratio (which yields D/H directly if the molecular fraction is small and if there are no isotope-selective processes significantly affecting the ratio) has been measured toward $\sim$ 50 stars, at distances from 2.6 pc to 2.7 kpc and with $N$(\mbox{H\,{\sc i}}) ranging from 4 $\times$ 10$^{17}$ to 1.5 $\times$ 10$^{21}$ cm$^{-2}$ (as compiled by Linsky et al. 2006, Ellison et al. 2007, and Lallement et al. 2008).
The local \mbox{D\,{\sc i}}/\mbox{H\,{\sc i}} ratio appears to be roughly constant, at 15--16 ppm, for sight lines with $N$(\mbox{H\,{\sc i}}) $\la$ 1.5 $\times$ 10$^{19}$ cm$^{-2}$ (mostly within 100 pc). 
There is considerable scatter in the ratio, however, along the more extended, higher $N$(\mbox{H\,{\sc i}}) sight lines, where \mbox{D\,{\sc i}}/\mbox{H\,{\sc i}} ranges from 5--23 ppm; values of about 22 ppm have been measured in several clouds in the lower Galactic halo (Sembach et al. 2004; Savage et al. 2007).
Measurements of the \mbox{D\,{\sc i}}/\mbox{H\,{\sc i}} ratio in a small number of low-metallicity damped Lyman-$\alpha$ systems (where the effects of astration should be minimal) suggest that the primordial D/H ratio was about 28 ppm (Pettini et al. 2008, and references therein) -- in excellent agreement with the predictions of standard Big Bang nucleosynthesis, as constrained by observations of the temperature fluctuations in the cosmic microwave background.

Several explanations for the scatter in the current Galactic \mbox{D\,{\sc i}}/\mbox{H\,{\sc i}} ratio have been proposed, with very different implications for Galactic chemical evolution:
(1) The scatter in the \mbox{D\,{\sc i}}/\mbox{H\,{\sc i}} ratio could be due to localized infall and incomplete mixing of less processed material (with higher D/H ratios) -- which would imply a relatively low ``true'' current value for D/H, and thus significant astration.
Observations of \mbox{O\,{\sc i}} absorption along many of the same sight lines, however, do not reveal the anticorrelation between the \mbox{O\,{\sc i}}/\mbox{H\,{\sc i}} and \mbox{D\,{\sc i}}/\mbox{H\,{\sc i}} ratios that would be expected from differences in astration (e.g., Steigman, Romano, \& Tosi 2007).
(2) Alternatively, the scatter in \mbox{D\,{\sc i}}/\mbox{H\,{\sc i}} could be due to variations in the selective depletion of deuterium into dust grains along the different lines of sight (Jura 1982; Tielens 1983; Draine 2006).
In this case, the true current value for the D/H ratio would be at least as high as the highest observed value -- which would imply fairly minimal astration.

If the scatter in the observed \mbox{D\,{\sc i}}/\mbox{H\,{\sc i}} ratios is due to variations in the depletion of the deuterium, we would expect to see correlations between the deuterium abundance and the abundances of other depleted species -- as is generally the case for pairs of refractory elements in the Galactic ISM (e.g., Jenkins 2009).
Such correlations have indeed been noted between \mbox{D\,{\sc i}}/\mbox{H\,{\sc i}} and \mbox{Ti\,{\sc ii}}/\mbox{H\,{\sc i}} (Prochaska et al. 2005; Ellison et al. 2007; Lallement et al. 2008) and between \mbox{D\,{\sc i}}/\mbox{H\,{\sc i}} and \mbox{Si\,{\sc ii}}/\mbox{H\,{\sc i}} and \mbox{Fe\,{\sc ii}}/\mbox{H\,{\sc i}} (Linsky et al. 2006).
The two low halo clouds with high \mbox{D\,{\sc i}}/\mbox{H\,{\sc i}} ratios also appear to be characterized by relatively mild depletions (Sembach et al. 2004; Savage et al. 2007).
However, the sight line samples employed in those studies are rather small, there are some sight lines which do not appear to follow the general trends, and the relative slopes of the correlations do not seem consistent with the known depletion behaviour of Si, Fe, and Ti (Linsky et al. 2006; Ellison et al. 2007; Lallement et al. 2008).

In principle, several factors might contribute to the scatter in those relationships and/or to the apparently inconsistent correlation slopes (e.g., Linsky 2009):

(1) The presence of ionized gas in some of the sight lines with relatively low $N$(\mbox{H\,{\sc i}}) might account for some of the outliers in the relationships with Si and Fe (as \mbox{Si\,{\sc ii}} and \mbox{Fe\,{\sc ii}} would be present in the ionized gas, but not \mbox{D\,{\sc i}}) and might contribute to some of the differences in the slopes of the relationships.
Ionized gas should not affect the relationship between \mbox{D\,{\sc i}}/\mbox{H\,{\sc i}} and \mbox{Ti\,{\sc ii}}/\mbox{H\,{\sc i}}, however, as both \mbox{D\,{\sc i}} and \mbox{Ti\,{\sc ii}} should be present only in the neutral gas.  

(2) As titanium is typically much more severely depleted than deuterium, different processes and/or sites may be involved in the depletions of the two elements, and the gas-phase \mbox{Ti\,{\sc ii}}/\mbox{D\,{\sc i}} ratio may exhibit considerable variations.
For example, if much of the Ti is locked in grain cores, but most of the D in more loosely bound mantles, there may be cases where much of the deuterium could be returned to the gas phase (e.g., via weak shocks) without significantly affecting the more tightly bound titanium.
In principle, such a scenario could account for sight lines with high \mbox{D\,{\sc i}}/\mbox{H\,{\sc i}} but appreciable depletions of Ti (e.g., HD 41161, WD1034+001; Ellison et al. 2007); information regarding the depletions of other mildly depleted species in those sight lines would be valuable.
On the other hand, subsequent destruction of the grain cores could significantly increase the gas-phase Ti without much increase in the gas-phase D.

(3) Many of the sight lines exhibit complex component structure in the available high-resolution optical/UV spectra, and there may be cases where the gas-phase D and Ti are distributed very differently among the various components (e.g., Prochaska et al. 2005).
The sight line toward $\mu$~Col, for example, appears to be somewhat anomalous, with low \mbox{D\,{\sc i}}/\mbox{H\,{\sc i}} but mild overall Ti depletion (Ellison et al. 2007; Lallement et al. 2008).
Examination of the profiles of both mildly and severely depleted species suggests that the bulk of the hydrogen (and presumably D) is in the components between about 10 and 35 km~s$^{-1}$, while the more refractory species (including Ti) have significant contributions from the component near 40~km~s$^{-1}$ -- which has a relatively low $N$(H$_{\rm tot}$) but very mild depletions (e.g., Howk, Savage, \& Fabian 1999; Lallement et al. 2008).
If only the lower velocity components are considered, then the $\mu$~Col sight line would not be as discrepant.

(4) The current relatively small sample of sight lines with data for both \mbox{D\,{\sc i}} and \mbox{Ti\,{\sc ii}} is not representative of the larger set of Galactic sight lines for which the Ti depletion is known.
For the small D-Ti sample, plots of D(Ti) versus both $N$(H$_{\rm tot}$) (as in Fig.~\ref{fig:tivsh}) and the mean sight line density $<n_{\rm H}>$ (e.g., fig.~5 in Ellison et al. 2007) indicate that the higher $N$(H$_{\rm tot}$) sight lines (all toward fairly distant stars) are all characterized by milder than average Ti depletions -- so that the slopes in each case are much shallower than those seen for the full Galactic \mbox{Ti\,{\sc ii}} sample.
There appear to be no sight lines with detections of \mbox{D\,{\sc i}} for which D(Ti) is less than $-$2 dex.

(5) The steeper than expected slopes of relationships involving the Si depletion (in the small D-Si sample) are strongly influenced by the rather uncertain data point for $\theta$~Car, which has the most severe Si depletion in the sample (e.g., fig.~5 in Ellison et al. 2007).
The \mbox{Si\,{\sc ii}} column density for $\theta$~Car is taken from the {\it Copernicus} study of Allen, Jenkins, \& Snow (1992), who ascribed about 70 per cent of the observed \mbox{Si\,{\sc ii}} to an ionized component in that sight line -- but the various other first ions (e.g., \mbox{C\,{\sc ii}}, \mbox{Mg\,{\sc ii}}, \mbox{S\,{\sc ii}}, \mbox{Fe\,{\sc ii}}) were all assigned entirely to the neutral component.
Increasing the \mbox{Si\,{\sc ii}} column density for $\theta$~Car by a factor of 2--3 would make the D-Si sample more consistent with the larger Galactic \mbox{Si\,{\sc ii}} sample.

Observations of \mbox{Ti\,{\sc ii}} in the rest of the sight lines of the \mbox{D\,{\sc i}} sample -- to increase the size of the D-Ti sample and minimize ionization effects -- should provide a clearer test of the deuterium depletion hypothesis.
In view of the observed trend in D(Ti) versus $N$(H$_{\rm tot}$) (Fig.~\ref{fig:tivsh}), \mbox{Ti\,{\sc ii}} may well be detectable even at the low hydrogen column densities characterizing some of those sight lines.

\subsection{\mbox{Ti\,{\sc ii}} in the Magellanic Clouds}
\label{sec-mcti2}

A primary motivation for studies of the interstellar \mbox{Ti\,{\sc ii}} absorption in the Magellanic Clouds is to explore the behaviour of elemental depletions and dust properties in lower metallicity galaxies -- i.e., to see whether the patterns and levels of depletion and/or the composition of the dust vary with metallicity.
The Magellanic Clouds are particularly useful for such studies because the total interstellar (gas plus dust) abundances of many elements may be inferred from analyses of the abundances in stellar atmospheres (for a variety of stellar types) and in gaseous nebulae at many locations within the two galaxies.
The current overall metallicities are sub-solar by factors of about 2 in the LMC and 4--5 in the SMC (except for C and N, which are somewhat more deficient).
The {\it relative} elemental abundances, however, are generally fairly similar to those found locally in our Galaxy (within the typical 0.1--0.2 dex uncertainties) -- at least for most of the elements commonly detected in the ISM (e.g., Russell \& Dopita 1992; appendices in Welty et al. 1997, 1999a).
For example, the current Ti/Fe ratios (in relatively young stars) are about $-$2.44 dex in the LMC and $-$2.49 dex in the SMC, versus $-$2.55 dex in the solar photosphere (Lodders 2003).
For predominantly neutral clouds in the SMC and LMC, depletions may thus be estimated by taking ratios of the column densities of the dominant ions of various refractory elements with respect to $N$(\mbox{S\,{\sc ii}}) and/or $N$(\mbox{Zn\,{\sc ii}}) (which are typically at most very mildly depleted) and comparing with the corresponding stellar/nebular ratios -- and those relative abundances may be fairly directly compared with Galactic values.

Although detailed interstellar abundance data are available for only a few sight lines in the Magellanic Clouds, there are some intriguing indications of differences in the depletion patterns, particularly for Si, Mg, and Ti in the SMC.
In several of the main components toward Sk~155, for example, ratios of the column densities of \mbox{Si\,{\sc ii}}, \mbox{Mg\,{\sc ii}}, and \mbox{Ti\,{\sc ii}} with respect to those of \mbox{Zn\,{\sc ii}} and \mbox{S\,{\sc ii}} (as for the [Ti/Zn] in Fig.~\ref{fig:mczn}) indicate that all three of those elements are much less severely depleted than would have been expected from the significant depletions of Fe and Ni in those components -- a pattern not seen (to this point) in any Galactic sight line (Welty et al. 2001 and in prep.; cf. Sofia et al. 2006). 
Milder than expected depletions of Si and Mg are also seen in the component at 220 km~s$^{-1}$ toward the LMC star Sk$-$70~115 (Welty et al. 2004 and in prep.) -- but in that case Ti is severely depleted (Fig.~\ref{fig:mczn}), consistent with the severe depletion of Fe in that component.
These differences in depletion patterns are most easily seen in components with fairly severe depletions of at least some of the more refractory elements. 
Unfortunately, few such components have been identified in the Magellanic Clouds; additional UV spectra of more reddened LMC and SMC sight lines are needed to gauge the prevalence of any ``anomalous'' depletions.
Because titanium is not a major constituent of the dust, any differences in the pattern of Ti depletion would primarily affect its utility as a depletion indicator.
If the very mild depletions of Si and Mg seen toward Sk~155 (and several other SMC stars) are typical of the SMC, however, there would be significant implications for the composition of the dust there, as most models for SMC dust rely heavily on silicates (and require most of the available silicon) (e.g., Cartledge et al. 2005; Li et al. 2006).

\subsubsection{Line profiles and \mbox{Ti\,{\sc ii}}/\mbox{Ca\,{\sc ii}} ratios}
\label{sec-mctica}

The \mbox{Ti\,{\sc ii}} data now available for the Milky Way, SMC, and LMC reveal both similarities and differences in the behaviour of titanium in the interstellar media of the three galaxies.
As in the Galactic ISM, the \mbox{Ti\,{\sc ii}} profiles in the Magellanic Clouds typically are relatively smooth and spread over a fairly wide velocity range -- more similar to the corresponding profiles of \mbox{Ca\,{\sc ii}} (and those of other dominant ions of depleted elements) than to the narrower, more structured profiles of \mbox{Na\,{\sc i}}, \mbox{K\,{\sc i}}, or the various molecular species (Cox et al. 2006, 2007; Welty et al. 2006; Welty \& Crowther, in prep.).
The absorption from both \mbox{Ti\,{\sc ii}} and \mbox{Ca\,{\sc ii}} often is much stronger, relative to that of \mbox{Na\,{\sc i}}, in the Magellanic Clouds.
The variations in the \mbox{Ti\,{\sc ii}}/\mbox{Ca\,{\sc ii}} ratio are again fairly smooth with velocity in the SMC and LMC, but the ratio itself is typically larger -- both for individual velocity components (or velocity intervals) and for the integrated sight line values -- which may be due to enhanced ionization of the \mbox{Ca\,{\sc ii}} in the (generally) stronger radiation fields there.
There are, however, still cases of low \mbox{Ti\,{\sc ii}}/\mbox{Ca\,{\sc ii}} ratios in both the SMC and LMC, such as in the highest velocity components toward stars at the southwestern end of the main ``bar'' of the SMC (e.g., Sk~40, AV~47) and in the components near 220 km~s$^{-1}$ toward Sk$-$70~115 in the LMC2 region.

\subsubsection{\mbox{Ti\,{\sc ii}} depletions}
\label{sec-mcdepl}

The most striking difference in the behaviour of titanium in the three galaxies is the much milder Ti depletion, for any given $N$(H$_{\rm tot}$), $E(B-V)$, or $f$(H$_2$), in the LMC and (especially) the SMC (Figs.~\ref{fig:tivsh}, \ref{fig:tivsebv}, and \ref{fig:tivsfh2}).
Because the depletions are computed with reference to the stellar titanium abundances in each galaxy, the D(Ti) in each case represent the fraction of Ti in the gas phase, relative to the total amount of titanium available (which is lower in the LMC and SMC; Table~\ref{tab:refab}).
In nearly all the sight lines (Galactic, LMC, and SMC), the \mbox{H\,{\sc i}} column densities have been derived from fits to the damped Lyman-$\alpha$ absorption profiles (not from 21 cm emission profiles), over the same path as the \mbox{Ti\,{\sc ii}} absorption.
In addition, the molecular fractions are typically fairly small in the SMC and LMC -- so that $N$(H$_{\rm tot}$) is generally not much greater than $N$(\mbox{H\,{\sc i}}).
The higher gas-phase abundances of titanium (and inferred milder Ti depletions) in the Magellanic Clouds thus are not due either to higher overall titanium abundances or to significantly underestimated total hydrogen column densities [even where $N$(H$_2$) is not known].

One factor which might contribute to differences in the overall level of the integrated sight line depletions in the Milky Way, LMC, and SMC is the ratio of cold, dense gas (with more severe depletions) to warmer, more diffuse gas (with milder depletions) sampled in each galaxy [as discussed above regarding the scatter in D(Ti) at any given $N$(H) within each of the three galaxies].
Wong et al. (2009) have recently conjectured, for example, that regions in the LMC with high \mbox{H\,{\sc i}} column densities but no associated CO emission might be characterized by a higher fraction of warm, diffuse gas.
Most of the Galactic sight lines in the current sample lie primarily within the disc, where the colder, denser gas is more likely to be found. 
The sight lines that probe significant lengths through the Galactic halo generally exhibit less severe overall depletions than the average Galactic value [at any given $N$(H$_{\rm tot}$)].
The Milky Way portion of the absorption toward the Magellanic Clouds targets, for example, has $N$(H) $\sim$ 4 $\times$ 10$^{20}$ cm$^{-2}$ (SMC) or 4--8 $\times$ 10$^{20}$ cm$^{-2}$ (LMC) (estimated from 21 cm emission profiles; e.g., Staveley-Smith et al. 2003), with an average D(Ti) $\sim$ $-$1.6 dex -- very similar to the values seen toward Galactic halo stars at $|z|$ $\ga$ 1~kpc (e.g., Lipman \& Pettini 1995).
The Magellanic Clouds sight lines, however, all probe gas in the outer ``halo'' regions of the LMC or SMC, with (perhaps) some contribution from gas in the main body of the galaxy (depending on the location of the target star within the galaxy).
Many of the targets in runs V02 and V04, for example, were chosen to be relatively lightly reddened, and may thus lie preferentially in front of much of the interstellar material in the SMC and LMC along those sight lines.
For a number of the LMC sight lines, for example, fits to the Lyman-$\alpha$ absorption profiles yield only upper limits for LMC $N$(\mbox{H\,{\sc i}}) (Table~\ref{tab:lmccd}; Welty, in prep.), and little, if any \mbox{Na\,{\sc i}} is detected (Fig.~\ref{fig:specm1}).

Differences in the ratios of cold, dense gas to warm, diffuse gas probably cannot entirely account for the observed differences in Ti depletions among the three galaxies, however.
Most of the SMC and LMC sight lines observed in runs V01 and V03, for example, exhibit significant column densities of \mbox{Na\,{\sc i}} and/or H$_2$ (Tumlinson et al. 2002; Welty et al. 2006; Welty \& Crowther, in prep.); several have detections of CN and/or CO absorption (Andr\'{e} et al. 2004; Welty et al. 2006).
Such sight lines presumably probe appreciable amounts of colder, denser gas in the Magellanic Clouds, as the kinetic temperatures inferred from the H$_2$ rotational excitation are similar to those found for Galactic sight lines with comparable $N$(H$_2$) (Tumlinson et al. 2002; Cartledge et al. 2005).
Even those sight lines, however, generally exhibit significantly milder Ti depletions than those found for Galactic sight lines with comparable $N$(H$_{\rm tot}$) and $f$(H$_2$).
Toward Sk~143 (SMC), with log[$N$(CN)] = 12.47 and log[$N$(CH) = 13.60 (Welty et al. 2006), D(Ti) = $-$1.68 dex for the sight line as a whole -- versus $-$2.0 to $-$3.1 dex for Galactic sight lines with similar column densities of CN or CH.
If most of the SMC $N$(H$_{\rm tot}$) toward Sk~143 is assumed to be in the component near 133 km~s$^{-1}$ (which contains the CH, CN, and most of the \mbox{Na\,{\sc i}}), then D(Ti) in that main component would be of order $-$2.3 dex -- again less severe than for comparable individual Galactic components. 
As noted above, however, Ti is severely depleted ([Ti/Zn] $\sim$ $-$2.8 dex) in the main component(s) toward Sk$-$70~115 (LMC).

\subsubsection{Depletions and local densities}
\label{sec-mcdnh}

While no clear relationship was discerned between the Ti depletion and the local hydrogen density in the Galactic sight lines, it is still worth examining corresponding data for the Magellanic Clouds.
Analyses of the \mbox{C\,{\sc i}} fine-structure excitation, in the handful of Magellanic Clouds sight lines with suitable UV spectra, indicate that the thermal pressures ($n_{\rm H}T$) in the main LMC or SMC components appear to be factors of 4--5 higher than those typically found in the local Galactic ISM (e.g., Jenkins \& Tripp 2001; Andr\'{e} et al. 2004; Welty et al., in prep.).
Such elevated pressures are consistent with theoretical models of interstellar clouds (e.g., Wolfire et al. 1995), which predict that higher pressures are needed to obtain stable cold, neutral clouds in environments characterized by lower dust-to-gas ratios and stronger ambient radiation fields.
As the kinetic temperatures inferred from analyses of the H$_2$ rotational excitation in those sight lines (50--80 K; Tumlinson et al. 2002; Welty et al., in prep.) are generally similar to those found in Galactic sight lines, the local hydrogen densities in the main LMC and SMC components would thus be {\it higher} than those in otherwise comparable Galactic clouds.
The milder Ti depletions in the LMC and SMC thus apparently cannot be ascribed to generally lower local hydrogen densities there.

\subsection{Intermediate-velocity gas toward the LMC}
\label{sec-ivgas}

While the absorption seen for many neutral and ionized species at $v$ $\la$ 100 km~s$^{-1}$ and at $v$ $\ga$ 200 km~s$^{-1}$ toward the LMC may be reasonably ascribed to gas in the Milky Way (disc and halo) and in the LMC, respectively, the origin of the absorption seen at intermediate velocities has been somewhat controversial (de Boer, Morras, \& Bajaja 1990; Staveley-Smith et al. 2003).
This intermediate-velocity (IV) absorption is commonly seen for \mbox{Ca\,{\sc ii}} and in the stronger lines of the dominant ions (in either neutral or somewhat ionized gas) of various abundant elements (e.g., \mbox{Si\,{\sc ii}}, \mbox{Fe\,{\sc ii}}), but is seldom seen for trace neutral species or H$_2$ (Savage \& de Boer 1981; Wayte 1990; Welty et al. 1999a; Danforth et al. 2002; Tumlinson et al. 2002; Welty \& Crowther, in prep.).

Lehner et al. (2009) have analyzed sensitive \mbox{H\,{\sc i}} 21 cm emission spectra together with far-UV spectra obtained with {\it FUSE} in order to characterize the depletions and physical properties of this IV gas.
They detect \mbox{H\,{\sc i}} emission in about one-third of the 139 LMC sight lines examined, with $N$(\mbox{H\,{\sc i}}) ranging from about 2.5--16 $\times$ 10$^{18}$ cm$^{-2}$, and absorption from \mbox{O\,{\sc i}} and/or \mbox{Fe\,{\sc ii}} in more than two-thirds of the sight lines.
The column density ratios \mbox{O\,{\sc i}}/\mbox{H\,{\sc i}} and \mbox{O\,{\sc i}}/\mbox{Fe\,{\sc ii}} suggest that the gas has a metallicity of about one-third solar and is predominantly ionized; the relative abundances of \mbox{Si\,{\sc ii}}, \mbox{S\,{\sc ii}}, and \mbox{Fe\,{\sc ii}} suggest that some dust may be present.
Some highly ionized gas, traced by \mbox{O\,{\sc vi}}, is found in most of the sight lines.
Lehner et al. conclude that the IV gas is associated with the LMC.

Intermediate-velocity absorption from \mbox{Ca\,{\sc ii}} is detected for all the LMC sight lines in Table~\ref{tab:mcew} (most of which are included in the Lehner et al. 2009 sample); \mbox{Ti\,{\sc ii}} is detected in a few cases.
The \mbox{Ti\,{\sc ii}}/\mbox{Ca\,{\sc ii}} ratios in the IV gas range from about 0.1 to 0.5 -- somewhat lower than the values typical of the main LMC gas at $v$ $\ga$ 200 km~s$^{-1}$.
The \mbox{Ti\,{\sc ii}}/\mbox{H\,{\sc i}} and \mbox{Ti\,{\sc ii}}/\mbox{O\,{\sc i}} ratios (for which ionization corrections should be minimal) indicate very mild Ti depletions, of order $-$0.5 to $-$0.1 dex; slightly larger \mbox{Ti\,{\sc ii}} deficits relative to \mbox{Fe\,{\sc ii}} are consistent with the gas being predominantly ionized.
Very mild depletions, similar to those found for Galactic halo clouds, also were found for Mg, Al, Si, Cr, Mn, Fe, and Ni in the IV gas toward SN 1987A (Welty et al. 1999a).
 
\begin{figure}   
\includegraphics[width=84mm]{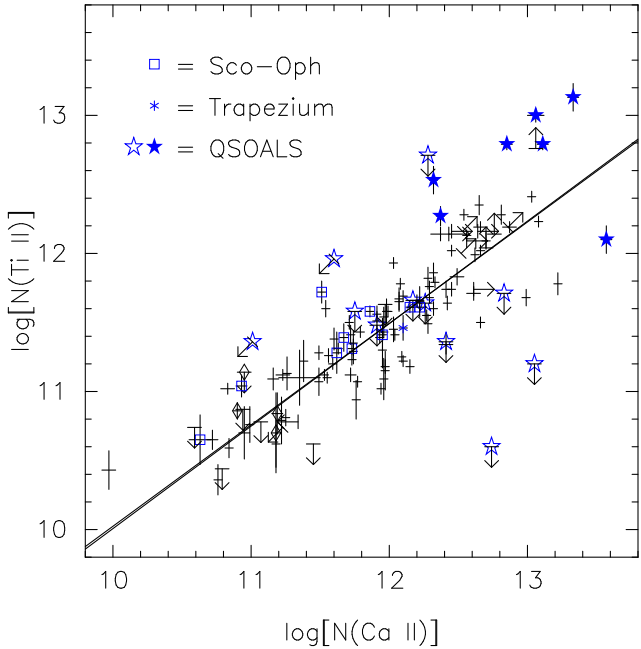}
\caption{$N$(\mbox{Ti\,{\sc ii}}) vs. $N$(\mbox{Ca\,{\sc ii}}) for both Galactic sight lines and QSO absorption-line systems.
Blue stars denote QSO absorbers (filled symbols for detections, open symbols for limits); 
blue open squares denote Sco-Oph sight lines; blue asterisks denote Orion Trapezium region sight lines; plus signs denote other Galactic sight lines (with size indicating $\pm$1$\sigma$ uncertainties).
The two (nearly indistinguishable) solid lines, with slopes $\sim$ 0.7, represent weighted and unweighted fits to the Galactic data (not including the Sco-Oph or Trapezium sight lines).}
\label{fig:tivscaq}
\end{figure}

\subsection{QSO absorption-line systems}
\label{sec-qso}

Measurements of the gas-phase abundances of various atomic species in the absorption-line systems seen toward distant quasars and gamma-ray bursts -- primarily the damped and sub-damped Lyman-$\alpha$ systems, with log $N$(\mbox{H\,{\sc i}}) $\ge$ 20.3 and from 19.0 to 20.3, respectively -- provide a key avenue for understanding the build-up of heavy elements and dust in galaxies, over a substantial fraction of the age of the Universe.
The mean metallicities inferred from large samples of damped Lyman-$\alpha$ systems (DLAs), over the redshift interval 0.1 $\la$ $z$ $\la$ 5, indicate (1) systematically lower values than those expected from models of galaxy evolution and other indications of star formation, (2) only modest evolution (slope $\sim$ $-$0.25 dex per unit redshift), and (3) a significantly subsolar value at the lowest redshifts (e.g., Prochaska et al. 2003b; Kulkarni et al. 2005).
While a small number of ``metal-rich'' DLAs have been identified (e.g., Herbert-Fort et al. 2006), a significant fraction of the ``missing'' metals may be in the sub-DLAs, which have both systematically higher mean metallicities than the DLAs and roughly solar mean metallicities at low $z$ -- at least for the current relatively small sub-DLA sample at $z$ $\la$ 3.2 (e.g., Khare et al. 2007; Kulkarni et al. 2007; Meiring et al. 2008; P\'{e}roux et al. 2008; Dessauges-Zavadsky, Ellison, \& Murphy 2009). 

Attempts to determine the nucleosynthetic history of the galaxies traced by the DLAs and sub-DLAs -- e.g., gauging the relative contributions of Type I and Type II supernovae by examining the relative gas-phase abundances of the so-called $\alpha$ and Fe-peak elements -- have been complicated by the generally poorly known effects of depletion, which can yield abundance ratios for some elements that are similar to those seen in the Galactic metal-poor stars that exemplify the Type II-dominated abundance pattern (e.g., Lauroesch et al. 1996; Lu et al. 1996; Prochaska \& Wolfe 2002; Vladilo 2002; Dessauges-Zavadsky et al. 2004). 
Positive values of the [Si/Fe] ratio, for example, could be consistent either with solar total abundances and some degree of depletion or with a Type II abundance pattern (with enhancements of the $\alpha$-elements Mg, Si, S, etc.) and no depletion.
Attempts to assess and correct for the effects of depletion (e.g., Vladilo 2002, 2004; Dessauges-Zavadsky et al. 2002, 2004; Jenkins 2009) generally assume that the depletion patterns in the absorbers are the same as those found in the Galactic ISM -- i.e., that the composition of the dust is independent of metallicity (or any other factors that may be different in the absorbers).
In view of the ambiguities in the interpretation of some of the more readily determinable ratios (e.g., Si/Fe), attention has been given to ratios such as Ti/Fe and Mn/Fe, where the effects of Galactic depletion and Galactic nucleosynthetic history act in opposition (e.g., Ledoux et al. 2002; Dessauges-Zavadsky et al. 2002).
Because Ti (a Type II product) is generally more severely depleted than Fe (primarily a Type I product) in the Galactic ISM, positive values of [Ti/Fe] would be suggestive of an underlying Type II abundance pattern (and negligible depletion), while negative values of [Ti/Fe] could be more consistent with solar total abundances and some degree of depletion.
The differences in the depletions of Mg, Si, and Ti found in some SMC and LMC sight lines suggest, however, that corrections for dust based on Galactic depletion patterns might not be appropriate, given the lower metallicities characterizing most of the higher-$z$ absorbers -- especially as LMC- or SMC-like UV extinction curves have typically been inferred for those systems (e.g., Pei, Fall, \& Bechtold 1991; Wild, Hewett, \& Pettini 2006; York et al. 2006; Vladilo, Prochaska, \& Wolfe 2008).
 
While the number of DLAs and sub-DLAs with detections of \mbox{Ti\,{\sc ii}} absorption is still relatively small, some preliminary comparisons can be made between the behaviour of \mbox{Ti\,{\sc ii}} in the absorbers, in our Galaxy, and in the Magellanic Clouds.
Table~\ref{tab:ti2qso} lists 79 absorption line systems for which \mbox{Ti\,{\sc ii}} column densities (35 detections, 44 limits) have been published, together with column densities for \mbox{H\,{\sc i}}, H$_2$, \mbox{Zn\,{\sc ii}}, and \mbox{Ca\,{\sc ii}}.
Where possible, the table also lists values for [Zn/H] (an indicator of the metallicity) and for [Ti/Zn] (an indicator of the Ti depletion).
(While [Ti/Zn] will underestimate the severity of the depletion if the $\alpha$-elements are enhanced in any particular system, such differences would be expected to be less than factors of 2--3, in view of the [Ti/Zn] ratios seen for Galactic metal-poor stars at comparable metallicities.)
For this sample, the system redshifts range from 0.43 to 3.77; the \mbox{H\,{\sc i}} column densities range from $<$10$^{18.8}$ to 10$^{21.9}$ cm$^{-2}$ (38 of the systems are DLAs, 19 are sub-DLAs); the metallicities (as gauged from [Zn/H]) range from about $-$1.6 to +0.7 dex.

Figure~\ref{fig:tivscaq} shows the relationship between $N$(\mbox{Ti\,{\sc ii}}) and $N$(\mbox{Ca\,{\sc ii}}) for the QSO absorbers (filled or open stars), with the Galactic data and trend lines also shown for reference (as in Fig.~\ref{fig:tivsca}).
While only six of the absorbers have detections of both \mbox{Ti\,{\sc ii}} and \mbox{Ca\,{\sc ii}}, five of those lie above the general Galactic relationship, in the region occupied by the LMC and SMC sight lines; the sixth absorber (and four others with upper limits for \mbox{Ti\,{\sc ii}}), however, lie well below the Galactic trend.
The \mbox{Ti\,{\sc ii}} and \mbox{Ca\,{\sc ii}} column densities derived from composite low-resolution Sloan Digital Sky Survey spectra of 27 systems selected for detectable \mbox{Mg\,{\sc ii}}, \mbox{Ca\,{\sc ii}}, and \mbox{Zn\,{\sc ii}} (Wild et al. 2006; see last three entries of Table~\ref{tab:ti2qso}) and of larger numbers of systems selected for various other properties (York et al. 2006; see their table A4) also lie somewhat above the Galactic trend.
Where both \mbox{Ti\,{\sc ii}} and \mbox{Ca\,{\sc ii}} are detected, the [Ti/Zn] ratios suggest very mild Ti depletions ([Ti/Zn] $\ga$ $-$0.5 dex); the four absorbers with upper limits on \mbox{Ti\,{\sc ii}} that lie significantly below the Galactic trend, however, all exhibit more severe Ti depletions ([Ti/Zn] $\la$ $-$1.3 dex).

\begin{figure}   
\includegraphics[width=84mm]{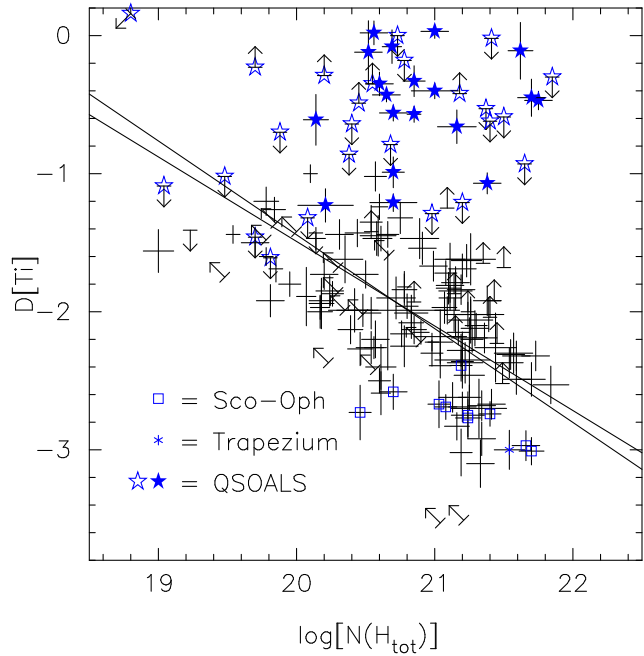}
\caption{Titanium depletions -- D(Ti) for Galactic sight lines and [Ti/Zn] for QSO absorption-line systems -- vs. $N$(H$_{\rm tot}$).
Blue stars denote QSO absorbers (filled symbols for detections, open symbols for limits); 
blue open squares denote Sco-Oph sight lines; blue asterisks denote Orion Trapezium region sight lines; plus signs denote other Galactic sight lines (with size indicating $\pm$1$\sigma$ uncertainties).
The depletions are relative to the solar Ti abundance in our Galaxy and relative to the observed zinc abundance in the QSO absorbers (assuming that zinc is undepleted).
The two solid lines, with slopes $\sim$ $-$0.65, represent weighted and unweighted fits to the Galactic data (not including the Sco-Oph or Trapezium sight lines).}
\label{fig:dtivshq}
\end{figure}

\begin{figure}   
\includegraphics[width=84mm]{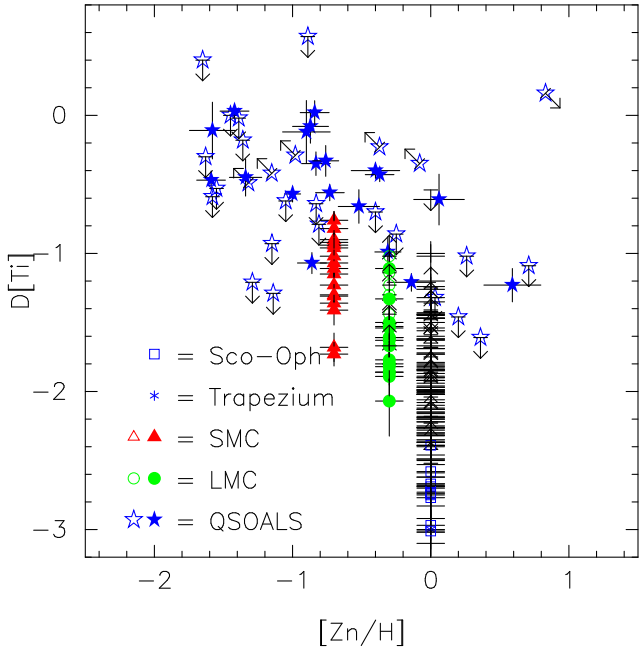}
\caption{Titanium depletions -- D(Ti) for Galactic sight lines and [Ti/Zn] for QSO absorption-line systems -- vs. metallicity -- 0.0 for Milky Way, $-$0.3 for LMC, $-$0.7 for SMC, [Zn/H] for QSO absorbers.
Blue stars denote QSO absorbers, green circles denote LMC sight lines, red triangles denote SMC sight lines (filled symbols for detections, open symbols for limits); 
blue open squares denote Sco-Oph sight lines; blue asterisks denote Orion Trapezium region sight lines; plus signs denote other Galactic sight lines (with size indicating $\pm$1$\sigma$ uncertainties; 0.1 dex assumed for Galactic, LMC, SMC metallicities).
The depletions are relative to the solar Ti abundance in our Galaxy and relative to the observed zinc abundance in the QSO absorbers (assuming that zinc is undepleted).}
\label{fig:dtivsmet}
\end{figure}

Figure~\ref{fig:dtivshq} shows the relationship between D(Ti) and $N$(H$_{\rm tot}$) for the absorbers, again with the Galactic data and trend lines for comparison (as in Fig.~\ref{fig:tivsh}).
Although there is considerable scatter in D(Ti) at any $N$(H$_{\rm tot}$), most of the absorber data lie well above the region occupied by the Galactic sight lines, in many cases at slightly higher (less negative) D(Ti) and somewhat lower $N$(H$_{\rm tot}$) than the SMC sight lines.
All the absorbers with log[$N$(H$_{\rm tot}$)] $<$ 20.3 and D(Ti) $<$ $-$0.9 dex -- which overlap with the lower column density end of the Galactic distribution -- are metal-rich sub-DLAs, with [Zn/H] ranging from 0.0 to +0.7 dex.
The rest of the absorbers exhibit a range in metallicity ([Zn/H]) for any given D(Ti) or $N$(H$_{\rm tot}$); there may be a tendency for lower metallicities at higher $N$(H$_{\rm tot}$), as seen for the sub-DLAs and DLAs in general (e.g., Khare et al. 2007).
While the data for the Milky Way, LMC, and SMC (Fig.~\ref{fig:tivsh}) suggest trends of increasingly severe Ti depletion at higher $N$(H$_{\rm tot}$) for sight lines within any individual system (at some constant metallicity), the ensemble of QSO absorbers (single sight lines through systems spanning a range of metallicities) in Fig.~\ref{fig:dtivshq} exhibits generally less severe Ti depletions at higher $N$(H$_{\rm tot}$) -- a consequence of trends for (1) lower metallicities at higher $N$(H$_{\rm tot}$) and (2) more severe depletions (on average) at higher metallicities (see next).

Figure~\ref{fig:dtivsmet} shows the relationship between D(Ti) and metallicity (which is assumed to be 0.0 dex for the Galactic sight lines, $-$0.3 dex for the LMC sight lines, $-$0.7 dex for the SMC sight lines, and [Zn/H] for the DLAs and sub-DLAs).
The uncertainties on the metallicity are arbitrarily assumed to be 0.1 dex for the Galactic, LMC, and SMC sight lines -- to give a rough indication of the density of points as a function of D(Ti).
As noted in previous studies of DLA abundances (e.g., Ledoux, Petitjean, \& Srianand 2003; Noterdaeme et al. 2008), there appears to be a trend of increasingly severe depletion at higher metallicities -- which is not due to the presence of the \mbox{Zn\,{\sc ii}} column density in both quantities, as the uncertainties in $N$(\mbox{Zn\,{\sc ii}}) are generally less than 0.2 dex.
In the metallicity range from $-$0.7 to 0.0 dex, the QSO absorbers generally exhibit somewhat milder Ti depletions than the SMC, LMC, and Milky Way -- perhaps an indication that most of the absorber sight lines contain a larger fraction of warm, diffuse gas (as might be expected on cross-section considerations).
Given the lower metallicities characterizing most DLAs at $z$ $\ga$ 1.6, Kanekar et al. (2009) conclude that those higher-$z$ DLAs are dominated by warm gas, based on an apparent anti-correlation between metallicity and the spin temperature derived from observations of 21~cm absorption.
The two absorbers in this sample with known significant column densities of H$_2$ [F0812+32 (z=2.6263) and Q1331+170 (z=1.7765)] both have more severe Ti depletions than those seen for other systems at comparable metallicities -- consistent with the general trend for systems with higher $f$(H$_2$) (e.g., Noterdaeme et al. 2008).
While the subset of absorbers with strong \mbox{Ca\,{\sc ii}} absorption is characterized by a higher dust content (reddening) than the general absorber population (e.g., Wild \& Hewett 2005), there is no apparent trend in the severity of Ti depletion with $N$(\mbox{Ca\,{\sc ii}}) (see also Nestor et al. 2008).


\section{Summary / Conclusions}
\label{sec-summ}

We have combined several new sets of absorption-line data for interstellar \mbox{Ti\,{\sc ii}} (for 60 sight lines in our Galaxy and 43 sight lines in the lower metallicity Magellanic Clouds) with previously reported data for \mbox{Ti\,{\sc ii}} and other atomic and molecular species, in order to explore the behaviour of \mbox{Ti\,{\sc ii}} in a wide variety of interstellar environments.
The new data include (most of) the highest resolution \mbox{Ti\,{\sc ii}} spectra yet reported (FWHM $\sim$ 1.3--1.5 km~s$^{-1}$, for 17 of the Galactic sight lines), a significantly expanded sample of \mbox{Ti\,{\sc ii}} spectra for more heavily reddened Galactic sight lines (29 with $E(B-V)$ $\ge$ 0.35), and the most extensive set of \mbox{Ti\,{\sc ii}} spectra obtained for sight lines in the Magellanic Clouds (15 in the SMC, 28 in the LMC).

Comparisons of the highest-resolution Galactic \mbox{Ti\,{\sc ii}} spectra with corresponding high-resolution spectra of \mbox{Na\,{\sc i}}, \mbox{K\,{\sc i}}, and \mbox{Ca\,{\sc ii}} directly reveal differences in the distribution of those species -- which reflect differences in their ionization and depletion behaviour and in the physical properties of the various interstellar clouds probed. 
In general, the \mbox{Ti\,{\sc ii}} profiles are relatively smoothly varying, and resemble most the profiles seen for \mbox{Ca\,{\sc ii}} and for the dominant ions of various other refractory elements -- with little indication of the narrow absorption components (likely due to relatively cold, moderately dense gas) seen for both \mbox{Ca\,{\sc ii}} and various trace neutral species.
Both the line profiles and the \mbox{Ti\,{\sc ii}} column densities -- obtained from detailed fits to the line profiles and/or by integrating the ``apparent'' optical depths over the line profiles -- suggest that the observed gas-phase \mbox{Ti\,{\sc ii}} traces primarily the warmer, more diffuse gas, and that Ti is severely depleted into the dust in the colder, denser clouds where the trace neutral and molecular species are most abundant.
The total Galactic (disc plus halo) \mbox{Ti\,{\sc ii}} column densities toward stars in the Magellanic Clouds are consistent with a Galactic \mbox{Ti\,{\sc ii}} scale height $\la$ 1500 pc.

The Galactic column densities of \mbox{Ti\,{\sc ii}} and \mbox{Ca\,{\sc ii}} are well correlated, with $N$(\mbox{Ti\,{\sc ii}}) varying roughly as $N$(\mbox{Ca\,{\sc ii}})$^{0.7}$.
The two species also appear to be fairly well correlated in the SMC and LMC, but the slope of the relationship is somewhat steeper and the average \mbox{Ti\,{\sc ii}}/\mbox{Ca\,{\sc ii}} ratio is somewhat higher (for the current sample of sight lines).
Variations in the \mbox{Ti\,{\sc ii}}/\mbox{Ca\,{\sc ii}} ratio may reflect differences in the relative depletions of Ti and Ca and/or changes in the calcium ionization balance.
While the \mbox{Ca\,{\sc ii}}/\mbox{Ti\,{\sc ii}} ratio might (in principle) be used to estimate the electron density in diffuse clouds where \mbox{Ca\,{\sc ii}} is a trace ion, calibration of that ratio is complicated by differences in the spatial distributions of \mbox{Ca\,{\sc i}}, \mbox{Ca\,{\sc ii}}, and \mbox{Ti\,{\sc ii}} and by differences in the temperature and calcium ionization balance between the diffuse clouds and the denser clouds where \mbox{Ca\,{\sc i}} is detected.

In the currently available sample of Galactic sight lines, the titanium depletion ranges from about $-$1.0 dex (toward the low halo star HD~93521) to about $-$3.1 dex (toward the moderately reddened HD~27778).
The depletions in individual components or small velocity intervals [gauged via reference to high-resolution UV spectra of \mbox{Zn\,{\sc ii}} and/or by assuming that most of the total $N$(H) is in the strongest components seen in the trace neutral species] span a somewhat wider range -- from $\la$ $-$3.5 dex (for the main components in moderately reddened sight lines with significant molecular abundances) to $\ga$ $-$0.3 dex (for some halo components).
In general, the titanium depletions become increasingly severe with increasing mean sight line density, total hydrogen column density, or colour excess; the considerable scatter at any given $N$(H$_{\rm tot}$) or $E(B-V)$ may be due to differences in the proportions of cold, dense gas and warmer, more diffuse gas along the various sight lines.
Despite the fairly tight anti-correlation between D(Ti) and $<n_{\rm H}>$, there does not appear to be a clear dependence of the Ti depletion on the local hydrogen density (as determined from \mbox{C\,{\sc i}} fine-structure excitation or C$_2$ rotational excitation) -- at least for the relatively small sample of sight lines with estimates for both D(Ti) and $n_{\rm H}$ currently available.
The titanium depletion is roughly constant, at about $-$1.8 dex (with much scatter) for molecular fractions less than 0.1, but becomes increasingly severe for $f$(H$_2$) $\ga$ 0.1; some sight lines with $f$(H$_2$) $>$ 0.5 have D(Ti) $<$ $-$3.0 dex.
Some apparent regional differences in D(Ti) are seen; many of the stars in the Sco-Oph region, for example, exhibit somewhat more severe depletions than are found for most other Galactic sight lines with comparable $N$(H$_{\rm tot}$), $E(B-V)$, or $f$(H$_2$).

While the deuterium depletion hypothesis appears to provide a reasonable explanation for the range in \mbox{D\,{\sc i}}/\mbox{H\,{\sc i}} ratios abserved in the local Galactic ISM, other factors -- such as the presence of ionized gas, complex component structures, and differences in the severity of the depletions of different elements -- may conspire to obscure the expected correlations between the \mbox{D\,{\sc i}}/\mbox{H\,{\sc i}} ratio and the depletions of various other elements. 
Additional data for both \mbox{Ti\,{\sc ii}} and other depleted species, in sight lines where \mbox{D\,{\sc i}} has been measured, would be valuable to better understand those other factors -- and thus to better understand the current Galactic D/H ratio.

While the titanium depletions in the Magellanic Clouds exhibit very similar trends with total hydrogen column density, colour excess, and molecular fraction to those seen in our Galaxy, the depletions generally are much less severe in the LMC and (especially) the SMC.
For a given $N$(H$_{\rm tot}$), for example, D(Ti) is less severe (on average) by factors of $\sim$ 5 in the LMC and $\sim$ 20 in the SMC, relative to the mean Galactic value.
In the current Magellanic Clouds sample, D(Ti) ranges from about $-$1.1 to $-$2.1 dex in the LMC and from about $-$0.8 to $-$1.7 dex in the SMC.
As in our Galaxy, a somewhat broader range is found for D(Ti) in individual components (for the three Magellanic Clouds sight lines with moderately high resolution spectra of both \mbox{Ti\,{\sc ii}} and \mbox{Zn\,{\sc ii}}); some individual components do exhibit fairly severe Ti depletions.
Very mild Ti depletions, ranging from $-$0.5 to $-$0.1 dex, are found for the ``intermediate-velocity'' (100 km~s$^{-1}$ $\la$ $v$ $\la$ 200 km~s$^{-1}$) gas seen toward several LMC targets.

While some of the differences in average D(Ti) may be due to sampling larger fractions of warmer, more diffuse gas in the LMC and SMC, even the Magellanic Clouds sight lines with strong absorption from \mbox{Na\,{\sc i}} and H$_2$ (and, in several cases, absorption from CN and/or CO) -- which presumably probe appreciable amounts of colder, denser gas -- exhibit milder Ti depletions than those found for comparable Galactic sight lines. 
For example, the most severe Ti depletions in the Magellanic Clouds are found toward Sk~143 (SMC) and Sk$-$67~2 (LMC) -- two moderately reddened sight lines with the highest molecular fractions and sole detections of CN absorption in the current sample -- but even there the depletions still are much less severe than those in Galactic sight lines with similar $N$(H$_{\rm tot}$) or $f$(H$_2$).
The local hydrogen densities in diffuse, neutral clouds in the Magellanic Clouds may be somewhat higher (on average) than those in the local Galactic ISM, so the milder \mbox{Ti\,{\sc ii}} depletions in the Magellanic Clouds are not due to lower $n_{\rm H}$ there.

Several of the main components toward Sk~155 (SMC) previously noted to have anomalously mild depletions of Mg and Si also exhibit much less severe Ti depletions than would have been expected from the depletions of Cr, Mn, Fe, and Ni in those components.
The main components toward Sk$-$70~115 (LMC), also with somewhat milder than expected depletions of Mg and Si, have very severe Ti depletions, however -- consistent with the severe depletions seen for Fe and Ni.
High-resolution UV spectra of additional SMC and LMC stars -- including the absorption lines from dominant ions of both severely and mildly depleted elements -- are needed to better understand the differences in depletion behaviour seen for Mg, Si, and Ti in the Magellanic Clouds.

Although \mbox{Ti\,{\sc ii}} has been detected in relatively few QSO absorption-line systems (DLAs and sub-DLAs), comparisons with corresponding column density data for \mbox{H\,{\sc i}}, \mbox{Ca\,{\sc ii}}, and \mbox{Zn\,{\sc ii}} suggest some similarities in the Ti depletion behaviour to that in the Magellanic Clouds.
Like the LMC and SMC sight lines, most (though not all) of the absorbers exhibit somewhat higher \mbox{Ti\,{\sc ii}}/\mbox{Ca\,{\sc ii}} ratios than those found for most Galactic sight lines.
And while there is considerable scatter in D(Ti) (as gauged from [Ti/Zn]) at any given $N$(H$_{\rm tot}$) or metallicity, the Ti depletions in the absorbers are (on average) even less severe than those in the LMC and SMC -- perhaps indicative of an even higher fraction of warm, diffuse gas in the absorbers.
The Ti depletions found for some metal-rich sub-DLAs, however, are similar to those found for Galactic sight lines with comparable $N$(H$_{\rm tot}$).
In (only) apparent contrast to the trends seen within the Milky Way, LMC, and SMC, the ensemble of absorber \mbox{Ti\,{\sc ii}} data exhibits generally milder Ti depletion at higher $N$(H$_{\rm tot}$) -- a consequence of trends for lower metallicities at higher $N$(H$_{\rm tot}$) and less severe depletions at lower metallicities.
As LMC- or SMC-like UV extinction has been inferred for DLAs and sub-DLAs, and as differences in the depletions of Mg, Si, and Ti have been noted in at least some LMC and SMC sight lines, the use of Galactic depletion patterns to interpret the gas-phase abundances in the QSO absorbers may not be entirely reliable.


\section*{Acknowledgments}

We thank J. Fowler for stepping in at the last minute to perform the 1994 KPNO run, 
A. Ritchey and J. Thorburn for processing the individual UVES exposures from 2000 and 2003 (respectively),
D. Willmarth (KPNO) and M. Rejkuba and M.-R. Cioni (ESO/Paranal) for their assistance in obtaining the observations,
and S. Federman, L. Hobbs, E. Jenkins, and K. Sembach for constructive comments on the original version of this paper.
This work was begun under support from NASA Long-Term Space Astrophysics grants NAG5-3228 and NAG5-11413 to the University of Chicago,
and has made use of NASA's ADS bibliographic services, the SIMBAD database (operated at CDS, Strasbourg, France), and the catalogues compiled by B. Skiff.


\appendix

\section{Spectra, equivalent widths, and column densities}
\label{sec-append}

Figure~\ref{fig:specg1} shows the normalized line profiles for \mbox{Ti\,{\sc ii}} ($\lambda$3383), \mbox{Ca\,{\sc ii}} ($\lambda$3933), and \mbox{Na\,{\sc i}} ($\lambda$3302 or $\lambda$5895) or \mbox{K\,{\sc i}} ($\lambda$7698) toward the Galactic targets.
As described in the text, the Galactic spectra were obtained from different sources, at different resolutions; the source of each spectrum is indicated by the letter code at the far right, just above the continuum.
Most of the higher resolution (FWHM $\la$ 2 km~s$^{-1}$) \mbox{Na\,{\sc i}}, \mbox{K\,{\sc i}}, and \mbox{Ca\,{\sc ii}} spectra are from Welty et al. (1994, 1996), Welty \& Hobbs (2001), or Welty, Snow, \& Morton (in prep.).
The few exceptions are the \mbox{Ti\,{\sc ii}} spectrum for $\zeta$~Per (Hobbs 1979), the \mbox{Ca\,{\sc ii}} and \mbox{K\,{\sc i}} spectra for HD~110432 (generated from component structures reported by Crawford 1995), and the \mbox{K\,{\sc i}} spectrum for HD~183143 (McCall et al. 2002).
The tick marks above the spectra indicate the locations of individual components identified in detailed fits to the higher resolution line profiles. 
Solid dots above the \mbox{Na\,{\sc i}} or \mbox{K\,{\sc i}} tick marks indicate components detected in CN absorption.
Some of the CN velocities were determined from fits to the CN lines near 3579 and/or 3875 \AA\ in the UVES spectra discussed in this paper or in unpublished higher resolution spectra obtained with the Kitt Peak coud\'{e} feed; the rest were taken from other published studies (Gredel, van Dishoeck, \& Black 1991; Welty \& Fowler 1992; Crawford et al. 1994; Crawford 1995; McCall et al. 2002).
Note the differences in total velocity interval for the Galactic sight lines; the vertical dotted lines are separated by 10 km~s$^{-1}$ in each case.
Note also the expanded vertical scales for the weaker absorption features.

Figure~\ref{fig:specm1} shows the spectra of the SMC and LMC targets, with the \mbox{Ti\,{\sc ii}} profile offest by 0.2 above the \mbox{Ca\,{\sc ii}} profile and the \mbox{Na\,{\sc i}} $\lambda$3302 profile offset by 0.2 above the \mbox{Na\,{\sc i}} $\lambda$5895 profile in each case; all profiles are on the same (unexpanded) vertical scale.
The \mbox{Ti\,{\sc ii}} spectra from runs V02 and V04 have been smoothed by summing adjacent pixels, to obtain more nearly optimal sampling for those lower resolution spectra.
Stellar absorption from \mbox{S\,{\sc ii}} and \mbox{Ca\,{\sc ii}} is indicated by the dotted lines superposed on several of the \mbox{Ca\,{\sc ii}} profiles.
For a number of the LMC sight lines, the LMC \mbox{Na\,{\sc i}} $\lambda$5889 (D2) absorption, at about $-$303 km~s$^{-1}$ relative to the $\lambda$5895 (D1) absorption, is blended with the Galactic D1 absorption near $v$ $\sim$ 0 km~s$^{-1}$. 

Table~\ref{tab:comps} lists the component structures (heliocentric velocities, $b$-values, and column densities) for Galactic sight lines for which high-resolution (FWHM $\la$ 2 km~s$^{-1}$) \mbox{Ca\,{\sc ii}} spectra are available (e.g., Welty et al. 1996; Price et al. 2001; Pan et al. 2004).
Fits to the \mbox{Ca\,{\sc ii}} profiles determined the number of components, the relative component velocities, and the $b$-values (in nearly all cases); fits to the \mbox{Ti\,{\sc ii}} profiles then determined the component column densities and an overall velocity offset.
Values for $b$ given in square braces were fixed in the fits, but are fairly well determined; values in parentheses were also fixed, but are less well determined.
The last column gives the column density ratio \mbox{Ti\,{\sc ii}}/\mbox{Ca\,{\sc ii}} for each component.

Table~\ref{tab:allwids} lists 314 Galactic sight lines for which \mbox{Ti\,{\sc ii}} data ($\lambda$3383 equivalent widths and/or column density estimates) are available.
While this table provides a more extensive compilation than the similar table in Jenkins (2009), only the values deemed ``reliable'' (based on considerations of resolution, S/N ratio, and/or listed uncertainty) have been used in the various correlations discussed in the text.
The 1-$\sigma$ uncertainties on the equivalent widths are those listed in the original references, but they were not all calculated in the same way (some include both photon noise and continuum uncertainties, others just photon noise).
There appears to be even less uniformity in the published upper limits; we have adjusted the limits in Albert et al. (1993) and Hobbs (1984) by a factor of 1.5 and the limit for $\kappa$~Vel in Dunkin \& Crawford (1999) by a factor of 3 in an attempt to make all the limits (roughly) 3-$\sigma$. 
Even so, comparisons with detections of other species (e.g., \mbox{Ca\,{\sc ii}}) may be complicated by differences in velocity extent between the detections and the limits (which in most cases refer to a single unresolved component).
Where the original reference did not list uncertainties (denoted by $\pm$x.x for the equivalent widths and/or by $\pm$0.xx for the column densities), we have assumed a typical value of 0.07 dex; for the correlations, we have assumed a minimum uncertainty of 10 per cent for $N$(\mbox{Ti\,{\sc ii}}).
 
In most cases, the equivalent widths for a given sight line from different references agree within the mutual uncertainties.
Some of the disagreements are likely due to differences in continuum fitting (especially for weak absorption spread over a broad velocity range) and/or in the treatment of blends with stellar \mbox{Ti\,{\sc ii}} absorption features (for stellar types later than about B5); temporal variations might be responsible in a few cases (e.g., toward disturbed regions such as the Vela supernova remnant).
Two estimates for the total sight line \mbox{Ti\,{\sc ii}} column density are listed.
The first estimate was obtained either directly from the equivalent width (values in parentheses) or by integrating the ``apparent'' optical depth (AOD) over the line profile -- in either case yielding a firm lower limit to $N$(\mbox{Ti\,{\sc ii}}).
The second estimate was obtained via a detailed multi-component fit to the line profile, which (in principle) can account for any saturation effects (if the $b$-values are well determined).
In most cases, the values from the detailed fits are not more than 0.1 dex larger than those estimated from the equivalent widths or AOD integrations -- suggesting that saturation effects are generally fairly small for the \mbox{Ti\,{\sc ii}} $\lambda$3383 line.
All column densities from other references have been adjusted for any differences from the $f$-value adopted here (0.358; Morton 2003).

Table~\ref{tab:ti2qso} lists selected column density data for 79 QSO absorption-line systems for which \mbox{Ti\,{\sc ii}} column densities (or upper limits) are available.
The last three entries are for the averages of 27, 13, and 14 \mbox{Ca\,{\sc ii}}-selected systems seen in lower resolution, lower S/N spectra from the Sloan Digital Sky Survey (Wild et al. 2006).
Most of the column densities have been derived from moderate-resolution spectra (FWHM $\sim$ 5--8 km~s$^{-1}$) obtained with Keck/HIRES, VLT/UVES, or Magellan/MIKE within the past decade.
In all cases, the column densities are for the entire system (regardless of the potentially different velocity extents covered by the lines measured for the various species); differences in velocity extent are likely an issue for the listed upper limits as well.
The relative abundances [Zn/H] and [Ti/Zn] are with reference to the adopted solar abundances (Ti/H = $-$7.08 dex; Zn/H = $-$7.37 dex; Lodders 2003).
Significant amounts of ionized gas may well be present for the systems with $N$(\mbox{H\,{\sc i}}) $\la$ 3 $\times$ 10$^{19}$ cm$^{-2}$.

\clearpage

\begin{figure*}
\begin{minipage}{180mm}
\includegraphics[width=180mm]{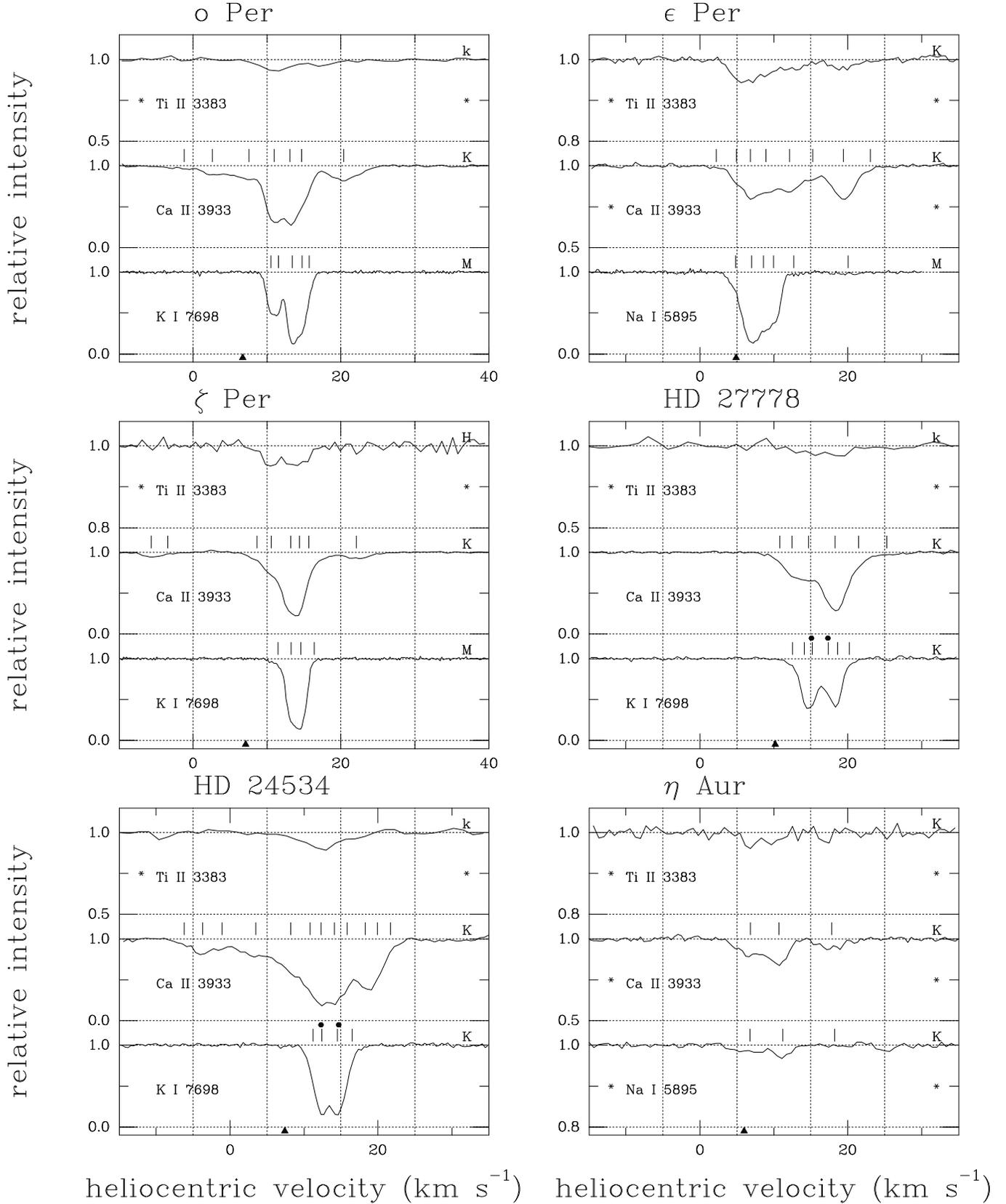}
\caption{The interstellar \mbox{Ti\,{\sc ii}}, \mbox{Ca\,{\sc ii}}, and \mbox{K\,{\sc i}} or \mbox{Na\,{\sc i}} profiles toward the Galactic targets.
The source of the spectra has been noted at the right, just above the continuum [A,c = AAT UHRF (FWHM = 0.3--0.4 km~s$^{-1}$; E = ESO 3.6m/CES (FWHM = 1.2--2.0 km~s$^{-1}$); H = McDonald 2.7m coud\'{e} (FWHM = 1.4 km~s$^{-1}$); K = KPNO coud\'{e} feed (FWHM = 1.2--1.5 km~s$^{-1}$); k = KPNO coud\'{e} feed (FWHM = 2.9--3.6 km~s$^{-1}$); M = McDonald 2.7m coud\'{e} (FWHM = 0.5--0.6 km~s$^{-1}$); m = McDonald 2.7m coud\'{e} (FWHM = 1.8 km~s$^{-1}$); V = VLT/UVES (FWHM = 3.8--8.7 km~s$^{-1}$)].
Tick marks above the spectra indicate the components found in fitting the profiles; solid dots indicate components detected in CN.
Solid triangles at bottom of each plot indicate zero point for LSR velocities.}
\label{fig:specg1}
\end{minipage}
\end{figure*}
 
\begin{figure*}
\begin{minipage}{180mm}
\includegraphics[width=180mm]{figa1b.eps}
\contcaption{}
\end{minipage}
\end{figure*}
 
\begin{figure*}
\begin{minipage}{180mm}
\includegraphics[width=180mm]{figa1c.eps}
\contcaption{}
\end{minipage}
\end{figure*}
 
\begin{figure*}
\begin{minipage}{180mm}
\includegraphics[width=180mm]{figa1d.eps}
\contcaption{}
\end{minipage}
\end{figure*}
 
\begin{figure*}
\begin{minipage}{180mm}
\includegraphics[width=180mm]{figa1e.eps}
\contcaption{}
\end{minipage}
\end{figure*}
 
\begin{figure*}
\begin{minipage}{180mm}
\includegraphics[width=180mm]{figa1f.eps}
\contcaption{}
\end{minipage}
\end{figure*}
 
\begin{figure*}
\begin{minipage}{180mm}
\includegraphics[width=180mm]{figa1g.eps}
\contcaption{}
\end{minipage}
\end{figure*}
 
\begin{figure*}
\begin{minipage}{180mm}
\includegraphics[width=180mm]{figa1h.eps}
\contcaption{}
\end{minipage}
\end{figure*}
 
\begin{figure*}
\begin{minipage}{180mm}
\includegraphics[width=180mm]{figa1i.eps}
\contcaption{}
\end{minipage}
\end{figure*}
 
\begin{figure*}
\begin{minipage}{180mm}
\includegraphics[width=180mm]{figa1j.eps}
\contcaption{}
\end{minipage}
\end{figure*}
 
\clearpage

\begin{figure*}
\begin{minipage}{180mm}
\includegraphics[width=160mm]{figa2a.eps}
\caption{The interstellar \mbox{Ti\,{\sc ii}}, \mbox{Ca\,{\sc ii}}, and \mbox{Na\,{\sc i}} profiles toward the SMC and LMC.
The source of the spectra has been noted at the right, just above the continuum [V = VLT + UVES (FWHM = 4.5--8.7 km~s$^{-1}$)].
The profiles of the weak \mbox{Na\,{\sc i}} $\lambda$3302.3 line are offset by 0.2 above the profiles of the stronger $\lambda$5895.9 line (on the same vertical scale).
Dotted lines indicate stellar S~II and \mbox{Ca\,{\sc ii}} lines adopted in the \mbox{Ca\,{\sc ii}} profile fits; estimates for the stellar radial velocities derived from other lines are noted by asterisks (literature) and/or S (Welty \& Crowther, in prep.).
Note that a wider velocity range is covered than for the Galactic profiles.
For a number of the LMC sight lines, the LMC \mbox{Na\,{\sc i}} $\lambda$5889 (D2) absorption, at about $-$303 km~s$^{-1}$ relative to the $\lambda$5895 (D1) absorption, is blended with the Galactic D1 absorption near $v$ $\sim$ 0 km~s$^{-1}$.}
\label{fig:specm1}
\end{minipage}
\end{figure*}
 
\begin{figure*}
\begin{minipage}{180mm}
\includegraphics[width=160mm]{figa2b.eps}
\contcaption{}
\end{minipage}
\end{figure*}
 
\begin{figure*}
\begin{minipage}{180mm}
\includegraphics[width=160mm]{figa2c.eps}
\contcaption{}
\end{minipage}
\end{figure*}
 
\begin{figure*}
\begin{minipage}{180mm}
\includegraphics[width=160mm]{figa2d.eps}
\contcaption{}
\end{minipage}
\end{figure*}
 
\begin{figure*}
\begin{minipage}{180mm}
\includegraphics[width=160mm]{figa2e.eps}
\contcaption{}
\end{minipage}
\end{figure*}

\begin{figure*}
\begin{minipage}{180mm}
\includegraphics[width=160mm]{figa2f.eps}
\contcaption{}
\end{minipage}
\end{figure*}

\clearpage

\begin{table}
\caption{\mbox{Ti\,{\sc ii}} component structures.}
\label{tab:comps}  
\begin{tabular}{@{}rrcrr}   
\hline
\multicolumn{1}{l}{Star}&  
\multicolumn{1}{c}{$v$}&  
\multicolumn{1}{c}{$b$}&
\multicolumn{1}{c}{$N_{10}$}&  
\multicolumn{1}{c}{\mbox{Ti\,{\sc ii}}/\mbox{Ca\,{\sc ii}}}\\
\multicolumn{1}{c}{Comp}&  
\multicolumn{1}{c}{(km s$^{-1}$)}&   
\multicolumn{1}{c}{(km s$^{-1}$)}&
\multicolumn{1}{c}{(cm$^{-2}$)}&   
\multicolumn{1}{c}{ }\\   
\hline
HD 24398 & H79 &     &  [2.1]\\ 
  1 & $-$5.2 & [1.4] &$<$1.8 &$<$0.86 \\
  2 & $-$2.9 & (2.0) &$<$2.1 &$<$2.63 \\
  3 &    9.1 & (1.2) &$<$1.8 &$<$0.56 \\
  4 &   11.1 & [0.95]&   3.6 &   0.44 \\
  5 &   13.7 & [1.3] &   2.8 &   0.06 \\
  6 &   14.9 & [0.8] &   1.3 &   0.07 \\
  7 &   16.1 & (2.0) &   1.9 &   0.20 \\
  8 &   22.6 & [2.5] &$<$2.1 &$<$0.43 \\
 & \\
HD 24760 & K94 &     &  [1.3]\\ 
  1 &    2.6 & [1.0] &$<$0.6 &$<$1.50 \\
  2 &    5.4 & [0.9] &   3.2 &   1.07 \\
  3 &    7.3 & [0.95]&   3.2 &   0.56 \\
  4 &    9.4 & [1.4] &   2.9 &   0.47 \\
  5 &   12.5 & [1.9] &   2.8 &   0.33 \\
  6 &   15.7 & [1.2] &   0.6 &   0.26 \\ 
  7 &   19.8 & [2.0] &   2.8 &   0.23 \\
  8 &   23.5 & [0.5] &$<$0.6 &$<$2.00 \\
 & \\
HD 32630 & K95 &     &  [2.9]\\ 
  1 &    6.6 & [2.2] &   3.3 &   0.52 \\
  2 &   10.5 & [1.5] &   1.8 &   0.26 \\
  3 &   17.6 & [2.2] &   1.1 &   0.31 \\ 
 & \\
HD 36486 & K95 &     &  [1.3]\\ 
  1 & $-$5.1 & (2.0) &$<$0.8 &$<$0.73 \\
  2 & $-$0.9 & (1.6) &$<$0.7 &$<$0.70 \\
  3 &    2.6 & [1.6] &$<$0.8 &$<$0.50 \\
  4 &    6.6 & [2.3] &   0.6 &   0.15 \\ 
  5 &    9.9 & [1.4] &   0.5 &   0.13 \\ 
  6 &   12.1 & [1.2] &   0.6 &   0.20 \\ 
  7 &   15.7 & [2.8] &   1.8 &   0.18 \\
  8 &   21.0 & (1.7) &   2.9 &   0.49 \\
  9 &   21.7 & [0.6] &$<$1.7 &$<$0.18 \\
 10 &   24.2 & (1.8) &   1.7 &   0.45 \\
 11 &   26.5 & [1.0] &   0.7 &   0.39 \\
 12 &   29.4 & [1.95]&   1.4 &   0.50 \\
 13 &   35.0 & [2.8] &   0.4 &   0.14 \\ 
 14 &   37.5 & [0.5] &$<$0.6 &$<$1.50 \\
 15 &   40.5 & (1.6) &$<$0.7 &$<$1.17 \\
 & \\
HD 37043 & K95 &     &  [1.3]\\ 
  1 & $-$4.2 & [2.5] &   0.4 &   0.27 \\ 
  2 &    1.0 & [2.1] &   0.7 &   0.25 \\
  3 &    5.5 & (2.4) &$<$1.2 &$<$1.00 \\
  4 &    8.3 & [0.65]&   0.6 &   0.06 \\ 
  5 &   12.0 & (2.8) &$<$1.2 &$<$0.26 \\
  6 &   17.8 & (2.8) &   1.8 &   0.43 \\
  7 &   23.2 & (2.0) &   1.8 &   0.60 \\
  8 &   25.2 & [1.4] &   4.8 &   0.60 \\
  9 &   28.7 & [2.7] &   5.7 &   0.36 \\
 10 &   33.2 & [1.2] &   6.0 &   0.22 \\
 11 &   35.7 & (2.0) &   1.7 &   0.61 \\
 12 &   40.5 & [2.4] &   1.0 &   0.26 \\
 & \\
HD 37128 & K94 &     &  [1.0]\\ 
  1 & $-$0.4 & [2.5] &   1.0 &   0.33 \\
  2 &    0.3 & (1.0) &$<$0.9 &$<$1.00 \\
  3 &    2.0 & [0.8] &$<$0.9 &$<$0.60 \\ 
  4 &    2.6 & [0.4] &$<$1.1 &$<$0.25 \\ 
  5 &    3.4 & (1.0) &$<$1.1 &$<$0.44 \\
\hline
\end{tabular}
\end{table}

\begin{table}
\contcaption{}
\begin{tabular}{@{}rrcrr}
\hline
\multicolumn{1}{l}{Star}&
\multicolumn{1}{c}{$v$}&
\multicolumn{1}{c}{$b$}&
\multicolumn{1}{c}{$N_{10}$}&
\multicolumn{1}{c}{\mbox{Ti\,{\sc ii}}/\mbox{Ca\,{\sc ii}}}\\
\multicolumn{1}{c}{Comp}&
\multicolumn{1}{c}{(km s$^{-1}$)}&
\multicolumn{1}{c}{(km s$^{-1}$)}&
\multicolumn{1}{c}{(cm$^{-2}$)}&
\multicolumn{1}{c}{ }\\
\hline
  6 &    4.8 & [1.0] &   1.3 &   0.35 \\
  7 &    6.3 & [0.5] &$<$0.7 &$<$1.75 \\
  8 &    8.5 & [2.5] &   2.1 &   0.23 \\
  9 &   10.0 & [1.0] &$<$1.5 &$<$0.37 \\
 10 &   10.4 & [0.35]&$<$1.5 &$<$0.56 \\
 11 &   11.1 & [0.7] &   0.8 &   0.11 \\
 12 &   12.1 & [0.4] &   1.0 &   0.40 \\
 13 &   12.7 & [1.2] &$<$1.1 &$<$0.17 \\
 14 &   14.4 & [0.6] &   0.6 &   0.50 \\ 
 15 &   16.2 & [1.4] &$<$1.1 &$<$0.37 \\
 16 &   16.9 & [0.5] &$<$1.0 &$<$0.53 \\ 
 17 &   18.0 & [1.5] &   1.0 &   0.17 \\
 18 &   20.1 & (1.5) &   0.8 &   0.31 \\
 19 &   24.0 & [1.8] &   1.9 &   0.19 \\
 20 &   24.6 & [0.65]&   1.4 &   0.24 \\
 21 &   25.7 & [0.8] &   2.2 &   0.42 \\
 22 &   27.7 & [1.15]&   3.9 &   0.20 \\
 23 &   29.4 & (1.0) &   1.4 &   0.78 \\
 24 &   36.5 & [1.5] &   1.0 &   0.71 \\
 25 &   40.0 & [2.3] &   1.1 &   1.10 \\
 & \\
HD 37742 & K95 &     &  [1.6]\\ 
  1 & $-$7.8 & [1.5] &$<$0.9 &$<$0.75 \\
  2 & $-$4.7 & [1.7] &   1.0 &   0.09 \\
  3 & $-$1.4 & [2.25]&$<$1.2 &$<$0.09 \\
  4 &    2.7 & [1.8] &$<$1.0 &$<$0.43 \\
  5 &    6.4 & (1.8) &$<$1.1 &$<$0.50 \\
  6 &   10.3 & [2.15]&$<$1.4 &$<$0.21 \\
  7 &   11.9 & [0.55]&$<$1.0 &$<$1.11 \\
  8 &   14.4 & [2.15]&   1.4 &   0.22 \\
  9 &   18.0 & (1.5) &$<$2.0 &$<$1.25 \\
 10 &   18.7 & [0.75]&   0.9 &   0.27 \\
 11 &   19.9 & [1.0] &   0.9 &   0.23 \\
 12 &   21.9 & [1.2] &   2.8 &   0.55 \\
 13 &   23.4 & [0.75]&   1.9 &   0.15 \\
 14 &   24.9 & (1.5) &   4.4 &   0.31 \\
 15 &   26.7 & [0.4] &$<$1.6 &$<$0.40 \\
 16 &   27.3 & [0.6] &$<$1.5 &$<$0.83 \\
 17 &   29.2 & [0.85]&$<$0.8 &$<$0.80 \\
 18 &   37.5 & [2.4] &$<$1.2 &$<$0.38 \\
 & \\
HD 47839 & K94,K95 &  &  [2.4]\\ 
  1 &$-$14.8 & [2.1] &$<$1.4 &$<$1.17 \\
  2 & $-$9.6 & [1.0] &$<$1.2 &$<$1.00 \\
  3 & $-$8.3 & [0.5] &   0.8 &   0.18 \\ 
  4 & $-$4.6 & [2.4] &   2.3 &   0.29 \\
  5 & $-$1.8 & [0.95]&$<$1.3 &$<$0.24 \\
  6 & $-$0.1 & (1.5) &   1.5 &   0.75 \\
  7 &    2.6 & (2.5) &   1.2 &   0.15 \\
  8 &    6.1 & [1.8] &   1.0 &   0.06 \\ 
  9 &    7.1 & [0.8] &   0.7 &   0.28 \\ 
 10 &   11.6 & [2.9] &   8.0 &   0.16 \\
 11 &   14.6 & (1.5) &   4.5 &   0.90 \\
 12 &   17.7 & [1.5] &   2.5 &   0.35 \\
 13 &   19.1 & [1.5] &   5.9 &   0.35 \\
 14 &   21.4 & [1.0] &   5.2 &   0.42 \\
 15 &   23.3 & [1.0] &   4.9 &   0.47 \\
 16 &   24.3 & [1.0] &$<$3.0 &$<$0.45 \\
 17 &   25.3 & [1.3] &   6.4 &   0.45 \\
 18 &   27.5 & [1.3] &   4.1 &   0.45 \\
 19 &   29.7 & (1.6) &   3.7 &   0.46 \\
 20 &   32.9 & [1.6] &   6.1 &   0.53 \\
\hline
\end{tabular}
\end{table}

\begin{table}
\contcaption{}
\begin{tabular}{@{}rrcrr}
\hline
\multicolumn{1}{l}{Star}&
\multicolumn{1}{c}{$v$}&
\multicolumn{1}{c}{$b$}&
\multicolumn{1}{c}{$N_{10}$}&
\multicolumn{1}{c}{\mbox{Ti\,{\sc ii}}/\mbox{Ca\,{\sc ii}}}\\
\multicolumn{1}{c}{Comp}&
\multicolumn{1}{c}{(km s$^{-1}$)}&
\multicolumn{1}{c}{(km s$^{-1}$)}&
\multicolumn{1}{c}{(cm$^{-2}$)}&
\multicolumn{1}{c}{ }\\
\hline
 21 &    36.2 & [1.8] &   2.4 &   0.33 \\
 22 &    39.1 & (0.9) &$<$1.0 &$<$1.00 \\
 & \\
HD 62542 & V03 &      &  [6.4]\\ 
  1 &     7.1 & [2.5] &$<$5.7 &$<$0.85 \\
  2 &    11.2 & [1.5] &   5.5 &   0.30 \\
  3 &    13.9 & [1.5] &   5.8 &   0.21 \\
  4 &    17.7 & (2.5) &   4.4 &   0.30 \\
  5 &    22.7 & (2.0) &   6.8 &   0.69 \\
  6 &    27.0 & [2.0] &   7.7 &   0.57 \\
  7 &    32.0 & (2.0) &$<$6.4 &$<$4.92 \\
 & \\
HD 73882 & V03 &      &  [6.0]\\ 
  1 &  $-$3.8 & (3.0) &$<$6.0 &$<$2.00 \\
  2 &     2.0 & (3.0) &$<$6.0 &$<$1.62 \\
  3 &     7.0 & (3.0) &   5.8 &   0.62 \\
  4 &    12.7 & (3.0) &  14.8 &   0.46 \\
  5 &    18.2 & (3.0) &  20.8 &   0.24 \\
  6 &    23.2 & (2.5) &  30.8 &   0.33 \\
  7 &    27.1 & (2.5) &  38.8 &   0.47 \\
  8 &    32.1 & (2.5) &  12.4 &   0.61 \\
  9 &    35.9 & (2.5) &$<$8.1 &$<$0.27 \\
 10 &    41.2 & (2.5) &  13.4 &   0.35 \\
 11 &    45.7 & (2.5) &   9.6 &   0.20 \\
 12 &    51.6 & [2.2] &$<$5.7 &$<$0.52 \\
 & \\ 
HD 91316 & K95 &      &  [0.8]\\ 
  1 & $-$14.2 & [2.2] &   9.1 &   0.73 \\
  2 & $-$12.7 & (1.5) &   4.5 &   0.56 \\
  3 & $-$10.9 & [1.25]&   5.0 &   0.36 \\
  4 &  $-$8.9 & (1.5) &   8.4 &   0.71 \\
  5 &  $-$7.2 & [0.6] &   2.7 &   1.04 \\
  6 &  $-$4.7 & [2.45]&  11.5 &   0.75 \\
  7 &  $-$1.2 & [1.15]&   2.3 &   0.33 \\
  8 &     1.0 & [1.0] &   2.0 &   0.71 \\
  9 &     4.2 & (1.6) &   1.6 &   1.33 \\
 10 &     7.6 & [1.2] &   1.4 &   1.40 \\
 11 &    11.0 & [2.6] &   7.0 &   1.71 \\
 12 &    15.3 & [1.3] &   9.9 &   2.15 \\
 13 &    17.6 & (1.4) &   6.3 &   1.17 \\
 14 &    18.3 & [0.55]&$<$1.4 &$<$0.56 \\
 15 &    19.4 & [1.0] &   2.4 &   0.96 \\
 16 &    21.0 & [0.7] &   1.8 &   1.13 \\
 17 &    22.8 & (1.5) &   3.4 &   1.48 \\
 18 &    24.1 & [1.0] &   2.2 &   1.29 \\
 19 &    25.8 & [1.0] &   3.1 &   1.29 \\
 & \\
HD116658 & K95 &      &  [0.6]\\ 
  1 & $-$11.8 & [3.3] &   1.2 &   0.41 \\
  2 &  $-$3.4 & [1.3] &   1.1 &   0.38 \\
 & \\
HD141637 & K95 &      &  [8.0]\\ 
  1 & $-$33.5 & [1.5] &   9.8 &   3.63 \\
  2 & $-$28.9 & [2.3] &   6.2 &   2.70 \\
  3 & $-$25.8 & [1.9] &   7.4 &   3.52 \\
  4 & $-$17.3 & (1.0) &$<$5.1 &$<$3.40 \\
  5 & $-$16.3 & [0.5] &$<$4.5 &$<$1.80 \\
  6 & $-$14.9 & [0.6] &$<$3.3 &$<$3.30 \\
  7 & $-$12.5 & [1.1] &   2.8 &   2.00 \\ 
  8 &  $-$9.2 & [1.0] &   6.4 &   3.37 \\
  9 &  $-$8.4 & [0.45]&$<$9.0 &$<$2.31 \\
 10 &  $-$7.7 & [0.8] &   8.6 &   1.18 \\
\hline
\end{tabular}
\end{table}

\begin{table}
\contcaption{}
\begin{tabular}{@{}rrcrr}
\hline
\multicolumn{1}{l}{Star}&
\multicolumn{1}{c}{$v$}&
\multicolumn{1}{c}{$b$}&
\multicolumn{1}{c}{$N_{10}$}&
\multicolumn{1}{c}{\mbox{Ti\,{\sc ii}}/\mbox{Ca\,{\sc ii}}}\\
\multicolumn{1}{c}{Comp}&
\multicolumn{1}{c}{(km s$^{-1}$)}&
\multicolumn{1}{c}{(km s$^{-1}$)}&
\multicolumn{1}{c}{(cm$^{-2}$)}&
\multicolumn{1}{c}{ }\\
\hline
 11 &  $-$6.5 & [0.45]&$<$6.6 &$<$1.50 \\
 12 &  $-$5.8 & [1.3] &   8.4 &   5.60 \\
 & \\
HD143018 & K95 &      &  [1.1]\\ 
  1 & $-$34.5 & [1.5] &   1.3 &   2.17 \\
  2 & $-$29.7 & [2.5] &   4.9 &   1.88 \\
  3 & $-$23.6 & [2.3] &   3.2 &   1.28 \\
  4 & $-$16.3 & [1.6] &$<$0.7 &$<$0.70 \\
  5 & $-$12.7 & [1.9] &$<$0.8 &$<$0.80 \\
  6 &  $-$7.4 & [1.0] &   1.2 &   1.50 \\
 & \\
HD144217 & K95 &      &  [1.2]\\ 
  1 & $-$31.9 & [2.2] &   2.8 &   3.50 \\
  2 & $-$28.5 & [1.9] &   4.9 &   1.88 \\
  3 & $-$24.1 & [2.2] &   3.3 &   2.54 \\
  4 & $-$20.5 & [2.1] &$<$0.9 &$<$0.41 \\
  5 & $-$14.9 & (1.8) &   0.9 &   1.00 \\
  6 & $-$10.5 & (1.8) &$<$1.8 &$<$0.25 \\
  7 & $-$10.0 & [0.8] &   1.7 &   0.18 \\
  8 &  $-$8.7 & [0.7] &$<$1.4 &$<$0.07 \\
  9 &  $-$8.0 & (1.6) &   4.4 &   0.44 \\
 & \\
HD147165 & K95 &      &  [1.5]\\ 
  1 & $-$30.5 & [1.4] &   2.3 &   1.53 \\
  2 & $-$27.5 & (2.0) &   2.4 &   1.41 \\
  3 & $-$21.1 & (2.5) &   1.5 &   1.00 \\
  4 & $-$16.5 & (2.5) &   3.7 &   2.31 \\
  5 & $-$11.9 & (2.5) &   1.6 &   0.44 \\
  6 &  $-$8.0 & [2.0] &   5.5 &   0.51 \\
  7 &  $-$6.1 & [0.9] &   4.6 &   0.21 \\
  8 &  $-$4.7 & [1.1] &  13.4 &   0.56 \\
  9 &  $-$3.5 & (1.5) &   2.9 &   0.51 \\
 & \\
HD147888 & V00     &  &  [2.1]\\ 
  1 & $-$30.6 & [1.2] &   4.9 &$>$1.63 \\
  2 & $-$25.2 & [1.2] &   3.3 &$>$1.10 \\
  3 & $-$19.9 & [1.2] &   4.1 &$>$1.37 \\
  4 & $-$13.3 & (1.5) &   2.8 &$>$0.93 \\
  5 & $-$11.7 & (1.5) &$<$4.8 &$<$1.55 \\
  6 &  $-$8.8 & [1.0] & (20.6)&   0.22 \\
  7 &  $-$7.1 & [1.0] &  (5.4)&   0.08 \\
 & \\
HD147933 & V00     &  &  [1.6]\\ 
  1 & $-$29.7 & (1.5) &   4.1 &   2.41 \\
  2 & $-$25.6 & (1.5) &   2.7 &   1.97 \\
  3 & $-$20.3 & [1.2] &   3.8 &$>$2.24 \\
  4 & $-$15.7 & (1.5) &   1.3 &$>$0.62 \\
  5 & $-$13.5 & (1.2) &   2.0 &   0.53 \\
  6 &  $-$9.6 & [1.2] &   7.1 &   0.14 \\
  7 &  $-$7.9 & [1.0] &  20.1 &   0.24 \\
 & \\
HD148184 & V00     &  &  [1.8]\\ 
  1 & $-$31.4 & (1.5) &   4.8 &   2.29 \\
  2 & $-$28.0 & (1.5) &   6.8 &   0.64 \\
  3 & $-$25.3 & (1.5) &   9.3 &   0.90 \\
  4 & $-$22.3 & [1.4] &   6.3 &   0.85 \\
  5 & $-$18.7 & (3.0) &   6.4 &   0.89 \\
  6 & $-$12.7 & [1.6] &$<$1.8 &$<$0.21 \\
  7 & $-$10.7 & [1.2] &   2.8 &   0.08 \\
  8 &  $-$8.5 & [0.9] &$<$1.8 &$<$0.14 \\
  9 &  $-$5.7 & (1.5) &$<$1.2 &$<$0.38 \\
 & \\
\hline
\end{tabular}
\end{table}

\begin{table}
\contcaption{}
\begin{tabular}{@{}rrcrr}
\hline
\multicolumn{1}{l}{Star}&
\multicolumn{1}{c}{$v$}&
\multicolumn{1}{c}{$b$}&
\multicolumn{1}{c}{$N_{10}$}&
\multicolumn{1}{c}{\mbox{Ti\,{\sc ii}}/\mbox{Ca\,{\sc ii}}}\\
\multicolumn{1}{c}{Comp}&
\multicolumn{1}{c}{(km s$^{-1}$)}&
\multicolumn{1}{c}{(km s$^{-1}$)}&
\multicolumn{1}{c}{(cm$^{-2}$)}&
\multicolumn{1}{c}{ }\\
\hline
HD149757 & K94,K95 &  &  [1.3]\\ 
  1 & $-$35.4 & (2.0) &   0.9 &   --   \\
  2 & $-$31.6 & (1.0) &   1.1 &   1.22 \\
  3 & $-$28.9 & (1.6) &  11.6 &   1.66 \\
  4 & $-$26.3 & [0.9] &   2.5 &   1.25 \\
  5 & $-$22.7 & [1.5] &   3.1 &   1.19 \\ 
  6 & $-$18.8 & [1.5] &   0.8 &   0.15 \\
  7 & $-$16.5 & [0.8] &   0.4 &   0.06 \\ 
  8 & $-$14.6 & [0.95]&   2.1 &   0.08 \\
  9 & $-$12.6 & [0.7] &   2.0 &   0.30 \\
 10 & $-$10.0 & (1.0) &   0.4 &   0.25 \\ 
 11 &  $-$5.6 & (2.0) &   0.8 &   --   \\ 
 & \\
HD152236 & V00     &  &  [1.7]\\ 
  1 & $-$54.4 & (3.0) &$<$1.2 &$<$0.20 \\
  2 & $-$47.9 & (2.5) &$<$1.2 &$<$0.15 \\
  3 & $-$42.9 & (2.5) &   2.1 &   0.10 \\
  4 & $-$38.4 & (2.5) &   2.1 &   0.11 \\
  5 & $-$33.3 & (2.5) &$<$1.4 &$<$0.13 \\
  6 & $-$28.9 & (2.5) &   1.9 &   0.19 \\
  7 & $-$24.5 & (2.5) &   3.1 &   0.15 \\
  8 & $-$21.3 & (2.0) &  16.6 &   0.17 \\
  9 & $-$17.0 & (2.0) &  21.2 &   0.27 \\
 10 & $-$13.8 & (2.0) &  18.3 &   0.30 \\
 11 & $-$10.2 & (2.0) &  20.3 &   0.23 \\
 12 &  $-$6.9 & (2.0) &  36.5 &   0.25 \\
 13 &  $-$2.5 & (2.0) &  26.8 &   0.27 \\
 14 &     1.8 & (2.0) &  10.1 &   0.18 \\
 15 &     6.2 & (2.5) &   3.8 &   0.20 \\
 16 &    10.6 & (2.5) &   4.8 &   0.32 \\
 17 &    15.9 & (3.0) &   3.2 &   0.30 \\
 & \\
HD154368 & V00     &  &  [1.7]\\ 
  1 & $-$26.4 & (2.8) &   5.0 &   0.47 \\
  2 & $-$20.7 & (1.8) &   6.9 &   0.18 \\
  3 & $-$18.0 & (1.8) &   7.7 &   0.40 \\
  4 & $-$14.5 & [1.2] &   8.9 &   0.22 \\
  5 & $-$10.4 & (2.5) &   9.5 &   0.26 \\
  6 &  $-$5.4 & (2.0) &  26.1 &   0.26 \\
  7 &  $-$3.0 & (2.0) &  10.8 &   0.10 \\
  8 &     1.0 & (2.0) &   9.9 &   0.27 \\
  9 &     7.9 & [3.3] &  18.6 &   0.27 \\
 & \\
HD159561 & K94,K95 &  &  [1.5]\\ 
  1 & $-$32.3 & [2.4] &   1.1 &   0.41 \\
  2 & $-$25.5 & [2.7] &  11.3 &   0.39 \\
  3 & $-$21.4 & [1.2] &   1.0 &   0.42 \\
 & \\
HD197345 & K94 &      &  [1.8]\\ 
  1 & $-$23.1 & (1.2) &   2.0 &   --   \\
  2 & $-$21.7 & (1.2) &   0.9 &   0.13 \\
  3 & $-$20.0 & (1.2) &   2.9 &   2.90 \\
  4 & $-$15.4 & (0.7) &   0.6 &   0.50 \\ 
  5 & $-$13.4 &  0.6  &  14.9 &   0.22 \\
  6 &  $-$9.8 &  0.8  &  21.4 &   0.32 \\
  7 &  $-$7.7 & [1.2] &   3.2 &   0.18 \\
  8 &  $-$5.7 & [0.65]&   1.4 &   0.11 \\
  9 &  $-$4.0 & [0.5] &   0.6 &   0.06 \\ 
 10 &  $-$2.8 & [0.5] &$<$0.9 &$<$0.15 \\
 11 &  $-$1.2 & (0.8) &$<$0.9 &$<$1.50 \\
\hline
\end{tabular}
\end{table}

\begin{table}
\contcaption{}
\begin{tabular}{@{}rrcrr}
\hline
\multicolumn{1}{l}{Star}&
\multicolumn{1}{c}{$v$}&
\multicolumn{1}{c}{$b$}&
\multicolumn{1}{c}{$N_{10}$}&
\multicolumn{1}{c}{\mbox{Ti\,{\sc ii}}/\mbox{Ca\,{\sc ii}}}\\
\multicolumn{1}{c}{Comp}&
\multicolumn{1}{c}{(km s$^{-1}$)}&
\multicolumn{1}{c}{(km s$^{-1}$)}&
\multicolumn{1}{c}{(cm$^{-2}$)}&
\multicolumn{1}{c}{ }\\
\hline
 12 &     1.5 & (0.5) &$<$0.6 &$<$0.60 \\
 13 &     5.4 & (0.5) &$<$0.9 &$<$0.50 \\
 14 &     6.5 & [0.7] &$<$0.9 &$<$0.90 \\
 & \\
HD212571 & V00 &      &  [1.2]\\ 
  1 & $-$21.1 & [1.45]&   0.7 &   0.20 \\
  2 & $-$18.0 & [1.0] &   1.6 &   0.59 \\
  3 & $-$16.7 & [0.55]&   2.4 &   0.37 \\
  4 & $-$13.9 & [2.0] &   9.4 &   0.36 \\
  5 & $-$11.4 & [0.75]&   4.2 &   0.30 \\
  6 & $-$10.1 & [0.65]&   3.1 &   0.12 \\
  7 &  $-$8.4 & (1.0) &   6.6 &   1.03 \\
  8 &  $-$6.9 & [0.5] &$<$6.9 &$<$1.60 \\ 
  9 &  $-$5.0 & [1.2] &   5.8 &   0.55 \\
 10 &  $-$3.0 & (1.0) &   4.3 &   0.98 \\
 11 &  $-$1.2 & [0.85]&   6.9 &   0.47 \\
 12 &     0.5 & [0.6] &   1.0 &   0.29 \\
 13 &     2.5 & [1.95]&   2.3 &   0.55 \\
 & \\
HD217675 & K94 &      &  [1.9]\\ 
  1 & $-$20.2 & [1.25]&$<$0.9 &$<$0.26 \\
  2 & $-$17.0 & (1.5) &$<$0.9 &$<$0.60 \\
  3 & $-$13.9 & [1.6] &   2.1 &   0.46 \\
  4 & $-$10.9 & [1.3] &   4.3 &   0.30 \\
  5 &  $-$8.4 & [0.75]&   3.7 &   0.23 \\
  6 &  $-$6.6 & [1.35]&   5.3 &   0.58 \\
  7 &  $-$3.9 & (1.0) &$<$0.9 &$<$0.90 \\
  8 &     4.1 & [0.9] &$<$0.7 &$<$0.25 \\
  9 &     6.9 & [1.0] &   0.9 &   1.00 \\
\hline
\end{tabular}
\medskip
~ ~ \\
Column densities are in units of 10$^{10}$ cm$^{-2}$; limits are 3-$\sigma$.  
Typical 3-$\sigma$ uncertainties (given in square braces on first line for each sight line) are for a single component with $b$ = 2.0 km~s$^{-1}$ and include both photon noise and continuum placement uncertainties.
Values in parentheses were fixed (somewhat arbitrarily) in the fits; values in square braces were also fixed in the fits, but are better determined.
\end{table}

\clearpage

\begin{table}
\caption{\mbox{Ti\,{\sc ii}} equivalent widths and column densities.}
\label{tab:allwids} 
\begin{tabular}{@{}rlrrcc}  
\hline
\multicolumn{1}{l}{HD}&
\multicolumn{1}{l}{Name}& 
\multicolumn{1}{l}{Ref}& 
\multicolumn{1}{c}{W$_{\lambda}$}&
\multicolumn{1}{c}{$N$(AOD)}&
\multicolumn{1}{c}{$N$(fit)}\\ 
\hline
   886 & $\gamma$ Peg    & 1 &   2.2$\pm$0.4 &   10.78$\pm$0.07 & \\
  2905 & $\kappa$ Cas    & 1 &  42.0$\pm$3.0 &   12.09$\pm$0.06 & \\
       &                 & 2 &  44.0$\pm$0.5 &  (12.08$\pm$0.01)& \\
  3379 &                 & 3 &$<$4.3         &$<$11.07          & \\
  4180 & o Cas           & 1 &   7.0$\pm$2.0 &   11.30$\pm$0.13 & \\
  5394 & $\gamma$ Cas    & 1 &   4.4$\pm$0.7 &   11.09$\pm$0.08 & \\
  6882 & $\zeta$ Phe     & 3 &$<$1.3         &$<$10.55          & \\
 10144 & $\alpha$ Eri    & 3 &$<$1.0         &$<$10.44          & \\
 10516 & $\phi$ Per      & 1 &   4.3$\pm$1.1 &   11.07$\pm$0.10 & \\
 14633 &                 & 4 &  68.0$\pm$x.x &   12.28$\pm$0.xx & \\
 15371 & $\kappa$ Eri    & 3 &  13.9$\pm$0.8 &  (11.58$\pm$0.03)& 11.65$\pm$0.06 \\
 16581 &                 & 5 &$<$9.0         &$<$11.39          & \\
 16582 &                 & 5 &   7.0$\pm$0.7 &  (11.29$\pm$0.04)& 11.30$\pm$0.04 \\
 18100 &                 & 5 &$<$15.0        &$<$11.63          & \\
 18216 &                 & 5 &$<$6.0         &$<$11.24          & \\
 18484 &                 & 2 &   5.2$\pm$x.x &  (11.18$\pm$0.xx)& \\
 18654 &                 & 2 &   3.0$\pm$x.x &  (10.93$\pm$0.xx)& \\
 21278 &                 & 6 &  14.0$\pm$0.5 &   11.59$\pm$0.02 & \\
 22192 & $\psi$ Per      & 1 &  13.0$\pm$2.0 &   11.56$\pm$0.06 & \\
 22586 &                 & 7 &  13.0$\pm$3.0 &  (11.55$\pm$0.11)& 11.65$\pm$0.11 \\
 22928 & $\delta$ Per    & 1 &   1.5$\pm$0.6 &   10.62$\pm$0.17 & \\
 22951 & 40 Per          & 1 &  20.0$\pm$2.0 &   11.77$\pm$0.07 & \\
 23180 & o Per           & 1 &   9.8$\pm$3.4 &   11.43$\pm$0.13 & \\ 
       &                 & 2 &   6.5$\pm$0.4 &  (11.25$\pm$0.03)& \\
       &                 & 8 &   5.0$\pm$x.x &   11.14$\pm$0.xx & \\
       &                 & 9 &   5.9$\pm$1.1 &   11.22$\pm$0.07 & \\
 23408 & 20 Tau          & 6 &$<$5.0         &$<$11.14          & \\
 23480 & 23 Tau          & 2 &   2.2$\pm$0.4 &  (10.78$\pm$0.08)& \\
       &                 & 2 &   1.8$\pm$x.x &  (10.68$\pm$0.xx)& \\
       &                 & 6 &   3.3$\pm$0.6 &   10.96$\pm$0.08 & \\
 23630 & $\eta$ Tau      & 1 &   2.5$\pm$1.0 &   10.84$\pm$0.15 & \\
 23850 & 27 Tau          & 1 &   1.8$\pm$1.7 &   10.70$\pm$0.29 & \\
 24398 & $\zeta$ Per     & 1 &   6.2$\pm$1.3 &   11.24$\pm$0.09 & \\
       &                 & 10&   3.3$\pm$0.2 &  (10.96$\pm$0.03)& \\
       &                 & 2 &   6.2$\pm$0.4 &  (11.23$\pm$0.03)& \\
       &                 & 2 &   4.7$\pm$x.x &  (11.13$\pm$0.xx)& \\
       &                 & 11&   5.5$\pm$0.2 &  (11.18$\pm$0.02)& \\
       &                 & 12&   3.0$\pm$x.x &   10.94$\pm$0.xx & \\
       &                 & 9 &   2.8$\pm$0.8 &   10.89$\pm$0.10 & \\
 24534 & X Per           & 9 &   7.8$\pm$1.6 &   11.34$\pm$0.07 & \\
 24760 & $\epsilon$ Per  & 1 &   4.6$\pm$0.6 &   11.11$\pm$0.07 & \\
       &                 & 13&   5.7$\pm$0.5 &   11.20$\pm$0.04 & 11.20$\pm$0.02 \\
 24912 & $\xi$ Per       & 1 &  13.0$\pm$1.0 &   11.57$\pm$0.05 & \\ 
       &                 & 2 &  12.8$\pm$0.3 &  (11.55$\pm$0.01)& \\
       &                 & 8 &  11.0$\pm$2.0 &   11.48$\pm$0.08 & \\
 25204 & $\lambda$ Tau   & 1 &   2.3$\pm$0.9 &   10.80$\pm$0.15 & \\
 27657 & $\theta$ Ret    & 3 &$<$3.7         &$<$11.01          & \\
 27778 & 62 Tau          & 9 &   5.0$\pm$2.0 &   11.15$\pm$0.04 & \\ 
 28497 &                 & 6 &   8.3$\pm$0.5 &   11.36$\pm$0.03 & \\
 29138 &                 & 14&  65.8$\pm$4.0 &   12.29$\pm$0.02 & 12.28$\pm$0.06 \\
 29248 & $\nu$ Eri       & 6 &   4.4$\pm$0.7 &   11.11$\pm$0.08 & \\
 30614 & $\alpha$ Cam    & 1 &  51.0$\pm$3.0 &   12.19$\pm$0.06 & \\
       &                 & 2 &  47.1$\pm$0.4 &  (12.11$\pm$0.01)& \\
       &                 & 15&  38.0$\pm$x.x &   12.04$\pm$0.xx & \\
 30677 &                 & 14&  52.8$\pm$5.1 &   12.19$\pm$0.02 & 12.19$\pm$0.07 \\
 30836 & $\pi^4$ Ori     & 1 &   4.8$\pm$1.4 &   11.13$\pm$0.12 & \\
 31726 &                 & 5 &   6.0$\pm$1.0 &  (11.22$\pm$0.06)& 11.30$\pm$0.06 \\
 32612 &                 & 5 &   6.0$\pm$1.0 &  (11.22$\pm$0.06)& 11.26$\pm$0.09 \\
 32630 & $\eta$ Aur      & 13&   1.9$\pm$0.7 &   10.75$\pm$0.15 & 10.79$\pm$0.17 \\
 33328 &                 & 14&  12.5$\pm$1.3 &   11.54$\pm$0.03 & 11.55$\pm$0.08 \\
 34078 & AE Aur          & 15&  27.0$\pm$x.x &   11.89$\pm$0.xx & \\
 34085 & $\beta$ Ori     & 3 &  15.0$\pm$1.5 &  (11.62$\pm$0.04)& 11.65$\pm$0.07 \\
       &                 & 6 &   1.0$\pm$0.2 &   10.47$\pm$0.07 & \\
\hline
\end{tabular}
\end{table}

\begin{table}
\contcaption{}
\begin{tabular}{@{}rlrrcc}  
\hline
\multicolumn{1}{l}{HD}&
\multicolumn{1}{l}{Name}& 
\multicolumn{1}{l}{Ref}& 
\multicolumn{1}{c}{W$_{\lambda}$}&
\multicolumn{1}{c}{$N$(AOD)}&
\multicolumn{1}{c}{$N$(fit)}\\ 
\hline
 34179 &                 & 5 &$<$13.5        &$<$11.65          & \\
 34748 &                 & 5 &   6.0$\pm$1.0 &  (11.22$\pm$0.06)& 11.16$\pm$0.09 \\
 34816 & $\lambda$ Lep   & 3 &   5.4$\pm$0.6 &  (11.17$\pm$0.05)& 11.18$\pm$0.05 \\
 35149 & 23 Ori          & 2 &   5.5$\pm$0.5 &  (11.18$\pm$0.03)& 11.23$\pm$0.03 \\
       &                 & 2 &   4.0$\pm$x.x &  (11.06$\pm$0.xx)& \\
       &                 & 9 &   6.6$\pm$1.3 &   11.27$\pm$0.07 & \\
       &                 & 16&   5.8$\pm$1.0 &   11.22$\pm$0.06 & \\
 35411 & $\eta$ Ori      & 1 &   3.2$\pm$1.1 &   10.94$\pm$0.14 & \\
 35468 & $\gamma$ Ori    & 1 &   2.3$\pm$0.3 &   10.81$\pm$0.06 & \\
 35575 &                 & 5 &$<$9.0         &$<$11.39          & \\  
 36486 & $\delta$ Ori    & 1 &   5.6$\pm$1.2 &   11.19$\pm$0.09 & \\
       &                 & 13&   3.9$\pm$0.8 &   11.03$\pm$0.09 & 11.04$\pm$0.04 \\
       &                 & 17&    --         &   11.15$\pm$0.04 & \\
 36646 &                 & 14&  13.3$\pm$1.7 &   11.56$\pm$0.04 & 11.58$\pm$0.09 \\
 36861 & $\lambda$ Ori   & 1 &  17.0$\pm$2.0 &   11.68$\pm$0.05 & \\
       &                 & 15&  21.0$\pm$x.x &   11.77$\pm$0.xx & \\
       &                 & 14&  19.0$\pm$1.5 &   11.74$\pm$0.03 & 11.74$\pm$0.05 \\ 
 37022 & $\theta^1$ Ori C& 9 &  10.2$\pm$2.4 &   11.46$\pm$0.09 & \\
 37043 & $\iota$ Ori     & 1 &  11.0$\pm$2.0 &   11.50$\pm$0.07 & \\
       &                 & 13&   8.0$\pm$0.6 &   11.35$\pm$0.03 & 11.39$\pm$0.03 \\
       &                 & 17&    --         &   11.31$\pm$0.03 & \\
 37055 &                 & 14&   6.6$\pm$1.0 &   11.34$\pm$0.07 & 11.28$\pm$0.09 \\
 37128 & $\epsilon$ Ori  & 1 &  11.0$\pm$2.0 &   11.47$\pm$0.06 & \\
       &                 & 13&   8.6$\pm$0.8 &   11.38$\pm$0.04 & 11.38$\pm$0.02 \\
       &                 & 17&    --         &   11.40$\pm$0.03 & \\
 37202 & $\zeta$ Tau     & 6 &  11.0$\pm$0.8 &   11.50$\pm$0.03 & \\
 37468 & $\sigma$ Ori    & 1 &  10.0$\pm$2.0 &   11.45$\pm$0.09 & \\ 
 37490 &                 & 14&  15.7$\pm$1.5 &   11.67$\pm$0.04 & 11.66$\pm$0.06 \\
 37742 & $\zeta$ Ori     & 1 &   5.5$\pm$0.8 &   11.18$\pm$0.06 & \\
       &                 & 13&   5.2$\pm$0.9 &   11.17$\pm$0.08 & 11.18$\pm$0.06 \\
 38666 & $\mu$ Col       & 3 &  22.0$\pm$3.5 &  (11.78$\pm$0.07)& 11.83$\pm$0.xx \\
       &                 & 6 &  14.0$\pm$0.5 &   11.57$\pm$0.02 & \\
       &                 & 18&    --         &$<$11.40          & \\
       &                 & 19&    --         &   --             & 11.50$\pm$0.02 \\
 38771 & $\kappa$ Ori    & 1 &   4.2$\pm$1.1 &   11.06$\pm$0.10 & \\
       &                 & 6 &   3.7$\pm$0.4 &   11.00$\pm$0.05 & \\
 40111 & 139 Tau         & 1 &  36.0$\pm$4.0 &   12.02$\pm$0.07 & \\
 40494 & $\gamma$ Col    & 3 &  11.1$\pm$1.0 &  (11.49$\pm$0.04)& 11.55$\pm$0.05 \\
 41117 & $\chi^2$ Ori    & 1 &  61.0$\pm$4.0 &   12.28$\pm$0.09 & \\
       &                 & 15&  71.0$\pm$x.x &   12.41$\pm$0.xx & \\
 41161 &                 & 18&   --          &   --             & 12.28$\pm$0.04 \\
 42933 & $\delta$ Pic    & 3 &   6.2$\pm$1.0 &  (11.23$\pm$0.07)& 11.49$\pm$0.06 \\
 43285 &                 & 14&   9.3$\pm$1.8 &   11.40$\pm$0.06 & 11.41$\pm$0.12 \\
 44743 & $\beta$ CMa     & 1 &   1.2$\pm$0.5 &   10.50$\pm$0.14 & \\
       &                 & 3 &$<$1.1         &$<$10.48          & \\
 45725 & $\beta$ Mon     & 14&   4.8$\pm$1.5 &   11.11$\pm$0.09 & 11.12$\pm$0.18 \\
 46185 &                 & 14&  13.9$\pm$1.9 &   11.62$\pm$0.03 & 11.61$\pm$0.09 \\
 47240 &                 & 15&  42.0$\pm$x.x &   12.08$\pm$0.xx & \\
 47670 & $\nu$ Pup       & 3 &$<$1.0         &$<$10.44          & \\
 47839 & 15 Mon A        & 1 &  23.0$\pm$4.0 &   11.82$\pm$0.09 & \\
       &                 & 15&  27.0$\pm$x.x &   11.87$\pm$0.xx & \\
       &                 & 13&  22.1$\pm$2.1 &   11.79$\pm$0.04 & 11.79$\pm$0.02 \\
 48099 &                 & 14&  43.3$\pm$2.0 &   12.12$\pm$0.01 & 12.11$\pm$0.03 \\
 49131 &                 & 14&   7.6$\pm$0.9 &   11.31$\pm$0.05 & 11.31$\pm$0.05 \\
 50013 & $\kappa$ CMa    & 1 &   6.1$\pm$2.5 &   11.22$\pm$0.15 & \\
 50896 &                 & 6 &  46.0$\pm$1.3 &   12.14$\pm$0.01 & \\
       &                 & 14&  37.5$\pm$3.2 &   12.07$\pm$0.05 & 12.05$\pm$0.07 \\
 52089 & $\epsilon$ CMa  & 1 &   3.0$\pm$1.2 &   10.92$\pm$0.16 & \\
       &                 & 3 &$<$0.8         &$<$10.34          & \\
 52918 & 19 Mon          & 14&   5.2$\pm$1.7 &   11.16$\pm$0.10 & 11.15$\pm$0.16 \\
 53138 & o$^2$ CMa       & 1 &  11.0$\pm$1.0 &   11.50$\pm$0.05 & \\
       &                 & 3 &  10.0$\pm$1.5 &  (11.44$\pm$0.07)& 11.49$\pm$0.06 \\
 53975 &                 & 18&   --          &   --             & 12.13$\pm$0.04 \\
\hline
\end{tabular}
\end{table}

\begin{table}
\contcaption{}
\begin{tabular}{@{}rlrrcc}  
\hline
\multicolumn{1}{l}{HD}&
\multicolumn{1}{l}{Name}& 
\multicolumn{1}{l}{Ref}& 
\multicolumn{1}{c}{W$_{\lambda}$}&
\multicolumn{1}{c}{$N$(AOD)}&
\multicolumn{1}{c}{$N$(fit)}\\ 
\hline
 54662 &                 & 15&  33.0$\pm$x.x &   11.97$\pm$0.xx & \\
 57060 & 29 CMa          & 1 &  10.0$\pm$3.0 &   11.45$\pm$0.11 & \\
 57061 & $\tau$ CMa      & 6 &  20.0$\pm$1.6 &   11.76$\pm$0.03 & \\
\multicolumn{2}{l}{Walker 67}& 16&  10.9$\pm$2.5 &   11.49$\pm$0.08 & \\
 58350 & $\eta$ CMa      & 1 &   9.2$\pm$1.3 &   11.41$\pm$0.07 & \\
 58343 &                 & 14&   8.4$\pm$2.0 &   11.40$\pm$0.04 & 11.38$\pm$0.12 \\
 58377 &                 & 14&   8.6$\pm$1.6 &   11.35$\pm$0.05 & 11.39$\pm$0.12 \\
 58978 &                 & 14&   7.9$\pm$0.8 &   11.28$\pm$0.04 & 11.35$\pm$0.05 \\
 60498 &                 & 14&  11.4$\pm$4.9 &   11.49$\pm$0.09 & 11.50$\pm$0.19 \\
 60848 &                 & 5 &  28.0$\pm$3.0 &  (11.89$\pm$0.04)& 11.91$\pm$0.04 \\
 61038 &                 & 5 &$<$6.0         &$<$11.24          & \\
 61429 &                 & 14&   4.5$\pm$0.9 &   11.17$\pm$0.05 & 11.11$\pm$0.08 \\
 62542 &                 & 16&  12.6$\pm$2.4 &   11.55$\pm$0.07 & 11.53$\pm$0.13 \\
 64972 &                 & 14&   8.7$\pm$1.3 &   11.34$\pm$0.04 & 11.41$\pm$0.10 \\
 65575 & $\chi$ Car      & 3 &$<$2.0         &$<$10.74          & \\
 66811 & $\zeta$ Pup     & 3 &  10.1$\pm$1.0 &  (11.44$\pm$0.04)& 11.50$\pm$0.06 \\
       &                 & 6 &   3.3$\pm$0.8 &   10.96$\pm$0.11 & \\
       &                 & 18&    --         &   --             & 11.07$\pm$0.04 \\
       &                 & 19&    --         &   --             & 11.16$\pm$0.06 \\
 67536 &                 & 14&  16.0$\pm$1.8 &   11.67$\pm$0.02 & 11.68$\pm$0.07 \\
 68273 & $\gamma^2$ Vel  & 3 &$<$2.0         &$<$10.74          & \\
       &                 & 18&    --         &   --             & 11.10$\pm$0.02 \\
       &                 & 19&    --         &   --             & 11.11$\pm$0.04 \\
 68761 &                 & 14&  28.6$\pm$3.0 &   11.92$\pm$0.06 & 11.91$\pm$0.08 \\
 72067 &                 & 14&   7.6$\pm$1.4 &   11.38$\pm$0.16 & 11.34$\pm$0.13 \\
 72089 &                 & 20&   9.1$\pm$x.x &   11.43$\pm$0.xx & \\
 72127A&                 & 16&  17.0$\pm$2.6 &   11.68$\pm$0.06 & 11.68$\pm$0.06 \\
       &                 & 20&  11.9$\pm$x.x &   11.54$\pm$0.xx & \\
 72127B&                 & 16&  17.2$\pm$1.5 &   11.69$\pm$0.03 & 11.78$\pm$0.08 \\
       &                 & 20&   9.1$\pm$x.x &   11.43$\pm$0.xx & \\
 72350 &                 & 20&  11.7$\pm$x.x &   11.52$\pm$0.xx & \\
 72648 &                 & 20&  33.2$\pm$x.x &   11.93$\pm$0.xx & \\
 73882 &                 & 16&  51.6$\pm$3.2 &   12.18$\pm$0.02 & 12.19$\pm$0.03 \\
 74195 & o Vel           & 3 &$<$2.3         &$<$10.80          & \\
 74280 & $\eta$ Hya      & 1 &   0.6$\pm$0.3 &   10.21$\pm$0.19 & \\
 74455 &                 & 20&  12.0$\pm$x.x &   11.47$\pm$0.xx & \\
 74575 & $\alpha$ Pyx    & 3 &   7.1$\pm$1.5 &  (11.29$\pm$0.09)& 11.72$\pm$0.07 \\
 74966 &                 & 14&   4.2$\pm$1.2 &   11.10$\pm$0.21 & 11.07$\pm$0.16 \\
 75821 &                 & 20&$<$5.4         &$<$11.30          & \\
 76131 &                 & 14&  19.1$\pm$1.5 &   11.75$\pm$0.02 & 11.75$\pm$0.06 \\
 76341 &                 & 14&  19.7$\pm$2.9 &   11.76$\pm$0.07 & 11.76$\pm$0.09 \\
 81188 & $\kappa$ Vel    & 21&$<$1.5         &$<$10.62          & \\
 85504 &                 & 5 &$<$3.0         &$<$10.94          & \\
 86440 & $\phi$ Vel      & 3 &  25.3$\pm$1.8 &  (11.84$\pm$0.03)& 11.95$\pm$0.06 \\
 88661 &                 & 14&  13.2$\pm$2.7 &   11.58$\pm$0.07 & 11.58$\pm$0.10 \\
 89688 & 23 Sex          & 5 &  14.0$\pm$2.0 &  (11.59$\pm$0.06)& 11.60$\pm$0.06 \\
 90882 &                 & 14&$<$4.4         &$<$11.08          &$<$11.08 \\
 90994 &                 & 5 &$<$4.5         &$<$11.11          & \\
 91316 & $\rho$ Leo      & 1 &  27.0$\pm$2.0 &   11.88$\pm$0.04 & \\
       &                 & 13&  29.7$\pm$0.7 &   11.93$\pm$0.01 & 11.92$\pm$0.01 \\
 92740 &                 & 14&  75.0$\pm$3.5 &   12.38$\pm$0.01 & 12.37$\pm$0.03 \\
 93030 & $\theta$ Car    & 3 &$<$3.0         &$<$10.92          & \\
       &                 & 18&    --         &   --             & 11.31$\pm$0.01 \\
       &                 & 19&    --         &   --             & 11.31$\pm$0.02 \\
 93130 &                 & 20& 109.5$\pm$x.x &   12.63$\pm$0.xx & \\
 93131 &                 & 14&  84.5$\pm$2.7 &   12.43$\pm$0.01 & 12.42$\pm$0.03 \\
 93160 &                 & 20& 172.1$\pm$x.x &   12.80$\pm$0.xx & \\
 93161E&                 & 20& 182.2$\pm$x.x &   12.82$\pm$0.xx & \\
 93161W&                 & 20& 203.2$\pm$x.x &   12.86$\pm$0.xx & \\
 93204 &                 & 20&  70.1$\pm$x.x &   12.43$\pm$0.xx & \\
 93205 &                 & 20& 144.3$\pm$x.x &   12.62$\pm$0.xx & \\
303308 &                 & 20& 164.5$\pm$x.x &   12.74$\pm$0.xx & \\
 93521 &                 & 22&  37.0$\pm$4.0 &  (12.01$\pm$0.05)& 12.02$\pm$0.03 \\
\hline
\end{tabular}
\end{table}

\begin{table}
\contcaption{}
\begin{tabular}{@{}rlrrcc}  
\hline
\multicolumn{1}{l}{HD}&
\multicolumn{1}{l}{Name}& 
\multicolumn{1}{l}{Ref}& 
\multicolumn{1}{c}{W$_{\lambda}$}&
\multicolumn{1}{c}{$N$(AOD)}&
\multicolumn{1}{c}{$N$(fit)}\\ 
\hline
 94910 &                 & 14&  99.6$\pm$7.3 &   12.50$\pm$0.03 & 12.50$\pm$0.05 \\
 94963 &                 & 14&  69.3$\pm$3.4 &   12.33$\pm$0.03 & 12.32$\pm$0.04 \\
 96917 &                 & 14&  99.3$\pm$4.9 &   12.52$\pm$0.02 & 12.52$\pm$0.03 \\
 97253 &                 & 14&  88.3$\pm$8.3 &   12.51$\pm$0.02 & 12.45$\pm$0.06 \\
 97277 & $\beta$ Crt     & 3 &$<$1.6         &$<$10.64          & \\
 97585 & 69 Leo          & 22&   6.8$\pm$3.8 &  (11.27$\pm$0.27)& 11.27$\pm$0.28 \\
 97991 &                 & 22&  47.0$\pm$9.0 &  (12.11$\pm$0.08)& 12.11$\pm$0.09 \\
100841 &                 & 14&   5.6$\pm$2.1 &   11.19$\pm$0.11 & 11.20$\pm$0.19 \\
104337 &                 & 1 &  17.0$\pm$2.0 &   11.69$\pm$0.05 & \\
105056 &                 & 14&  63.9$\pm$4.5 &   12.28$\pm$0.03 & 12.28$\pm$0.06 \\
105435 & $\delta$ Cen    & 3 &$<$1.6         &$<$10.64          & \\
106625 & $\gamma$ Crv    & 16&   1.4$\pm$0.1 &   10.59$\pm$0.03 & \\
107070 &                 & 5 &$<$4.5         &$<$11.11          & \\
108248 & $\alpha$ Cru    & 18&   --          &   --             & 11.12$\pm$0.02 \\
       &                 & 19&   --          &   --             & 11.12$\pm$0.04 \\
109860 &                 & 5 &$<$6.0         &$<$11.21          & \\
109867 &                 & 14&  57.1$\pm$5.0 &   12.22$\pm$0.04 & 12.23$\pm$0.07 \\
110432 &                 & 16&  10.4$\pm$0.4 &   11.46$\pm$0.01 & \\
111973 & $\kappa$ Cru    & 3 &  46.0$\pm$3.0 &  (12.10$\pm$0.03)& 12.18$\pm$0.08 \\
112244 &                 & 16&  31.6$\pm$0.5 &   11.96$\pm$0.02 & \\
112272 &                 & 14&  69.8$\pm$8.3 &   12.32$\pm$0.03 & 12.32$\pm$0.08 \\
112842 &                 & 14&  81.7$\pm$4.4 &   12.39$\pm$0.02 & 12.39$\pm$0.05 \\
113904 & $\theta$ Mus    & 16&  66.7$\pm$1.4 &   12.29$\pm$0.01 & \\
       &                 & 14&  68.6$\pm$8.7 &   12.30$\pm$0.03 & 12.31$\pm$0.10 \\
114213 &                 & 16&  93.1$\pm$1.2 &   12.45$\pm$0.01 & \\
115363 &                 & 14& 115.6$\pm$8.6 &   12.57$\pm$0.01 & 12.57$\pm$0.05 \\
115842 &                 & 14&  54.1$\pm$3.9 &   12.22$\pm$0.03 & 12.21$\pm$0.07 \\
116658 & $\alpha$ Vir    & 1 &   1.6$\pm$0.3 &   10.64$\pm$0.07 & \\
       &                 & 3 &$<$2.8         &$<$10.89          & \\
       &                 & 13&   0.8$\pm$0.2 &   10.36$\pm$0.11 & 10.36$\pm$0.11 \\
       &                 & 19&    --         &   --             & 10.49$\pm$0.13 \\
118246 &                 & 5 &$<$9.0         &$<$11.39          & \\
118716 & $\epsilon$ Cen  & 3 &$<$1.8         &$<$10.70          & \\
119608 &                 & 22& 108.0$\pm$20. &  (12.47$\pm$0.08)& 12.55$\pm$0.08 \\
       &                 & 5 &  75.0$\pm$8.0 &  (12.32$\pm$0.04)& 12.34$\pm$0.04 \\
120086 &                 & 5 &  25.0$\pm$6.0 &  (11.84$\pm$0.11)& 11.84$\pm$0.11 \\
120315 & $\eta$ UMa      & 1 &   1.0$\pm$0.4 &   10.43$\pm$0.14 & \\
121968 &                 & 23&  38.0$\pm$2.0 &  (12.02$\pm$0.03)& 12.06$\pm$0.03 \\
122451 & $\beta$ Cen     & 18&    --         &   --             & 11.05$\pm$0.04 \\
       &                 & 19&    --         &   --             & 10.99$\pm$0.04 \\
123884 &                 & 7 & 172.0$\pm$26. &  (12.68$\pm$0.07)& 12.75$\pm$0.07 \\
125924 &                 & 23&  46.0$\pm$5.0 &  (12.10$\pm$0.05)& 12.14$\pm$0.05 \\
135485 &                 & 5 &  47.0$\pm$9.0 &  (12.11$\pm$0.09)& 12.19$\pm$0.09 \\
136239 &                 & 14& 155.7$\pm$13.5&   12.69$\pm$0.02 & 12.68$\pm$0.06 \\
137569 &                 & 5 &  28.0$\pm$4.0 &  (11.89$\pm$0.07)& 11.97$\pm$0.07 \\
138485 &                 & 5 &   7.0$\pm$1.0 &  (11.29$\pm$0.07)& 11.28$\pm$0.07 \\
138527 &                 & 5 &$<$3.0         &$<$10.94          & \\
141637 & 1 Sco           & 13&  19.7$\pm$4.2 &   11.72$\pm$0.09 & 11.72$\pm$0.06 \\
142758 &                 & 14& 162.1$\pm$11.2&   12.70$\pm$0.01 & 12.71$\pm$0.06 \\
143018 & $\pi$ Sco       & 1 &   2.5$\pm$0.6 &   10.85$\pm$0.10 & \\
       &                 & 13&   4.0$\pm$0.7 &   11.04$\pm$0.08 & 11.03$\pm$0.02 \\
143275 & $\delta$ Sco    & 1 &   3.4$\pm$0.6 &   10.97$\pm$0.07 & \\
       &                 & 14&   6.9$\pm$1.5 &   11.28$\pm$0.07 & 11.29$\pm$0.15 \\ 
143448 &                 & 14&  17.2$\pm$2.5 &   11.64$\pm$0.07 & 11.70$\pm$0.10 \\
144217 & $\beta^1$ Sco   & 1 &   5.4$\pm$0.8 &   11.18$\pm$0.07 & \\
       &                 & 13&   7.2$\pm$0.7 &   11.31$\pm$0.04 & 11.30$\pm$0.02 \\
144470 & $\omega^1$ Sco  & 1 &   8.9$\pm$2.2 &   11.39$\pm$0.10 & \\
145482 &                 & 14&   5.7$\pm$1.4 &   11.16$\pm$0.10 & 11.21$\pm$0.14 \\
145502 & $\nu$ Sco       & 1 &   9.2$\pm$2.7 &   11.41$\pm$0.12 & \\
147165 & $\sigma$ Sco    & 1 &  10.0$\pm$2.0 &   11.47$\pm$0.08 & \\
       &                 & 13&  13.0$\pm$0.8 &   11.58$\pm$0.03 & 11.58$\pm$0.02 \\
147701 &                 & 16&  19.2$\pm$1.4 &   11.73$\pm$0.03 & \\
147888 & $\rho$ Oph D    & 16&  13.0$\pm$0.6 &   11.57$\pm$0.02 & 11.61$\pm$0.02 \\
\hline
\end{tabular}
\end{table}

\begin{table}
\contcaption{}
\begin{tabular}{@{}rlrrcc}  
\hline
\multicolumn{1}{l}{HD}&
\multicolumn{1}{l}{Name}& 
\multicolumn{1}{l}{Ref}& 
\multicolumn{1}{c}{W$_{\lambda}$}&
\multicolumn{1}{c}{$N$(AOD)}&
\multicolumn{1}{c}{$N$(fit)}\\ 
\hline
147889 &                 & 16&   9.5$\pm$0.7 &   11.43$\pm$0.03 & \\
147933 & $\rho$ Oph A    & 16&  14.0$\pm$0.6 &   11.60$\pm$0.02 & 11.61$\pm$0.02 \\
147971 & $\epsilon$ Nor  & 3 &$<$2.5         &$<$10.84          & \\
148184 & $\chi$ Oph      & 16&  13.1$\pm$0.6 &   11.57$\pm$0.02 & 11.58$\pm$0.02 \\
       &                 & 14&  14.4$\pm$2.2 &   11.67$\pm$0.03 & 11.62$\pm$0.08 \\ 
148379 &                 & 16&  69.8$\pm$0.8 &   12.32$\pm$0.01 \\
       &                 & 14&  70.1$\pm$5.0 &   12.10$\pm$0.04 & 12.33$\pm$0.05 \\ 
148688 &                 & 14&  50.4$\pm$2.9 &   12.20$\pm$0.02 & 12.18$\pm$0.04 \\
148937 &                 & 14&  46.7$\pm$3.8 &   12.13$\pm$0.05 & 12.13$\pm$0.07 \\
149038 & $\mu$ Nor       & 3 &  31.0$\pm$2.0 &  (11.93$\pm$0.03)& 11.99$\pm$0.05 \\
149363 &                 & 7 &  52.0$\pm$9.0 &  (12.16$\pm$0.08)& 12.22$\pm$0.08 \\
149404 &                 & 16&  52.5$\pm$1.0 &   12.20$\pm$0.01 & \\
149438 & $\tau$ Sco      & 1 &   1.6$\pm$0.8 &   10.65$\pm$0.18 & \\
149757 & $\zeta$ Oph     & 1 &   9.6$\pm$0.6 &   11.43$\pm$0.04 & \\
       &                 & 2 &   9.8$\pm$0.2 &  (11.43$\pm$0.01)& \\
       &                 & 3 &   9.7$\pm$0.6 &  (11.43$\pm$0.03)& 11.48$\pm$0.05 \\
       &                 & 15&   7.0$\pm$x.x &   11.29$\pm$0.xx & \\
       &                 & 13&   9.1$\pm$0.8 &   11.41$\pm$0.04 & 11.41$\pm$0.01 \\
       &                 & 16&   9.5$\pm$0.3 &   11.43$\pm$0.02 & \\
149822 &                 & 5 &$<$4.5         &$<$11.11          & \\
149881 &                 & 22&  84.0$\pm$13. &  (12.36$\pm$0.07)& 12.47$\pm$0.07 \\
150136 &                 & 16&  38.1$\pm$0.7 &   12.05$\pm$0.02 & \\
150483 &                 & 22&   6.5$\pm$4.6 &  (11.25$\pm$0.39)& 11.25$\pm$0.38 \\
151890 & $\mu^1$ Sco     & 6 &$<$2.4         &$<$10.87          & \\
151932 &                 & 14&  65.7$\pm$5.4 &   12.29$\pm$0.03 & 12.29$\pm$0.06 \\
152003 &                 & 14&  64.2$\pm$5.0 &   12.27$\pm$0.03 & 12.30$\pm$0.06 \\
152235 &                 & 14&  61.4$\pm$3.5 &   12.27$\pm$0.03 & 12.26$\pm$0.05 \\
152236 & $\zeta^1$ Sco   & 3 &  51.4$\pm$3.0 &  (12.15$\pm$0.03)& 12.24$\pm$0.07 \\
       &                 & 16&  56.9$\pm$1.0 &   12.19$\pm$0.01 & 12.23$\pm$0.01 \\
152270 &                 & 14&  60.6$\pm$5.5 &   12.25$\pm$0.03 & 12.25$\pm$0.07 \\
154368 &                 & 16&  36.1$\pm$0.8 &   12.02$\pm$0.02 & 12.02$\pm$0.02 \\
154811 &                 & 14&  47.2$\pm$4.8 &   12.15$\pm$0.02 & 12.14$\pm$0.08 \\
154873 &                 & 14&  34.8$\pm$2.8 &   12.00$\pm$0.04 & 12.01$\pm$0.06 \\
155806 &                 & 14&  37.2$\pm$5.1 &   12.04$\pm$0.04 & 12.04$\pm$0.10 \\
156385 &                 & 14&  43.3$\pm$2.8 &   12.12$\pm$0.02 & 12.10$\pm$0.04 \\
156575 &                 & 14&  44.7$\pm$3.4 &   12.12$\pm$0.02 & 12.12$\pm$0.05 \\
157042 & $\iota$ Ara     & 3 &  12.6$\pm$1.3 &  (11.54$\pm$0.04)& 11.59$\pm$0.05 \\
157056 & $\theta$ Oph    & 1 &   8.6$\pm$1.5 &   11.38$\pm$0.08 & \\
157246 & $\gamma$ Ara    & 3 &  11.4$\pm$0.9 &  (11.50$\pm$0.04)& 11.52$\pm$0.05 \\
158408 & $\upsilon$ Sco  & 1 &$<$2.0         &$<$10.74          & \\
158926 & $\lambda$ Sco   & 1 &$<$2.0         &$<$10.74          & \\
159561 & $\alpha$ Oph    & 1 &   5.3$\pm$0.9 &   11.17$\pm$0.08 & \\
       &                 & 13&   4.6$\pm$0.4 &   11.11$\pm$0.03 & 11.12$\pm$0.02 \\         
161056 &                 & 16&  20.9$\pm$0.7 &   11.78$\pm$0.02 & \\
163522 &                 & 23& 178.0$\pm$3.0 &  (12.69$\pm$0.02)& 12.76$\pm$0.02 \\
163745 &                 & 14&  21.1$\pm$2.9 &   11.77$\pm$0.03 & 11.81$\pm$0.08 \\
163758 &                 & 14& 114.1$\pm$8.1 &   12.56$\pm$0.02 & 12.56$\pm$0.05 \\
163800 &                 & 14&  36.9$\pm$3.2 &   12.05$\pm$0.02 & 12.04$\pm$0.05 \\
164353 & 67 Oph          & 6 &  21.0$\pm$0.9 &   11.78$\pm$0.02 & \\
164794 & 9 Sgr           & 3 &  17.8$\pm$2.0 &  (11.69$\pm$0.05)& 11.83$\pm$0.07 \\
       &                 & 15&  34.0$\pm$x.x &   11.99$\pm$0.xx & \\
       &                 & 14&  41.5$\pm$3.4 &   12.10$\pm$0.08 & 12.08$\pm$0.06 \\
165024 & $\theta$ Ara    & 3 &  14.2$\pm$0.8 &  (11.59$\pm$0.03)& 11.64$\pm$0.05 \\
166182 & 102 Her         & 1 &  17.0$\pm$3.0 &   11.67$\pm$0.08 & \\
166734 &                 & 16&  87.0$\pm$0.9 &   12.43$\pm$0.01 & \\
166937 & $\mu$ Sgr       & 1 &  33.0$\pm$3.0 &   11.99$\pm$0.06 & \\
167264 & 15 Sgr          & 6 &  20.0$\pm$1.3 &   11.77$\pm$0.03 & \\
       &                 & 14&  31.1$\pm$3.0 &   11.98$\pm$0.03 & 11.98$\pm$0.07 \\
167971 &                 & 16&  71.0$\pm$0.9 &   12.33$\pm$0.01 & \\
\multicolumn{2}{l}{BD$-$14~5037}& 16&  98.5$\pm$0.6 &   12.50$\pm$0.01 & \\
169454 &                 & 16& 105.2$\pm$1.0 &   12.53$\pm$0.01 & \\
       &                 & 14& 106.6$\pm$4.0 &   12.53$\pm$0.01 & 12.54$\pm$0.03 \\ 
170235 &                 & 14&  37.6$\pm$4.4 &   12.07$\pm$0.03 & 12.06$\pm$0.07 \\
\hline
\end{tabular}
\end{table}

\begin{table}
\contcaption{}
\begin{tabular}{@{}rlrrcc}
\hline
\multicolumn{1}{l}{HD}&
\multicolumn{1}{l}{Name}&
\multicolumn{1}{l}{Ref}&
\multicolumn{1}{c}{W$_{\lambda}$}&
\multicolumn{1}{c}{$N$(AOD)}&
\multicolumn{1}{c}{$N$(fit)}\\
\hline
170740 &                 & 16&  16.0$\pm$0.5 &   11.66$\pm$0.02 & \\
171432 &                 & 14&  55.6$\pm$3.0 &   12.23$\pm$0.04 & 12.23$\pm$0.04 \\
172028 &                 & 16&  23.8$\pm$0.7 &   11.84$\pm$0.02 & \\
173948 & $\lambda$ Pav   & 3 &  11.5$\pm$1.5 &  (11.50$\pm$0.06)& 11.57$\pm$0.07 \\
174638 & $\beta$ Lyr     & 1 &  13.0$\pm$3.0 &   11.58$\pm$0.09 & \\
175191 & $\sigma$ Sgr    & 1 &   1.6$\pm$0.3 &   10.65$\pm$0.07 & \\
176162 &                 & 3 &  22.6$\pm$1.8 &  (11.79$\pm$0.03)& 11.84$\pm$0.12 \\
179406 & 20 Aql          & 3 &  28.2$\pm$2.7 &  (11.89$\pm$0.04)& 12.93$\pm$0.xx \\
179407 &                 & 23&  83.0$\pm$4.0 &  (12.36$\pm$0.02)& 12.41$\pm$0.02 \\
182985 &                 & 16&   4.9$\pm$0.4 &   11.15$\pm$0.03 & \\
183143 &                 & 16&  67.4$\pm$0.9 &   12.31$\pm$0.01 & \\
184915 & $\kappa$ Aql    & 2 &  13.5$\pm$0.4 &  (11.57$\pm$0.01)& \\
       &                 & 3 &   5.4$\pm$0.8 &  (11.17$\pm$0.07)& 11.19$\pm$0.07 \\
186882 & $\delta$ Cyg    & 1 &   1.8$\pm$0.9 &   10.70$\pm$0.19 & \\
188209 &                 & 15&  54.0$\pm$x.x &   12.23$\pm$0.xx & \\
188220 &                 & 16&   5.9$\pm$0.4 &   11.22$\pm$0.02 & \\
188294 &                 & 14&   7.7$\pm$1.4 &   11.28$\pm$0.06 & 11.34$\pm$0.11 \\
189103 & $\theta^1$ Sgr  & 3 &   9.0$\pm$1.0 &  (11.39$\pm$0.05)& 11.30$\pm$0.06 \\
191877 &                 & 17&   --          &   12.24$\pm$0.02 & \\
193924 & $\alpha$ Pav    & 3 &$<$2.2         &$<$10.78          & \\
195455 &                 & 23&  49.0$\pm$3.0 &  (12.13$\pm$0.02)& 12.16$\pm$0.02 \\
       &                 & 7 &  28.0$\pm$5.0 &  (11.89$\pm$0.08)& 11.93$\pm$0.08 \\
195965 &                 & 17&   --          &   12.02$\pm$0.02 & \\
197345 & $\alpha$ Cyg    & 13&  15.4$\pm$0.7 &   11.66$\pm$0.02 & 11.68$\pm$0.01 \\
199081 & 57 Cyg          & 1 &   5.4$\pm$1.7 &   11.21$\pm$0.13 & \\
199579 &                 & 9 &  17.9$\pm$2.1 &   11.68$\pm$0.05 & \\         
200120 & 59 Cyg          & 1 &  12.0$\pm$3.0 &   11.55$\pm$0.11 & \\
       &                 & 6 &   7.0$\pm$0.8 &   11.28$\pm$0.05 & \\
202904 & $\upsilon$ Cyg  & 1 &   7.9$\pm$1.6 &   11.35$\pm$0.09 & \\
203064 & 68 Cyg          & 2 &  26.5$\pm$0.4 &  (11.86$\pm$0.01)& \\
       &                 & 6 &  27.0$\pm$0.9 &   11.89$\pm$0.01 & \\
203664 &                 & 5 &  40.0$\pm$4.0 &  (12.04$\pm$0.04)& 12.04$\pm$0.04 \\
203699 &                 & 5 &  46.0$\pm$7.0 &  (12.10$\pm$0.07)& 12.13$\pm$0.07 \\
204076 &                 & 7 &  31.0$\pm$5.0 &  (11.93$\pm$0.07)& 11.90$\pm$0.07 \\
204862 &                 & 5 &   8.0$\pm$2.0 &  (11.34$\pm$0.12)& 11.41$\pm$0.12 \\
205637 & $\epsilon$ Cap  & 16&  13.4$\pm$0.4 &   11.58$\pm$0.02 & \\
206144 &                 & 23&  51.0$\pm$4.0 &  (12.15$\pm$0.03)& 12.20$\pm$0.03 \\
       &                 & 7 &  28.0$\pm$14. &  (11.89$\pm$0.23)& 11.93$\pm$0.23 \\
206165 & 9 Cep           & 24&  36.0$\pm$x.x &   11.90$\pm$0.xx & \\
206267 &                 & 24&   9.0$\pm$x.x &   11.62$\pm$0.xx & \\
       &                 & 9 &  40.2$\pm$2.2 &   12.09$\pm$0.02 & \\
206773 &                 & 18&   --          &   --             & 12.16$\pm$0.02 \\
207198 &                 & 9 &  47.1$\pm$2.9 &   12.16$\pm$0.03 & \\  
207330 & $\pi^2$ Cyg     & 1 &  11.0$\pm$2.0 &   11.51$\pm$0.07 & \\
209008 & 18 Peg          & 22&  22.0$\pm$4.0 &  (11.78$\pm$0.08)& 11.78$\pm$0.08 \\
209339 &                 & 18&   --          &   --             & 12.14$\pm$0.03 \\
209522 &                 & 5 &$<$3.0         &$<$10.94          & \\
209684 &                 & 23&  51.0$\pm$4.0 &  (12.15$\pm$0.03)& 12.19$\pm$0.03 \\
209975 & 19 Cep          & 6 &  48.0$\pm$1.0 &   12.16$\pm$0.01 & \\
210121 &                 & 2 &  26.9$\pm$x.x &  (11.90$\pm$0.xx)& \\
       &                 & 16&  26.7$\pm$0.6 &   11.89$\pm$0.01 & 11.89$\pm$0.01 \\
210191 & 35 Aqr          & 5 &  34.0$\pm$6.0 &  (11.97$\pm$0.08)& 12.00$\pm$0.08 \\
210839 & $\lambda$ Cep   & 1 &  46.0$\pm$3.0 &   12.14$\pm$0.06 & \\
       &                 & 24&  55.0$\pm$x.x &   12.09$\pm$0.xx & \\
212076 & 31 Peg          & 22&  22.0$\pm$3.0 &  (11.78$\pm$0.06)& 11.81$\pm$0.07 \\
212571 & $\pi$ Aqr       & 22&  19.0$\pm$3.0 &  (11.72$\pm$0.07)& 11.71$\pm$0.07 \\
       &                 & 16&  17.0$\pm$0.3 &   11.68$\pm$0.01 & 11.69$\pm$0.05 \\
213236 &                 & 5 &$<$3.0         &$<$10.94          & \\
213420 & 6 Lac           & 1 &  23.0$\pm$3.0 &   11.82$\pm$0.08 & \\
214080 &                 & 22&  34.0$\pm$6.0 &  (11.97$\pm$0.08)& 12.03$\pm$0.07 \\
214680 & 10 Lac          & 1 &  19.0$\pm$2.0 &   11.74$\pm$0.06 & \\
214993 & 12 Lac          & 6 &  25.0$\pm$0.6 &   11.84$\pm$0.01 & \\
214930 &                 & 22&  44.0$\pm$7.0 &  (12.08$\pm$0.07)& 12.06$\pm$0.07 \\
\hline
\end{tabular}
\end{table}

\begin{table}
\contcaption{}
\begin{tabular}{@{}rlrrcc}
\hline
\multicolumn{1}{l}{HD}&
\multicolumn{1}{l}{Name}&
\multicolumn{1}{l}{Ref}&
\multicolumn{1}{c}{W$_{\lambda}$}&
\multicolumn{1}{c}{$N$(AOD)}&
\multicolumn{1}{c}{$N$(fit)}\\
\hline
215733 &                 & 22&  83.0$\pm$14. &  (12.36$\pm$0.07)& 12.35$\pm$0.07 \\
       &                 & 15&  52.0$\pm$x.x &   12.16$\pm$0.xx & \\
217675 & o And           & 1 &   5.0$\pm$1.1 &   11.14$\pm$0.08 & \\
       &                 & 13&   5.6$\pm$0.5 &   11.20$\pm$0.04 & 11.23$\pm$0.03 \\        
218376 & 1 Cas           & 1 &  23.0$\pm$3.0 &   11.83$\pm$0.07 & \\
       &                 & 2 &  20.0$\pm$0.5 &  (11.74$\pm$0.01)& \\    
218624 &                 & 2 &   7.0$\pm$x.x &  (11.30$\pm$0.xx)& \\
218700 & 58 Peg          & 22&   7.2$\pm$6.1 &  (11.30$\pm$0.55)& 11.29$\pm$0.56 \\
219188 &                 & 22&  25.0$\pm$4.0 &  (11.84$\pm$0.07)& 11.86$\pm$0.07 \\
219688 & $\psi^2$ Aqr    & 22&   3.5$\pm$1.8 &  (10.98$\pm$0.25)& 10.85$\pm$0.42 \\
       &                 & 16&   2.1$\pm$0.6 &   10.76$\pm$0.10 & \\
219832 &                 & 5 &$<$3.0         &$<$10.94          & \\
220172 &                 & 22&  43.0$\pm$7.0 &  (12.07$\pm$0.07)& 12.10$\pm$0.07 \\
220787 &                 & 5 &  15.0$\pm$5.0 &  (11.62$\pm$0.13)& 11.65$\pm$0.13 \\
       &                 & 23&  26.0$\pm$3.0 &  (11.85$\pm$0.05)& 11.88$\pm$0.05 \\
224572 & $\sigma$ Cas    & 2 &  14.5$\pm$0.3 &  (11.60$\pm$0.01)& \\
       &                 & 6 &  11.0$\pm$0.7 &   11.50$\pm$0.03 & \\
224990 & $\zeta$ Sc1     & 3 &$<$1.7         &$<$10.67          & \\
233622 &                 & 23&  27.0$\pm$3.0 &  (11.87$\pm$0.05)& 11.89$\pm$0.05 \\
       & JL 9            & 19&   --          &   --             & 11.88$\pm$0.10 \\
       & JL 212          & 23&  42.0$\pm$5.0 &  (12.06$\pm$0.06)& 12.09$\pm$0.06 \\
       & Feige 110       & 17&   --          &   11.59$\pm$0.03 & \\
       & LS 1274         & 19&   --          &   --             & 11.68$\pm$0.04 \\
\multicolumn{2}{l}{BD+28 4211}& 17&    --         &   11.08$\pm$0.08 & \\
\multicolumn{2}{l}{BD+39 3226}& 18&    --         &   --             & 11.49$\pm$0.08 \\
\multicolumn{2}{l}{WD1034+001}& 18&    --         &   --             & 11.10$\pm$0.20 \\
\hline
\end{tabular}
\medskip
~ ~ \\
Uncertainties (where available) are 1-$\sigma$; limits are 3-$\sigma$ (see text for details and caveats); uncertainties are given by $\pm$x.x or $\pm$0.xx where no values were given in the original references; $N$(AOD) in parentheses are derived from the equivalent widths, assuming the lines to be optically thin.
All column densities have been scaled to $f$(3383) = 0.358 (Morton 2003).\\
References: \\
1 = Stokes 1978 (3.7--5.5 km~s$^{-1}$; AOD) \\ 
2 = Magnani \& Salzer 1989, 1991 (12.4 km~s$^{-1}$) \\ 
3 = Welsh et al. 1997 (4.5 km~s$^{-1}$; fit)\\ 
4 = Hobbs 1983 \\ 
5 = Albert et al. 1993 (5.0 km~s$^{-1}$; fit) \\ 
6 = Hobbs 1984 (5.6 km~s$^{-1}$; AOD) \\ 
7 = Albert et al. 1994 (5.2 km~s$^{-1}$; fit) \\ 
8 = Chaffee 1974 \\ 
9 = new Kitt Peak coud\'{e} feed (3.4 km~s$^{-1}$; AOD, fit) \\ 
10 = Hobbs 1979 (1.4 km~s$^{-1}$) \\ 
11 = Meyer \& Roth 1991 \\ 
12 = Chaffee \& Lutz 1977 \\ 
13 = new Kitt Peak coud\'{e} feed (1.3--1.5 km~s$^{-1}$; AOD, fit) \\ 
14 = Hunter et al. 2006 (3.75 km~s$^{-1}$; AOD, fit) \\ 
15 = Wallerstein \& Goldsmith 1974 \\ 
16 = new ESO/VLT UVES (3.8--4.5 km~s$^{-1}$; AOD, fit) \\ 
17 = Prochaska et al. 2005 (4.5--6.0 km~s$^{-1}$; AOD) \\ 
18 = Ellison et al. 2007 (3.0--6.7 km~s$^{-1}$; fit) \\ 
19 = Lallement et al. 2008 (5.0 km~s$^{-1}$; fit) \\ 
20 = Wallerstein \& Gilroy 1992 \\ 
21 = Dunkin \& Crawford 1999 (0.32 km~s$^{-1}$) \\ 
22 = Albert 1983 (5.5 km~s$^{-1}$; fit) \\ 
23 = Lipman \& Pettini 1995 (5.3--6.6 km~s$^{-1}$; fit) \\  
24 = Chaffee \& Dunham 1979 \\ 
\end{table}

\clearpage

\begin{table*}
\begin{minipage}{160mm}
\caption{QSOALS \mbox{Ti\,{\sc ii}} column densities.}
\label{tab:ti2qso}
\begin{tabular}{@{}llrrrrrrrc}
\hline
\multicolumn{1}{l}{QSO}&
\multicolumn{1}{c}{$z$}&
\multicolumn{1}{c}{\mbox{H\,{\sc i}}}&
\multicolumn{1}{c}{H$_2$}&
\multicolumn{1}{c}{\mbox{Zn\,{\sc ii}}}&
\multicolumn{1}{c}{[Zn/H]}&
\multicolumn{1}{c}{\mbox{Ca\,{\sc ii}}}&
\multicolumn{1}{c}{\mbox{Ti\,{\sc ii}}}&
\multicolumn{1}{c}{[Ti/Zn]}&
\multicolumn{1}{c}{Ref}\\
\hline
Q0005$+$0524 & 0.8514 &    19.08 &     --   & $<$11.24 &$<-$0.47 &     --   & $<$11.03 &    --   & 1\\
Q0012$-$0122 & 1.3862 &    20.26 &     --   & $<$11.55 &$<-$1.34 &     --   & $<$11.96 &    --   & 1\\
Q0014$+$813  & 1.11   &     --   &     --   &    12.83 &    --   &     --   &    12.36 & $-$0.76 & 2\\
Q0027$-$1836 & 2.402  &    21.75 &    17.30 &    12.79 & $-$1.59 &     --   &    12.61 & $-$0.47 & 3,4\\
Q0058$+$019  & 0.612  &    20.14 &     --   &    12.83 &   +0.06 &     --   &    12.51 & $-$0.61 & 2\\
Q0100$+$130  & 2.3090 &    21.37 &     --   &    12.45 & $-$1.55 &     --   & $<$12.21 &$<-$0.53 & 5,6\\
Q0118$-$272  & 0.558  &     --   &     --   &     --   &    --   &    12.37 &    12.27 &    --   & 2\\
P0133$+$0400 & 3.7736 &    20.55 &     --   & $<$13.10 &$<-$0.08 &     --   &    13.04 &$>-$0.35 & 7\\
Q0138$-$0005 & 0.7821 &    19.81 &     --   &    12.80 &   +0.36 &     --   & $<$11.48 &$<-$1.61 & 8\\
Q0149$+$33   & 2.1408 &    20.50 &     --   &    11.48 & $-$1.65 &     --   & $<$12.17 &$<+$0.40 & 5,6\\
Q0153$+$0009 & 0.7714 &    19.70 &     --   & $<$11.96 &$<-$0.37 &     --   &    12.02 &$>-$0.23 & 8\\
Q0201$+$365  & 2.4628 &    20.38 &     --   &    12.76 & $-$0.25 &     --   & $<$12.19 &$<-$0.86 & 5,6\\
J0255$+$00   & 3.2529 &    20.70 &     --   &     --   &    --   &     --   & $<$12.81 &    --   & 6\\
J0256$+$0110 & 0.725  &    20.70 &     --   &    13.19 & $-$0.14 &     --   &    12.27 & $-$1.21 & 9\\
J0334$-$0711 & 0.5976 &     --   &     --   &    12.58 &    --   &     --   &    11.90 & $-$0.97 & 10\\
Q0347$-$382  & 3.0247 &    20.73 &    14.53 &    11.91 & $-$1.45 &     --   & $<$12.20 &$<+$0.00 & 3,11\\
Q0427$-$1302 & 1.5623 &    18.90 &     --   & $<$11.58 &$<+$0.05 &     --   & $<$12.47 &    --   & 12\\
Q0449$-$1645 & 1.0072 &    20.98 &     --   &    12.47 & $-$1.14 &     --   & $<$11.47 &$<-$1.29 & 8\\
Q0453$-$423  & 1.150  &     --   &     --   & $<$12.63 &    --   &     --   & $<$13.11 &    --   & 2\\
Q0454$+$039  & 0.8598 &    20.69 &     --   &    12.45 & $-$0.87 &     --   &    12.66 & $-$0.08 & 2\\
Q0458$-$020  & 2.0395 &    21.65 &     --   &    13.13 & $-$1.15 &     --   & $<$12.49 &$<-$0.93 & 5,6\\
HE0515$-$4414& 1.151  &    19.88 &    16.94 &    12.11 & $-$0.40 &    12.72 & $<$11.70 &$<-$0.70 & 13,14,15\\
J0517$-$441  & 1.1496 &     --   &     --   &    12.22 &    --   &    12.74 & $<$10.60 &$<-$1.91 & 10\\
Q0528$-$251  & 2.8110 &    21.35 &    18.22 &    13.09 & $-$0.89 &     --   & $<$13.95 &$<+$0.57 & 3,16\\
Q0551$-$364  & 1.962  &    20.70 &    17.42 &    13.02 & $-$0.31 &     --   &    12.32 & $-$0.99 & 3,5,2\\
Q0738$+$313  & 0.0912 &    21.18 &     --   & $<$12.66 &$<-$1.15 &    12.32 &    12.53 &$>-$0.42 & 17,18\\
Q0738$+$313  & 0.2210 &    20.90 &     --   & $<$12.83 &$<-$0.70 &    11.91 & $<$11.48 &    --   & 18\\   
F0812$+$32   & 2.0668 &    21.00 &     --   &    12.21 & $-$1.42 &     --   &    12.53 &   +0.03 & 5\\
F0812$+$32   & 2.6263 &    21.35 & $>$19.88 &    13.15 & $-$0.83 &     --   &    12.37 & $-$1.07 & 5,7,19\\ 
Q0826$-$2230 & 0.9110 &    19.04 &     --   &    12.35 &   +0.68 &    11.75 & $<$11.58 &$<-$1.06 & 12\\
Q0827$+$243  & 0.2590 &     --   &     --   &     --   &    --   & $<$11.60 & $<$11.96 &    --   & 18\\
Q0827$+$243  & 0.5247 &    20.30 &     --   & $<$12.80 &$<-$0.13 &     --   & $<$11.76 &    --   & 18\\
Q0836$+$113  & 2.4651 &    20.58 &     --   & $<$12.12 &$<-$1.09 &     --   & $<$12.54 &    --   & 6\\
Q0841$+$129  & 2.4762 &    20.78 &     --   &    12.05 & $-$1.36 &     --   & $<$12.16 &$<-$0.18 & 6\\
J0846$+$0529 & 0.7429 &     --   &     --   & $<$12.80 &    --   &    13.06 &    13.00 &$>-$0.09 & 10\\
Q0933$+$733  & 1.4790 &    21.62 &     --   &    12.67 & $-$1.58 &     --   &    12.85 & $-$0.11 & 17\\
Q0935$+$417  & 1.3726 &    20.52 &     --   &    12.25 & $-$0.90 &     --   &    12.42 & $-$0.12 & 2\\
J0953$+$0801 & 1.0232 &     --   &     --   &    12.25 &    --   &    13.57 &    12.10 & $-$0.44 & 10\\
J1005$+$1157 & 0.8346 &     --   &     --   & $<$12.80 &    --   &     --   & $<$12.40 &    --   & 10\\
J1007$+$2853 & 0.8839 &     --   &     --   &    13.49 &    --   &    13.33 &    13.13 & $-$0.65 & 20\\
Q1009$-$0026 & 0.8426 &    20.20 &     --   & $<$11.85 &$<-$0.98 &     --   &    11.85 &$>-$0.29 & 12\\
Q1009$-$0026 & 0.8866 &    19.48 &     --   &    12.36 &   +0.25 &    12.26 & $<$11.64 &$<-$1.01 & 12\\
Q1054$-$0020 & 0.9514 &    19.28 &     --   & $<$11.70 &$<-$0.21 &     --   & $<$12.36 &    --   & 21\\
Q1104$-$181  & 1.662  &    20.85 &     --   &    12.48 & $-$1.00 &     --   &    12.20 & $-$0.57 & 2,5\\
Q1107$+$0003 & 0.9547 &    20.26 &     --   & $<$12.08 &$<-$0.81 &     --   & $<$13.01 &    --   & 18\\
J1107$+$0048 & 0.7403 &    21.00 &     --   &    13.23 & $-$0.40 &     --   &    13.12 & $-$0.40 & 9,10\\ 
Q1122$-$168  & 0.682  &    20.45 &     --   & $<$11.76 &$<-$1.32 &     --   &    11.56 &$>-$0.49 & 2\\
J1129$+$0204 & 0.9650 &     --   &     --   &    12.80 &    --   &    13.11 &    12.79 & $-$0.30 & 10\\
Q1157$+$014  & 1.944  &    21.70 &     --   &    12.99 & $-$1.34 &     --   &    12.83 & $-$0.45 & 22\\
J1203$+$1028 & 0.7463 &     --   &     --   &    12.63 &    --   &    13.05 & $<$11.20 &$<-$1.72 & 10\\
Q1209$+$0919 & 2.5841 &    21.40 &     --   &    12.98 & $-$1.05 &     --   & $<$12.65 &$<-$0.62 & 5,7\\
Q1210$+$17   & 1.8918 &    20.60 &     --   &    12.40 & $-$0.83 &     --   &    12.34 & $-$0.35 & 6,22\\
Q1220$-$0040 & 0.9746 &    20.20 &     --   & $<$11.69 &$<-$1.14 &     --   & $<$12.47 &    --   & 21\\
Q1223$+$175  & 2.4661 &    21.50 &     --   &    12.55 & $-$1.58 &     --   & $<$12.25 &$<-$0.59 & 5,6\\
Q1228$+$1018 & 0.9376 &    19.41 &     --   & $<$11.67 &$<-$0.37 &     --   & $<$11.65 &    --   & 21\\
P1253$-$0228 & 2.7828 &    21.85 &     --   &    12.85 & $-$1.63 &     --   & $<$12.84 &$<-$0.30 & 5,7\\
J1323$-$0021 & 0.716  &    20.21 &     --   &    13.43 &   +0.59 &     --   &    12.49 & $-$1.23 & 23\\
Q1330$-$2056 & 0.8526 &    19.40 &     --   & $<$11.96 &$<-$0.07 &     --   & $<$11.45 &    --   & 21\\
Q1331$+$170  & 1.7765 &    21.18 &    19.65 &    12.54 & $-$1.27 &     --   & $<$11.62 &$<-$1.21 & 24,5,25\\
J1430$+$0149 & 1.2418 &     --   &     --   &    12.94 &    --   &    12.85 &    12.79 & $-$0.44 & 10\\
Q1436$-$0051 & 0.7377 &    20.08 &     --   &    12.74 &   +0.03 &    12.83 & $<$11.71 &$<-$1.32 & 21\\
Q1436$-$0051 & 0.9281 & $<$18.80 &     --   &    12.26 &$>+$0.83 &    12.28 & $<$12.71 &$<+$0.16 & 21\\
Q1455$-$0045 & 1.0929 &    20.08 &     --   & $<$11.91 &$<-$0.80 &     --   & $<$12.44 &    --   & 21\\
Q1622$+$238  & 0.656  &    20.36 &     --   &     --   &    --   &     --   &    12.35 &    --   & 2\\
\hline
\end{tabular}
\end{minipage}
\end{table*}

\begin{table*}
\begin{minipage}{160mm}
\contcaption{}
\begin{tabular}{@{}llrrrrrrrc}
\hline
\multicolumn{1}{l}{QSO}&
\multicolumn{1}{c}{$z$}&
\multicolumn{1}{c}{\mbox{H\,{\sc i}}}&
\multicolumn{1}{c}{H$_2$}&
\multicolumn{1}{c}{\mbox{Zn\,{\sc ii}}}&
\multicolumn{1}{c}{[Zn/H]}&
\multicolumn{1}{c}{\mbox{Ca\,{\sc ii}}}&
\multicolumn{1}{c}{\mbox{Ti\,{\sc ii}}}&
\multicolumn{1}{c}{[Ti/Zn]}&
\multicolumn{1}{c}{Ref}\\
\hline
Q1631$+$1156 & 0.9004 &    19.70 &     --   & $<$12.18 &$<-$0.15 &    12.17 & $<$11.66 &    --   & 1\\
J1727$+$5302 & 0.9449 &    21.16 &     --   &    13.27 & $-$0.52 &     --   &    12.90 & $-$0.66 & 17\\
J1727$+$5302 & 1.0341 &    21.41 &     --   &    12.65 & $-$1.39 &     --   & $<$12.92 &$<-$0.02 & 17\\
Q1946$+$766  & 1.7382 &     --   &     --   &    11.53 &    --   &     --   & $<$12.45 &$<+$0.63 & 16\\
Q2128$-$123  & 0.430  &    19.37 &     --   &     --   &    --   &     --   & $<$11.13 &    --   & 2\\
Q2206$-$199  & 0.752  &     --   &     --   &     --   &    --   &     --   &    12.20 &    --   & 2\\
Q2206$-$199  & 1.920  &    20.65 &     --   &    12.91 & $-$0.37 &     --   &    12.77 & $-$0.43 & 5,6\\
Q2230$+$025  & 1.8644 &    20.85 &     --   &    12.72 & $-$0.76 &     --   &    12.68 & $-$0.33 & 5,22\\
Q2231$-$002  & 2.0661 &    20.56 &     --   &    12.35 & $-$0.84 &     --   &    12.66 &   +0.02 & 5,26\\
J2240$-$0053 & 1.3606 &     --   &     --   &    12.62 &    --   &     --   & $<$12.71 &$<-$0.20 & 17\\
Q2318$-$1107 & 1.989  &    20.68 &    15.49 &    12.50 & $-$0.81 &     --   & $<$12.00 &$<-$0.79 & 3,4\\
Q2335$+$1501 & 0.6798 &    19.70 &     --   &    12.53 &   +0.20 &    12.41 & $<$11.36 &$<-$1.46 & 8\\
Q2343$+$123  & 2.4313 &    20.40 &    13.69 &    12.20 & $-$0.83 &     --   & $<$11.85 &$<-$0.64 & 3,4\\
Q2352$-$0028 & 0.8730 &    19.18 &     --   & $<$11.67 &$<-$0.14 & $<$11.01 & $<$11.36 &    --   & 1\\
Q2359$-$02   & 2.0951 &    20.70 &     --   &    12.60 & $-$0.73 &     --   &    12.33 & $-$0.56 & 5,6\\
 ~ & \\
WHP all      &   --   &     --   &     --   &    12.82 &    --   &    12.94 &    12.48 & $-$0.63 & 27\\
WHP high Ca  &   --   &     --   &     --   &    13.16 &    --   &    13.10 &    12.31 & $-$1.14 & 27\\
WHP low Ca   &   --   &     --   &     --   &    12.59 &    --   &    12.86 &    12.45 & $-$0.43 & 27\\
\hline
\end{tabular}
\medskip
~ ~ \\
References: 
1 = Meiring et al. 2009; 
2 = Ledoux et al. 2002 (and references therein);
3 = Noterdaeme et al. 2008; 
4 = Noterdaeme et al. 2007; 
5 = Prochaska et al. 2007; 
6 = Prochaska et al. 2001; 
7 = Prochaska et al. 2003a; 
8 = P\'{e}roux et al. 2008; 
9 = P\'{e}roux et al. 2006b; 
10 = Zych et al. 2009; 
11 = Levshakov et al. 2002; 
12 = Meiring 2009; 
13 = Reimers et al. 2003; 
14 = de la Varga et al. 2000; 
15 = Quast et al. 2008; 
16 = Lu et al. 1996; 
17 = Khare et al. 2004; 
18 = Meiring et al. 2006; 
19 = Jorgenson et al. 2009;
20 = Zhou et al. 2010;
21 = Meiring et al. 2008; 
22 = Dessauges-Zavadsky et al. 2006;
23 = P\'{e}roux et al. 2006a; 
24 = Cui et al. 2005; 
25 = Dessauges-Zavadsky et al. 2002; 
26 = Dessauges-Zavadsky et al. 2004; 
27 = Wild et al. 2006\\
\end{minipage}
\end{table*}

\end{document}